\documentclass[12pt]{iopart}
\usepackage{placeins}
\usepackage{xspace}
\usepackage{xcolor}
\usepackage{hyperref}
\usepackage{iopams}
\usepackage{graphicx}
\usepackage{booktabs}
\usepackage{dcolumn}
\usepackage{ragged2e}
\expandafter\let\csname equation*\endcsname\relax
\expandafter\let\csname endequation*\endcsname\relax

\usepackage{amsmath}
 \usepackage{caption}
\usepackage{subcaption}
\usepackage{float}
\usepackage{colortbl}
\usepackage{siunitx}

\newcommand{\spinunc}[1]{\ensuremath{{}^{+#1}_{-#1}}}

\makeatletter
\renewcommand\@makecaption[2]{%
  \vskip\abovecaptionskip
  \sbox\@tempboxa{\textbf{#1.} #2}%
  \ifdim \wd\@tempboxa >\hsize
    \textbf{#1.} #2\par
  \else
    \global \@minipagefalse
    \hb@xt@\hsize{\textbf{#1.} #2\hfil}%
  \fi
  \vskip\belowcaptionskip}
\makeatother

\begin{document}
\title{$t\bar{t}$ spin correlations from toponium production scenarios to non-relativistic resummation beyond the standard perturbative baseline}

\author{A. Cota Rodríguez$^{1}$,
Jesús A.V. Corral$^{2}$,
A. Paredes Sotelo$^{1}$,\\ J. A. Murillo Quijada$^{1}$}

\address{$^1$ Departamento de Investigaci\'{o}n en F\'{i}sica, Universidad de Sonora}%

\address{$^2$ Department of Physics and Astronomy, The University of Kansas}%

\begin{abstract}

{\justifying

$t\bar{t}$ spin correlations originating from theoretical models assuming on one side toponium bound state formation in proton-proton collisions at 13 TeV are compared against predictions from non-relativistic QCD ressumation effects via Green's function formalism that go beyond the perturbative $t\bar{t}$ baseline with no bound state but could justify extra yield of events in the $t\bar{t}$ mass spectrum recently observed by ATLAS and CMS Collaborations, still awaiting an explanation. Toponium production comprises two scenarios, a simplified Effective Field Theory (EFT) with $\eta_t\rightarrow t\bar{t}$ decays and an additional model with scalar and vector toponium states with $\eta_t\rightarrow HZ$ and $J_t\rightarrow ZW^+W^-$ decays. Experimental measurements on $t\bar{t}$ system are added to the comparisons. Fitted diagonal spin correlation C$_{kk}$, C$_{rr}$, C$_{nn}$ matrix terms show a relative increase in value ranging from 200\%-2330\% for toponium $t\bar{t}$ decays with respect the perturbative NLO baseline. Relative contribution from resummation effect ranges from 0.6\% to 10.4\% and its combination with perturbative baseline improves agreement with Data distributions, with $\chi^2/N_\mathrm{dof}$ value reduced by $\sim$4.5\%-60.0\% . Equivalent fitted cross-section for resummation effect $\sigma_{res}$, ranges from 7.3 to 9.1 pb in consistency with excess yield reported by LHC experiments. $\sigma_{res}$ $>$ 50 pb range is excluded at 95\% C.L. A classifier performing at detector level achieves a ROC curve area of $0.91$ leading to toponium signal to background significances over 55. These results indicate that efficient isolation of observed $t\bar{t}$ excess events from background is feasible and a characterisation via spin correlation measurements would reveal their consistency with either toponium production or non-relativistic effects beyond the perturbative baseline.\\


}

\end{abstract}

\pagestyle{plain}

\clearpage


\justifying




\section{Introduction}

\noindent As the top quark, the heaviest fundamental particle in the Standard Model of particle physics (SM), possesses a quite restricted lifetime $\tau_t \sim 5 \times 10^{-25}\,\mathrm{s}$ smaller than hadronization time scales~\cite{Bernreuther2008TopReview}, its appearance within bound states was not previously considered to emerge in the context of high energy proton-proton collisions at the Large Hadron Collider (LHC) at CERN European Laboratory. Such possibility has become a matter of high interest within the scientific community as a candidate explanation for recent data excess observation near the lower threshold limit of the $t\bar{t}$ spectra achieved by CMS and ATLAS Collaborations at CERN during 2025~\cite{CMS:2025kzt, ATLAS:2026nrx}. Such observations were reported within the $t\bar{t}$ dileptonic decay channel with a two lepton $l^+l^-$ system in final state and have raised speculation on its origin from a toponium ($t\bar{t}$) quasibound state. The data excess significance has been strengthened further by more recent observation probes during 2026 that make use of the $t\bar{t}$ semileptonic decay channel~\cite{CMS-PAS-TOP-25-002}, with a single lepton in the final state, confirming either an additional decay process apart from $t\bar{t}$ perturbative production baseline or a yet to be considered missing correction to the SM  prediction so such data excess could be modelled with success.\\ 

\noindent Formation of a toponium bound state would match the required mass scale $\sim$344 GeV, for the observed data excess within the $t\bar{t}$ invariant mass spectra, around twice the top quark mass value. If such production is confirmed the toponium bound state would be the most massive compound state ever observed and would provide a mean to increase the top quark mass measurement precision than currently reported~\cite{Fu:2025yft}. It is yet important to confirm whether such excess events are consistent with toponium theoretical models prospects. Recent statistical tests performed over published experimental data favour the toponium scenario further~\cite{Fuks:2025toq}, though there still could be chance for interpretation in terms of non-relativistic QCD effects (NRQCD),~\cite{Fuks:2024yjj, Fuks:2025sxu, Fuks:2025wtq, Maltoni:2024csn} associated to Coulomb resummed $t\bar{t}$ interactions and additional dynamic effects, that could distort the threshold mass spectrum producing resonance-like spikes emulating a particle resonance. Toponium related spin correlation and quantum information signatures are previously discussed in this reference~\cite{Aoude:2026ToponiumSpin}.\\ 

\begin{figure}[!ht]
    \centering

    \makebox[\textwidth][c]{%
    \begin{subfigure}{0.85\textwidth}
        \centering
        \includegraphics[width=14cm]{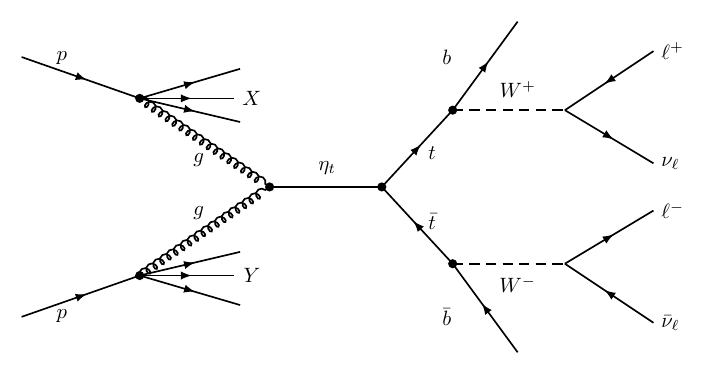}
        \caption{$gg \to \eta_t \to t\bar{t}$}
        \label{fig-toponium-eta}
    \end{subfigure}
    }

    \vspace{0.35cm}

    \begin{subfigure}{0.48\textwidth}
        \centering
        \includegraphics[width=\linewidth]{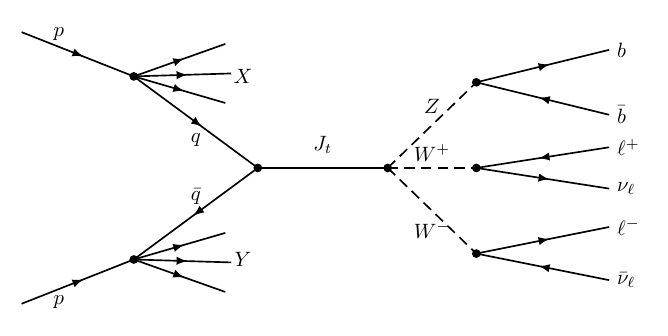}
        \caption{$q\bar{q} \to J_t \to ZW^-W^+$}
        \label{fig-toponium-Jt}
    \end{subfigure}
    \hfill
    \begin{subfigure}{0.48\textwidth}
        \centering
        \includegraphics[width=\linewidth]{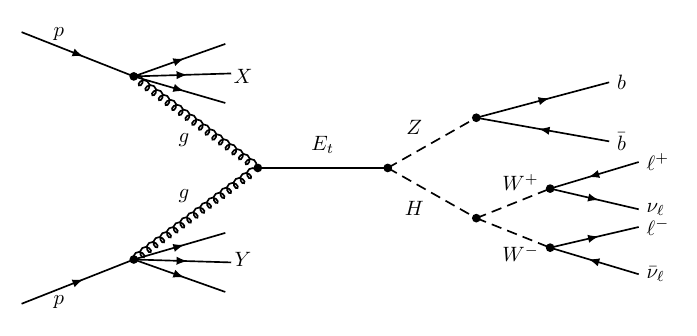}
        \caption{$gg \to E_t \to ZH$}
        \label{fig-toponium-Et}
    \end{subfigure}

    \caption{\justifying Leading order diagrams for toponium simplified model $\eta_t$ object production via color-singlet gluon fusion and decay to $t\bar{t}$ pair that subsequently decays in leptonic mode (top), and decays for scalar $\eta_t$ and vector $J_t$ states in the context of EFT model. }
    \label{fig-toponium-dia}
\end{figure}

\noindent Three independent theoretical models are used as reference for the present analysis. A first model makes use of a simplified Effective Field Theory (EFT) framework where a toponium field $\phi_{\eta_t}(x)$ is introduced with respective toponium couplings to top quark and gluon fields including NRQCD effects. Production happens with color-singlet gluon fusion as indicated in the top diagram in Figure~\ref{fig-toponium-dia} where $\eta_t$ decay to a $t\bar{t}$ pair is highlighted, producing a six-objects $bl^+\nu_ll^-\bar{\nu_l}\bar{b}$ dileptonic final state. Such events would resonate at near to the $m_{t\bar{t}}$ threshold limits. More details on this model are included in the following reference~\cite{Fuks:2021vtq}. \\

\noindent A second model includes scalar and pseudoscalar states $\eta_t(0^{-+})$ and vector states $J_t(1^{--})$ following a quarkonia approach that includes toponium mesons with spin J 0 and 1 and different polarization states.  Mesonic decays are not restricted to $t\bar{t}$ states within this model, but also final states that include Higgs (H), Z-boson (Z) and W-boson (W) heavy particles are included. Their resonances are not then suppressed by the 2$m_t$ lower limit, introducing another aspect to be inspected with precision at experimental level. Decays for scalar $\eta_t$ and vector $J_t$ states are shown at the bottom diagrams in Figure~\ref{fig-toponium-dia}. Since their final states also include a lepton $l^+l^-$ couple, a $b\bar{b}$ pair and neutrinos they would pass the event selections that are used for $t\bar{t}$ invariant mass studies. Mesonic states and model specifics are described with more detail in the following reference~\cite{Fu:2025toponium}.\\

\noindent Finally a third theoretical approach that makes use of Green's function formalism of non-relativistic QCD is introduced instead on top of the perturbative calculations with no toponium state formation. It makes consideration that extreme top quark properties make the bound state formation possibility non feasible scenario. It highlights the potential large contributions from non-relativistic corrections that would enhance the modelling capacity beyond the perturbative baseline capabilities. Details are described in the following reference~\cite{Sjostrand:2026TopPairThreshold}. Such formalism was first introduced from $e^+e^-$ to hadron colliders in the following references~\cite{Fadin:1987wz, Fadin:1988fn, Fadin:1989Moriond, Fadin:1990wx}.\\

\noindent In this work, aforementioned theoretical models addressing different alternatives to probe for data excess at near to the $m_{t\bar{t}}$ threshold region are implemented within Monte Carlo (MC), technique so resulting phenomenology features, cross-sections and decays are studied in order to identify discriminating observables against $t\bar{t}$ background events that dominate the lowest spectrum limits. A precise isolation would enable precise characterizations against toponium and NRQCD models.\\

\noindent Additionally, feasibility of characterizations via measurement of spin correlation observables from top and antitop quark pair products, is analysed. Spin correlation prospects from toponium decay $t\bar{t}$ objects are compared with those from data measurements with standard $t\bar{t}$ production events~\cite{CMS:2019nrx}  and those with NRQCD Green's function formalism effect added up so a complete perspective is set towards incoming Data characterisations and interpretations. Such comparison will determine if  angular probes would be effective to characterise the nature of the excess data compare to the standard perturbative $t\bar{t}$ baseline. Where a different behaviour would indicate that indeed such $t\bar{t}$ products are produced through a different mechanism, possibly from the decay of a new mesonic state. Spin correlation measurements on the other hand have not yet taken into account the NRQCD Green's function formalism effect, so we explore by how much the theoretical reference improves its agreement with Data measurements.\\ 

\noindent Finally, a machine learning training methodology using deep neural networks based on TensorFlow is produced and implemented seeking to identify the most discriminating observables to isolate the toponium events at the resonance more efficiently from the perturbative $t\bar{t}$ using detector level information.\\


\section{Monte Carlo production and cross section}

\noindent Monte Carlo packages \texttt{mg5\_amc} (MG),~\cite{Alwall:2014hca} and PYTHIA8~\cite{Bierlich:2022pfr} are implemented for signal and background samples production. A signal $\eta_t$ sample consistent with MC reference included in CMS observation paper~\cite{CMS:2025kzt} was produced, that makes use of first toponium model described in the introduction and will be regarded as `Simplified EFT $\eta_t$'. A second signal MC production was done implementing the second model with scalar and vector mesons, it will be regarded as `Scalar E$_t$' and `Vector J$_t$' states standing for newly introduced scalar and vector toponium mesons. Finally a $pp\rightarrow t\bar{t}$ production including NRQCD Green's function formalism effect was generated using PYTHIA8 that has just recently implemented all the formalism described in~\cite{Sjostrand:2026TopPairThreshold}. Apart from NR correction from Green's function formalism we also add a partial correction using also the Coulomb approach.\\

\begin{figure}[!ht]
    \centering
    \includegraphics[width=0.48\linewidth]{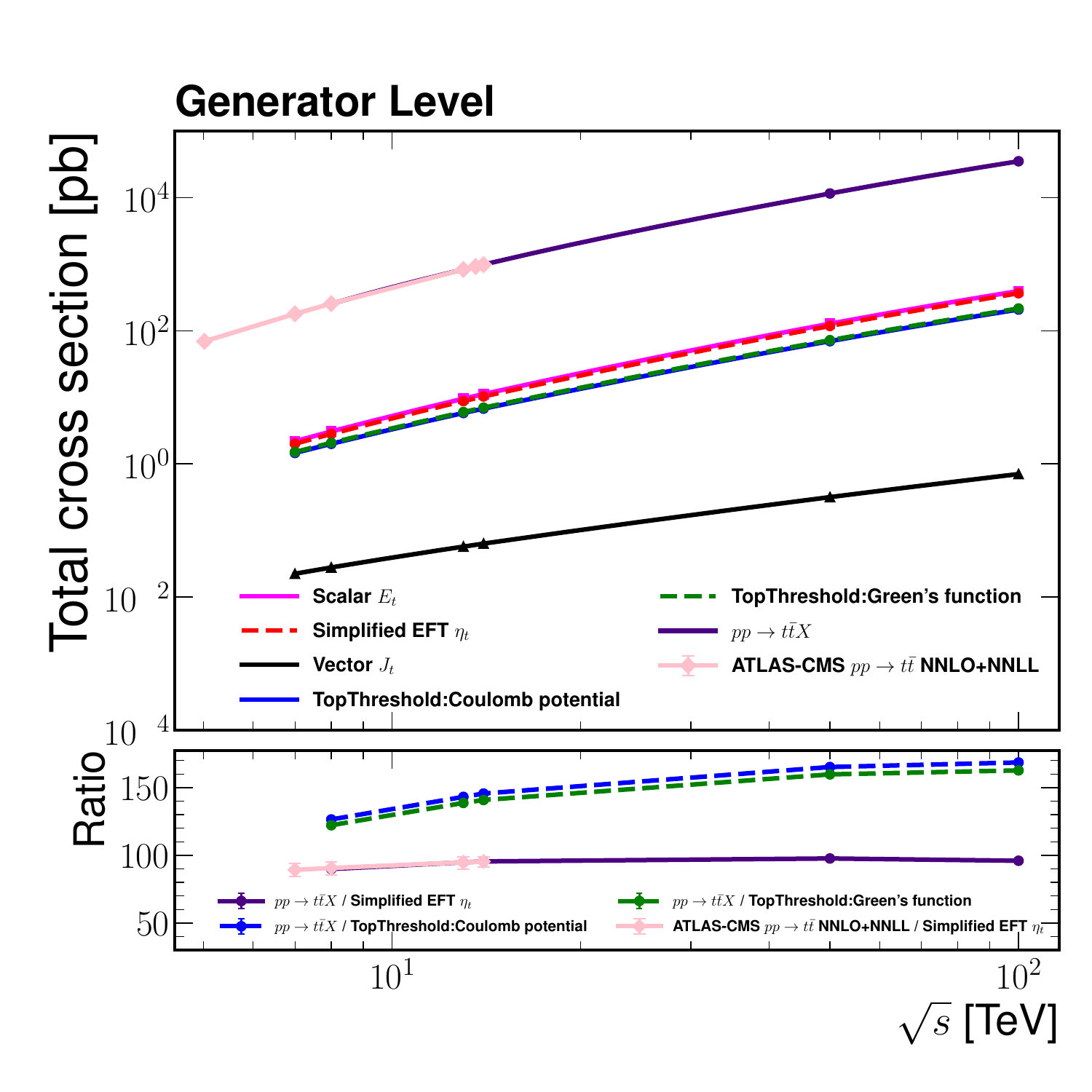}
        \includegraphics[width=0.48\linewidth]{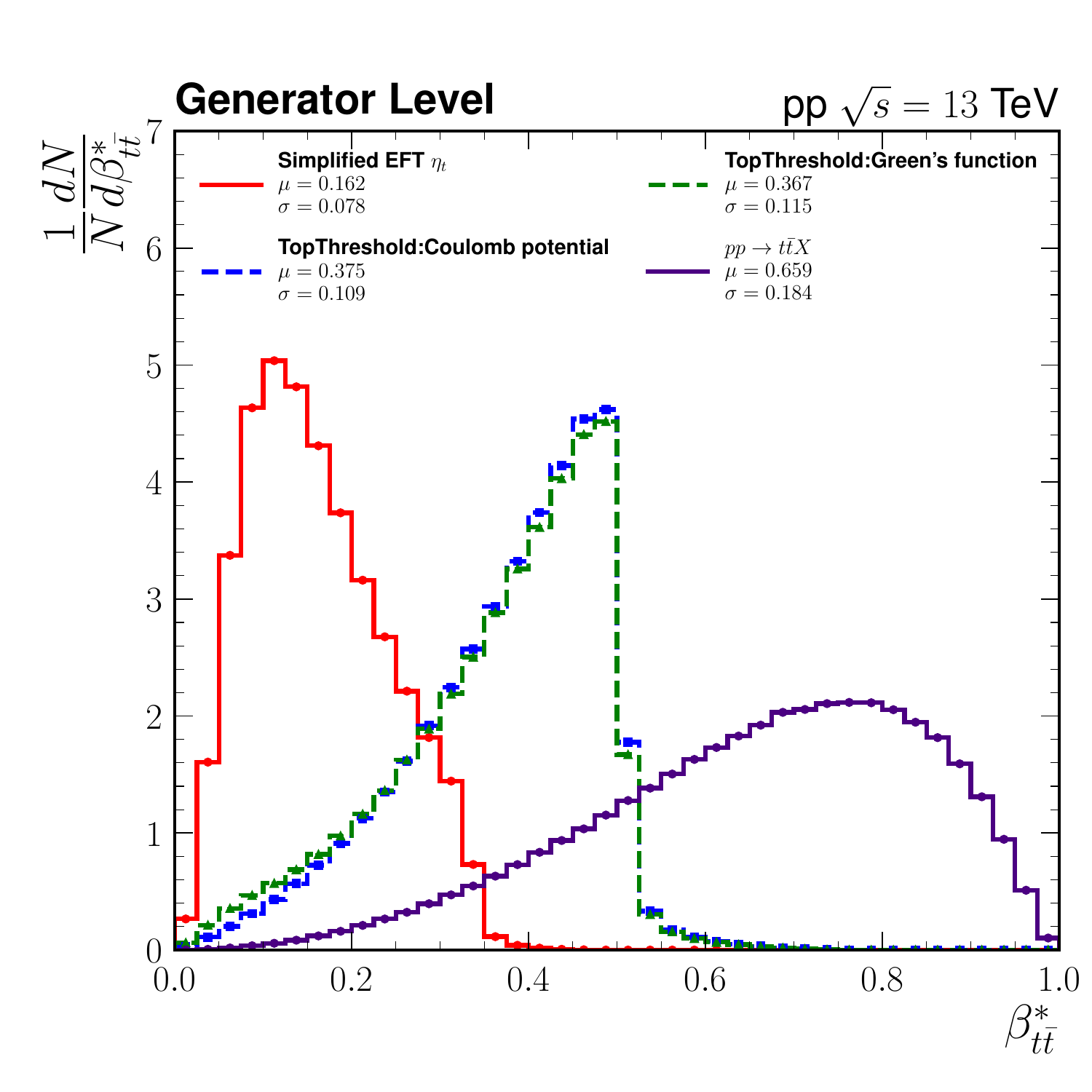}

    \caption{\justifying Production cross-section evolution as functions of centre-of-mass pp collision energy for $pp \rightarrow t\bar{t}$ and toponium production process, 100 TeV scenario is at the level of FCC accelerator scale (left) and $\beta^{*}_t$ variable for top quark in $t\bar{t}$ rest frame (right).}
    \label{fig_xs_vs_energy}
\end{figure}

\noindent Figure~\ref{fig_xs_vs_energy} shows at the left the dependence for inclusive production cross sections as functions of proton-proton collision energies for inclusive $pp \to t\bar t$ production and $\eta_t$ dileptonic production, covering the center of mass range $8 \leq \sqrt{s} \leq 100~\mathrm{TeV}$. $\eta_t$ production cross section increases from $0.408~\mathrm{pb}$ to $52.9~\mathrm{pb}$ over 8-100 TeV energy interval a factor about 130 times and about a factor of 137 times corresponding increase for inclusive $pp \to t\bar t$ production. Although the cross section values increase steeply with energy, their relative magnitudes remain stable at the per-mille level, $\sigma(\eta_t)/\sigma(t\bar t) \simeq (0.18\text{--}0.20)\%$, corresponding to a suppression factor $\sigma(t\bar t)/\sigma(\eta_t) \approx 5.4\times 10^{2}$ across the full energy range. A power-law fit of the form $\sigma \sim s^{n}$ yielded exponents $n_{t\bar t} \approx 1.93$ and $n_{\eta_t} \approx 1.91$, indicating that both processes exhibit nearly identical scaling behaviour with centre-of-mass energy. This stability of the ratio suggests that $\eta_t$ production follows the same underlying partonic luminosity growth as inclusive $t\bar t$ production, while remaining uniformly suppressed by approximately three orders of magnitude over the considered energy range.\\

\section{Toponium models phenomenology at generator level}

\noindent Signal $\eta_t$ and background $pp \rightarrow t\bar{t}$ (perturbative baseline) processes stochastic kinematic and angular features are studied at generator level. It is of particular interest to compare variables from $t\bar{t}$ pairs produced either from pp interaction or from $\eta_t$ state decay. A subsequent $bW^+W^-\bar{b}$ topology will be produced in both cases from their respective $t\bar{t}$ final states. Such final state originates all hadronic, semileptonic and dileptonic decay topologies depending on the W boson decay combinations. Figure~\ref{fig_xs_vs_energy} (right) shows $\beta_t^*=|\vec p_t^{\,*}|/E_t^*$, the top-quark speed in the $t\bar t$ rest frame. For equal on-shell masses, $\beta_t^*=\sqrt{1-4m_t^2/m_{t\bar t}^2}$. The concentration of the simplified toponium $\eta_t$ sample at $\beta_t^*<0.4$ characterizes near-threshold production opposed to the continuum $t\bar{t}$ background with back-to-back alignment closer to $\beta_t^*\sim1.0$  \cite{CMS-PAS-TOP-25-002}. Contribution from NRQCD effect from Green function and Coulomb potential methodologies center at an intermediate range in the Figure indicated with blue and green coloured lines with peak at $\beta_t^*\sim0.5$.\\

\begin{figure}[!h]
    \centering

    \includegraphics[width=0.46\linewidth]{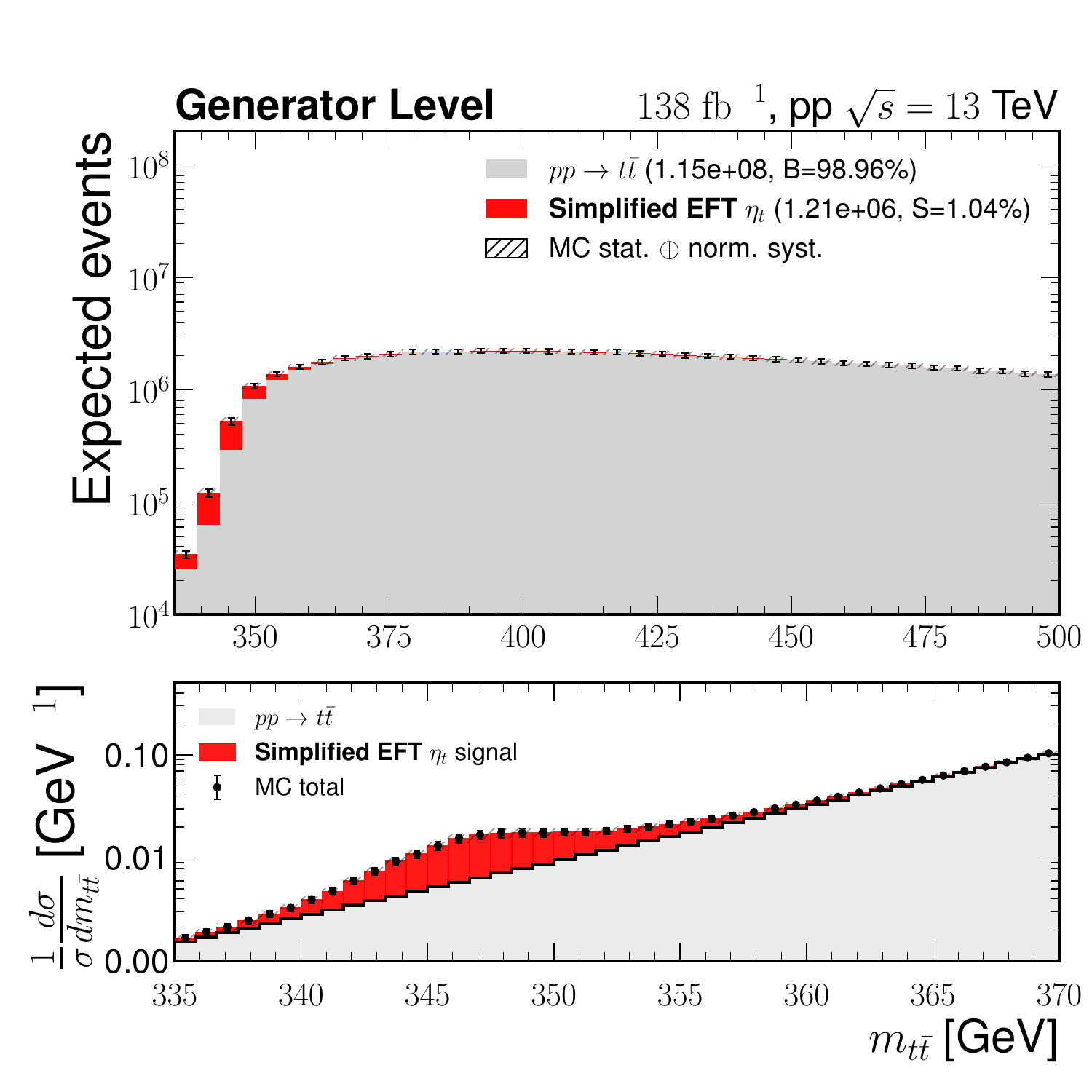}
    \includegraphics[width=0.46\linewidth]{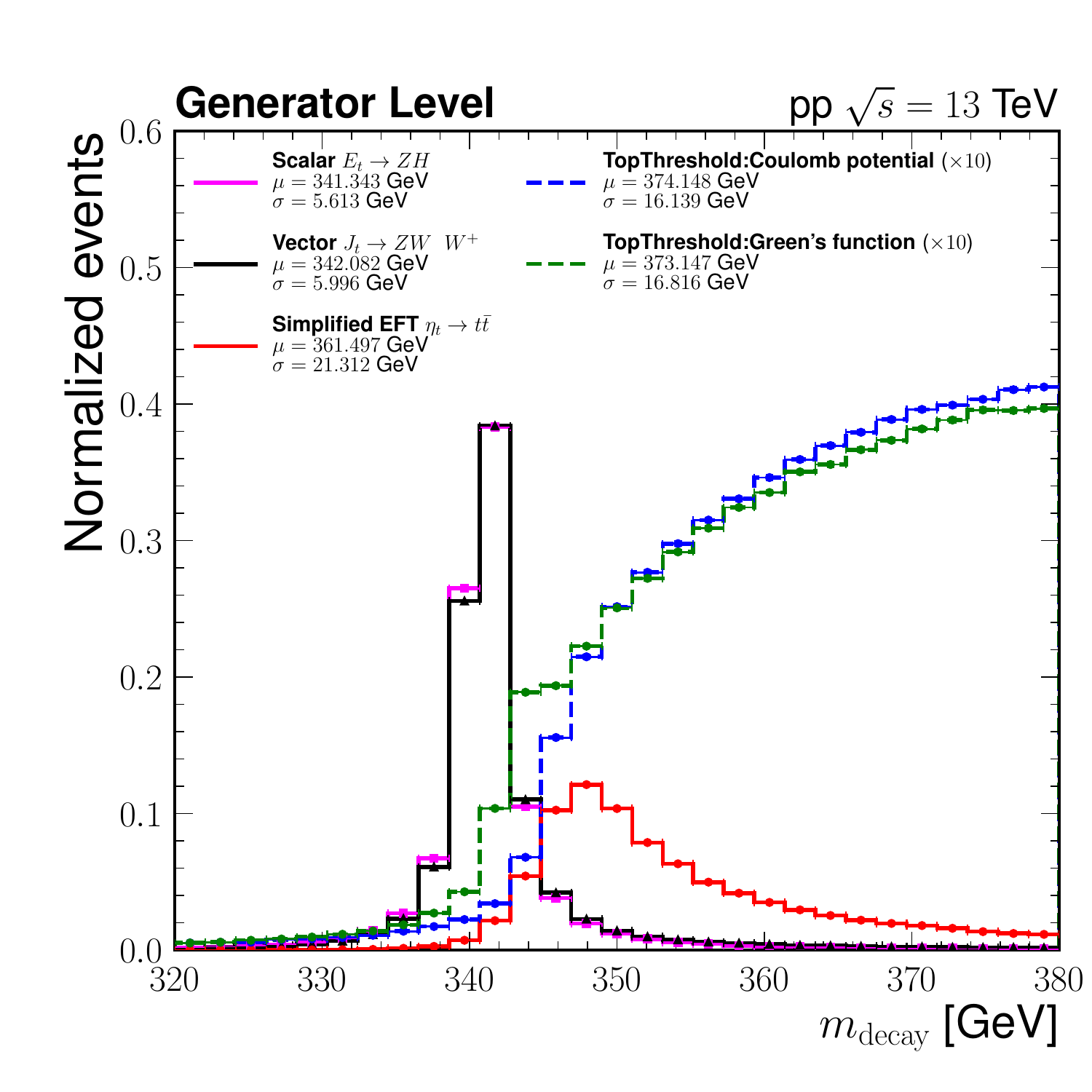}
       
    \caption{\justifying $\eta_t$ invariant mass and background $t\bar{t}$ production continuum at truth level (left) and $\eta_t$ compound state with decays to $t\bar{t}$ compared with $J_t$ and $E_t$ resonances that decay to ZH and ZWW products (right). 
    }
    \label{fig:gen_mtt_signal_background}
\end{figure}


\begin{figure*}[!ht]
    \centering
    \begin{subfigure}[t]{0.32\textwidth}
        \includegraphics[width=\linewidth]{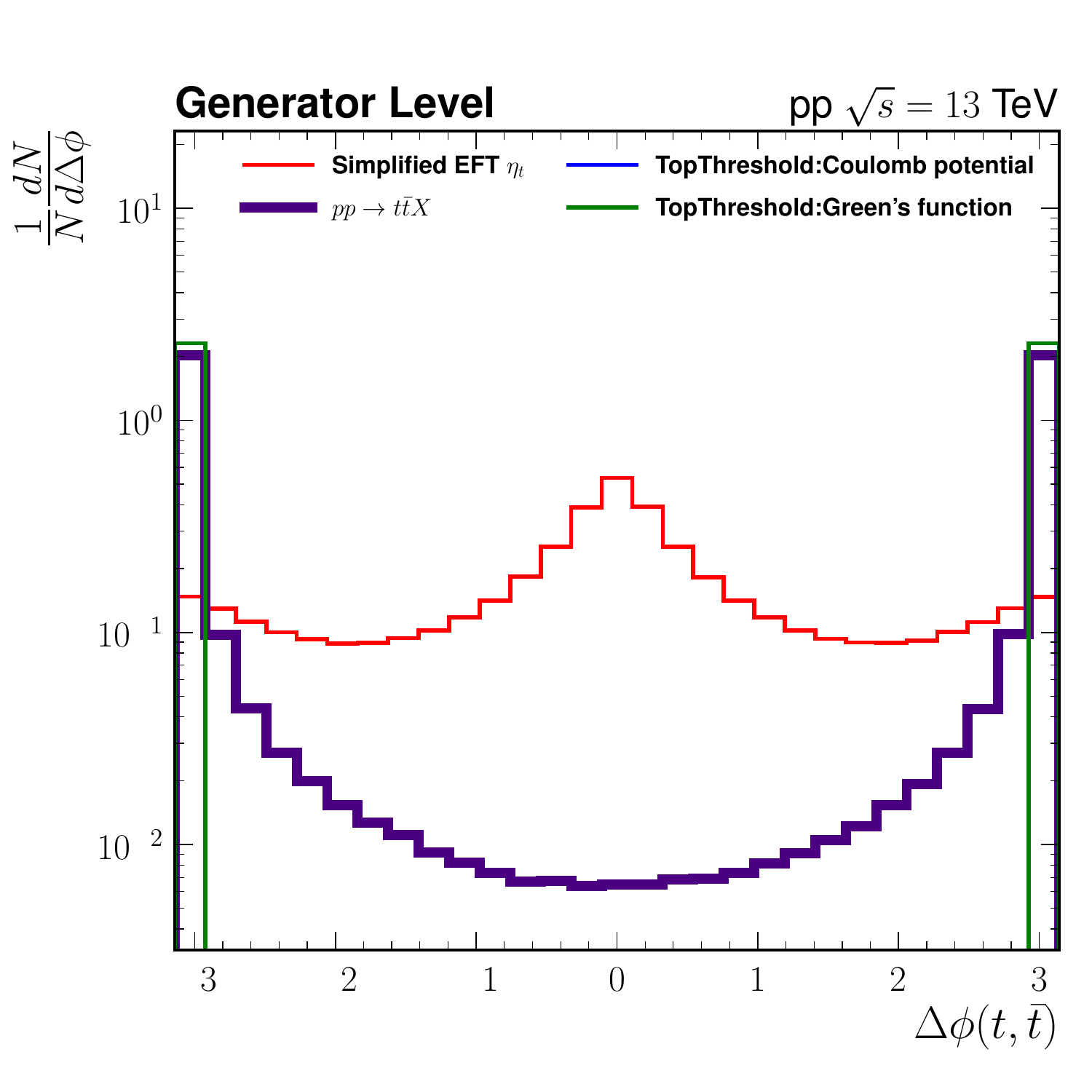}
    \end{subfigure}\hfill
    \begin{subfigure}[t]{0.32\textwidth}
        \includegraphics[width=\linewidth]{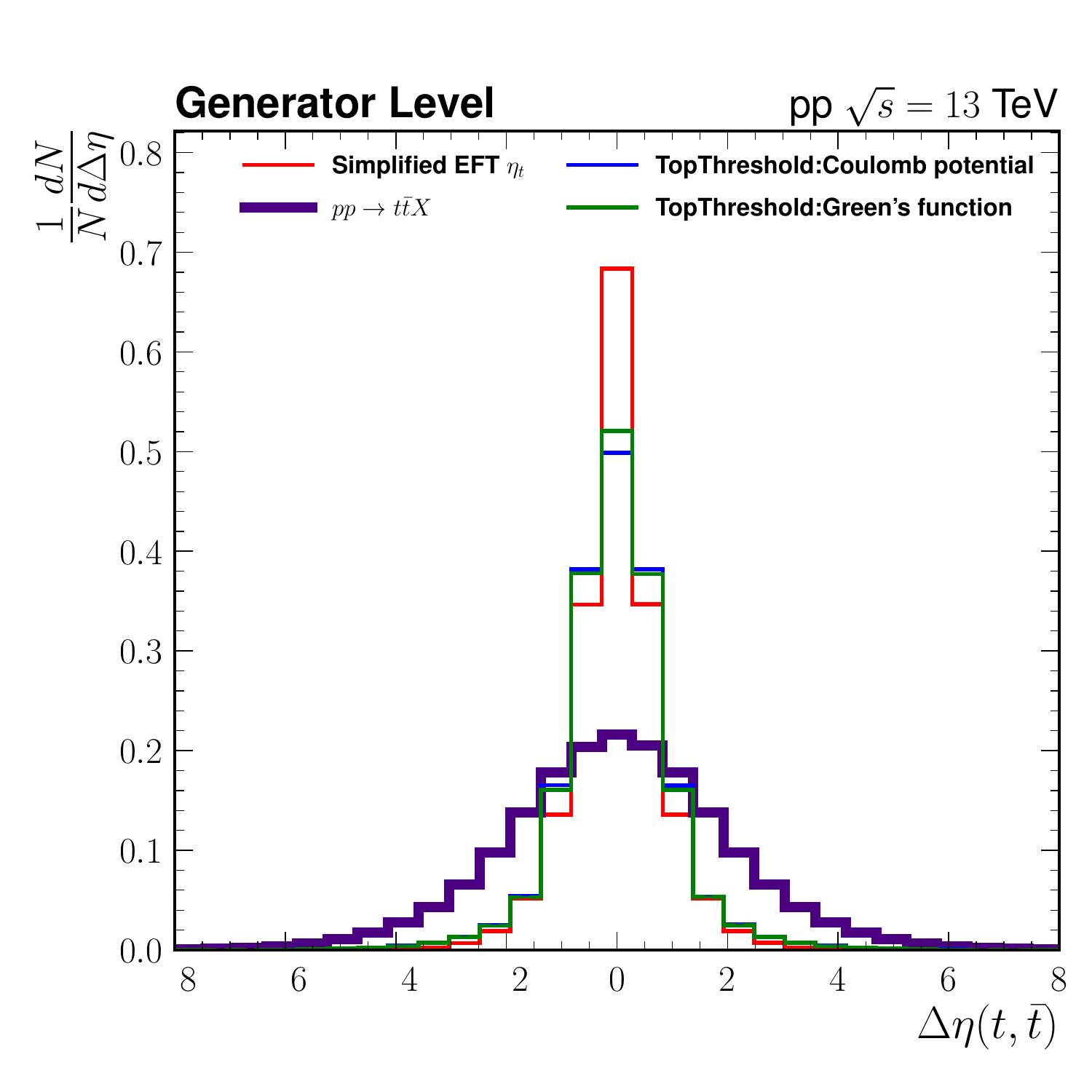}
    \end{subfigure}\hfill
    \begin{subfigure}[t]{0.32\textwidth}
        \includegraphics[width=\linewidth]{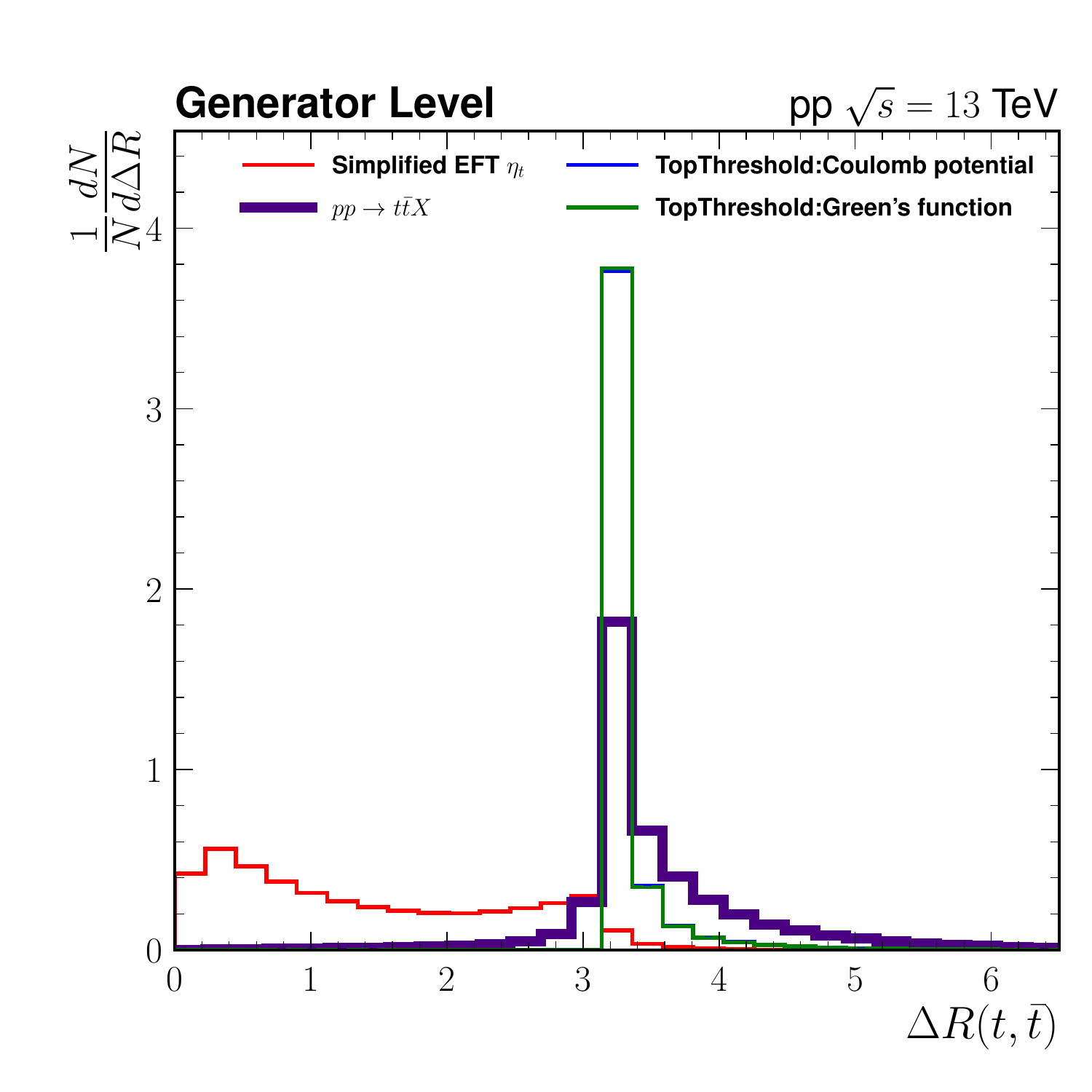}
    \end{subfigure}

    \vspace{1em}

    \begin{subfigure}[t]{0.32\textwidth}
        \includegraphics[width=\linewidth]{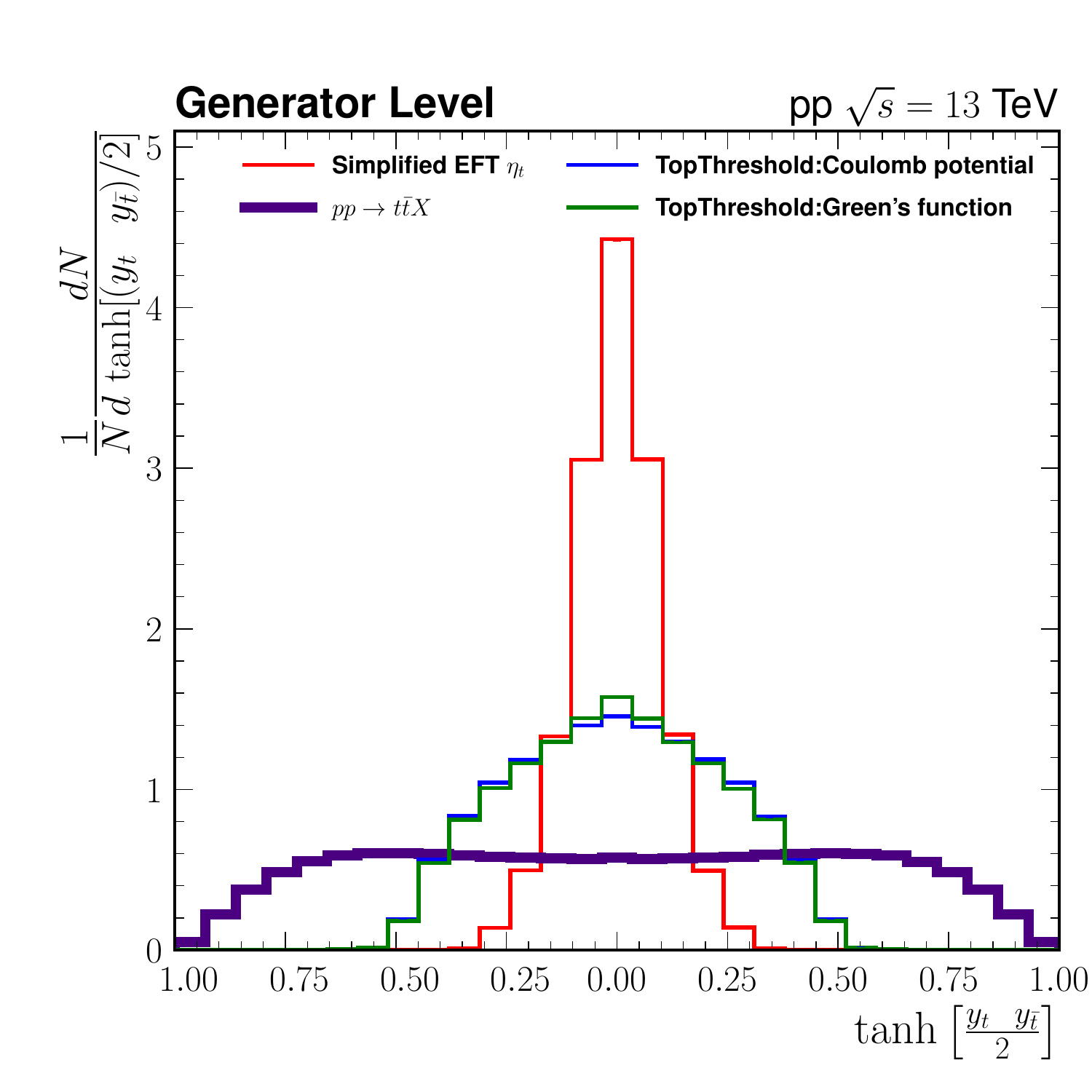}
    \end{subfigure}\hfill
    \begin{subfigure}[t]{0.32\textwidth}
        \includegraphics[width=\linewidth]{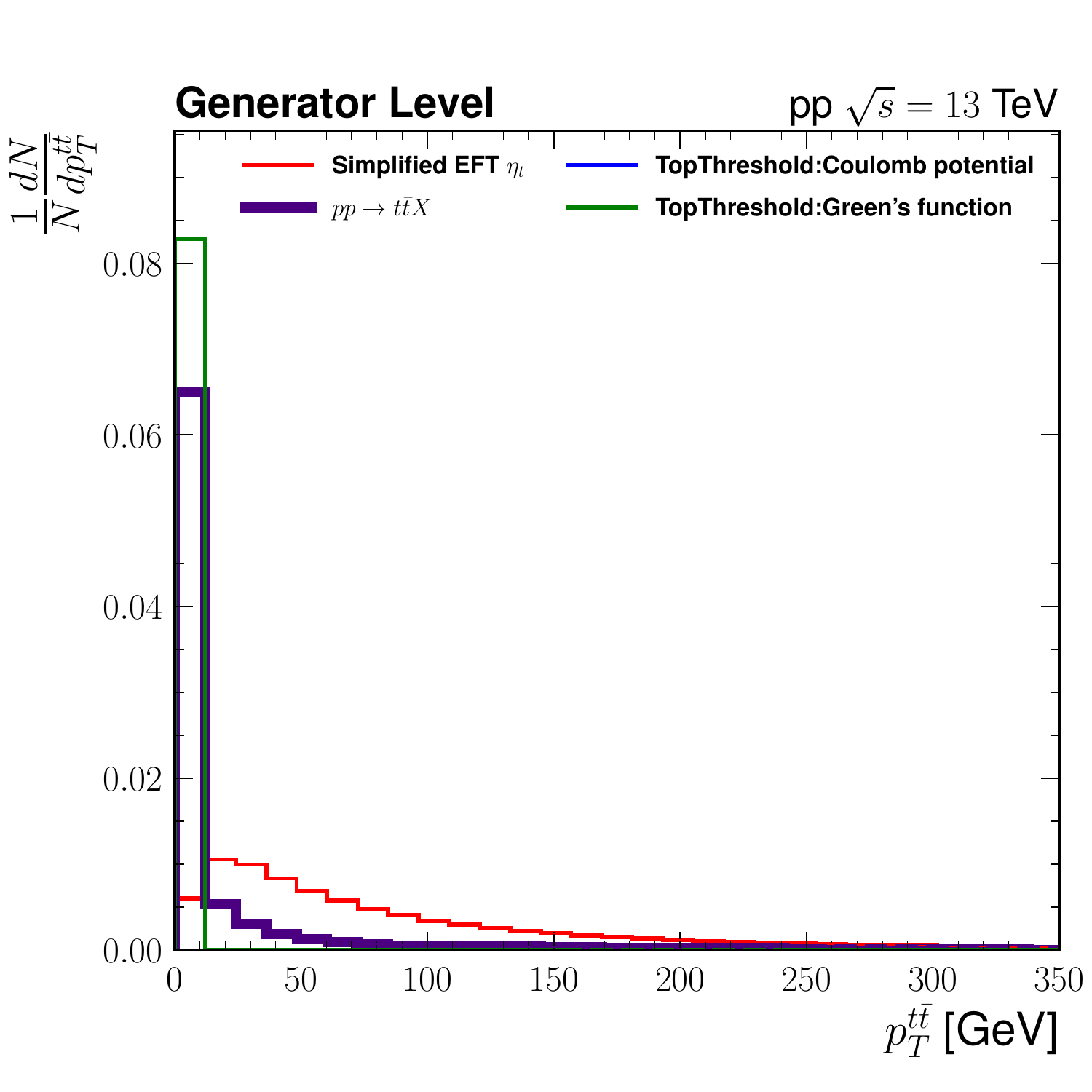}
    \end{subfigure}\hfill
    \begin{subfigure}[t]{0.32\textwidth}
        \includegraphics[width=\linewidth]{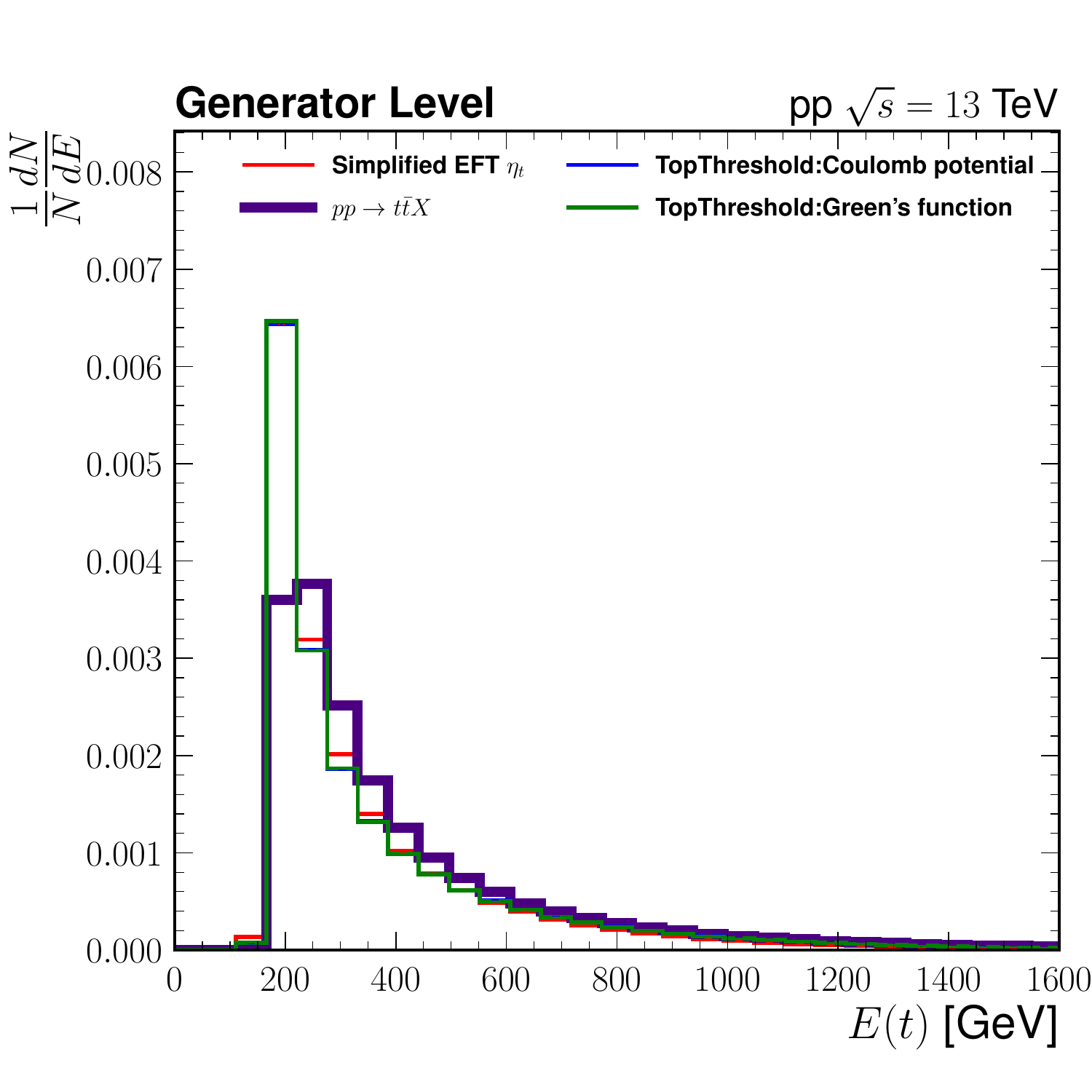}
    \end{subfigure}

    \caption{\justifying Truth-level kinematic and angular distributions comparing the toponium simplified model signal $\eta_t$, $t\bar{t}$ decay products
    to the central $pp \rightarrow t\bar{t}X$ background.
    Shown from top to bottom and left to right are:
    $\Delta\eta(t,\bar{t})$, $\Delta\phi(t,\bar{t})$, $\Delta R(t,\bar{t})$,
    the top-quark transverse momentum $p_T^t$, pseudorapidity $\eta^t$,
    azimuthal angle $\phi^t$, total energy $E^t$,
    longitudinal momentum $p_z^t$, and polar angle $\theta^t$.
    All distributions are normalized to unity.
    }

    \label{fig:truth_top_variables_ttbar_topponium}
\end{figure*}

\noindent Figure~\ref{fig:gen_mtt_signal_background} shows at the left the invariant mass for perturbative $t\bar{t}$ contributions indicated with gray coloured histogram along with $\eta_t$ resonance from EFT model shown with red filled histogram. It replicates the scenario from recent experimental probes at the $t\bar{t}$ spectrum left threshold limit. $\eta_t$ resonance has mean around 344 GeV and width about 7.0 GeV as recommended in this reference~\cite{Fuks:2021wpp}. The $t\bar{t}$ background in gray extends over its characteristic broad continuum spectrum beginning at the kinematic threshold $m_{t\bar{t}} \approx 2m_t$. As shown in the plot the mass spectrum surpasses 3000 GeV at current LHC energies with mean value around $\mu$ $\sim$ 536 GeV. Such spectrum is governed by QCD production dynamics and parton luminosities.\\ 

\begin{figure*}[!ht]
    \centering
    \begin{subfigure}[t]{0.33\textwidth}
        \includegraphics[width=\linewidth]{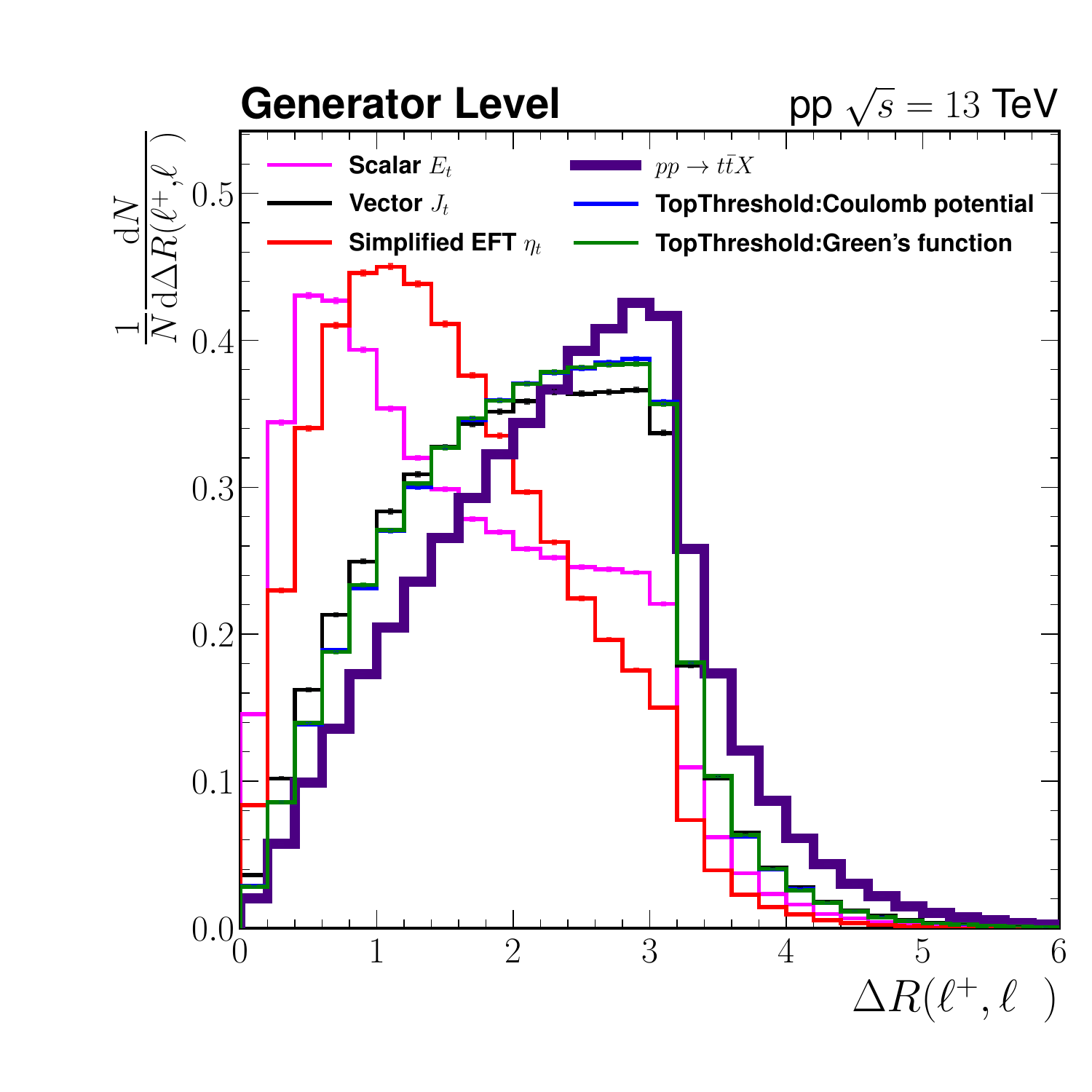}
    \end{subfigure}\hfill
    \begin{subfigure}[t]{0.33\textwidth}
        \includegraphics[width=\linewidth]{ 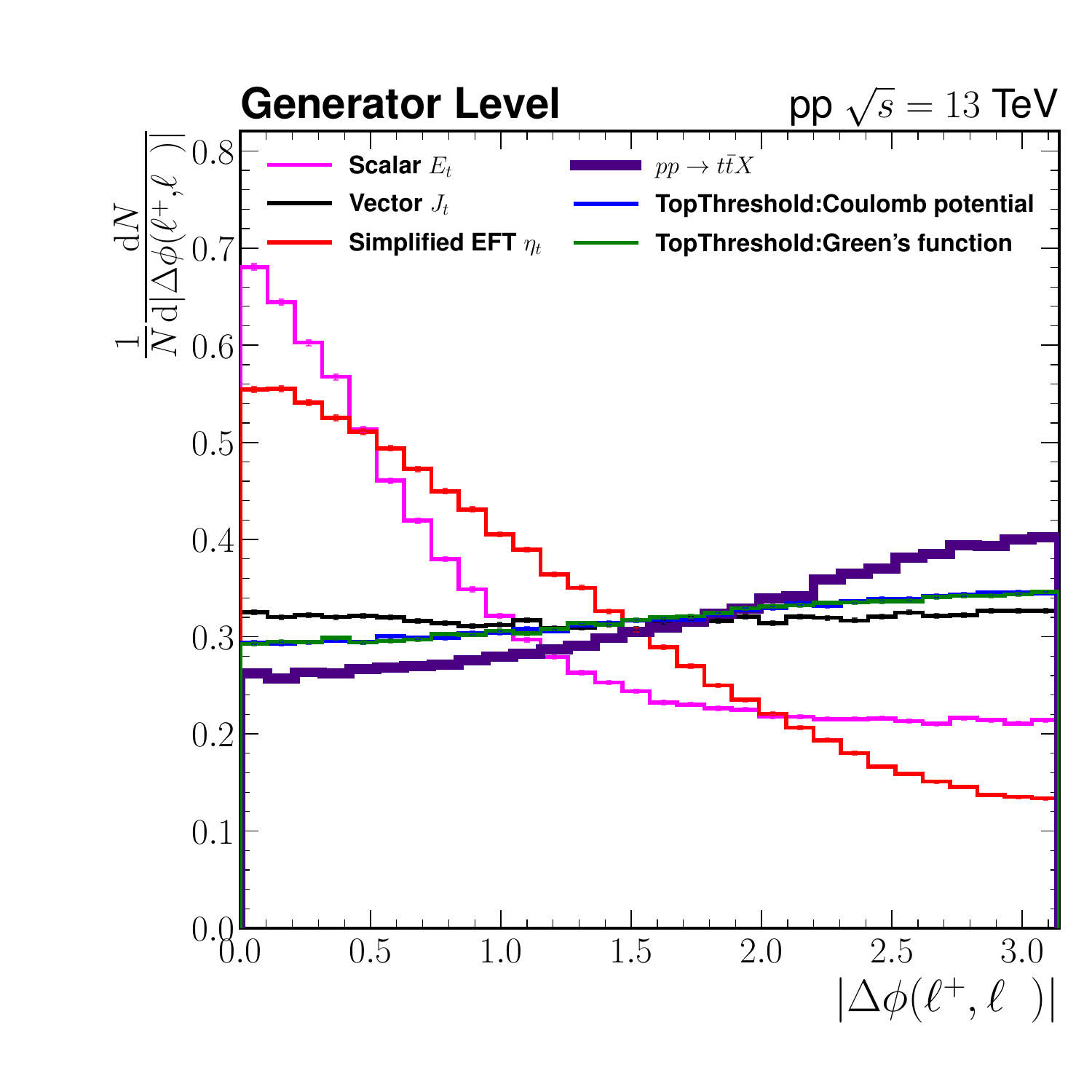}
    \end{subfigure}\hfill
    \begin{subfigure}[t]{0.33\textwidth}
        \includegraphics[width=\linewidth]{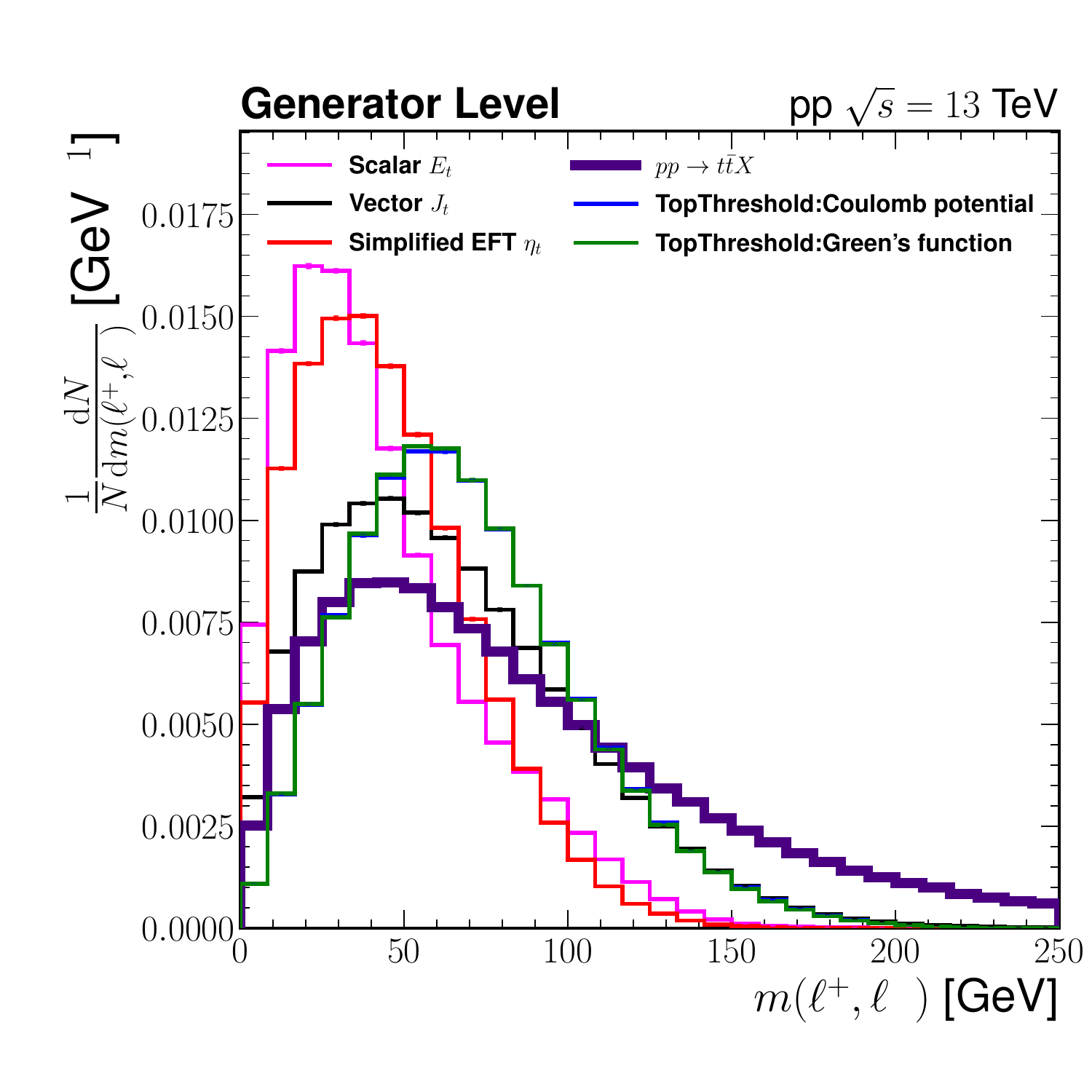}
    \end{subfigure}

    \vspace{1em}

    \begin{subfigure}[t]{0.33\textwidth}
        \includegraphics[width=\linewidth]{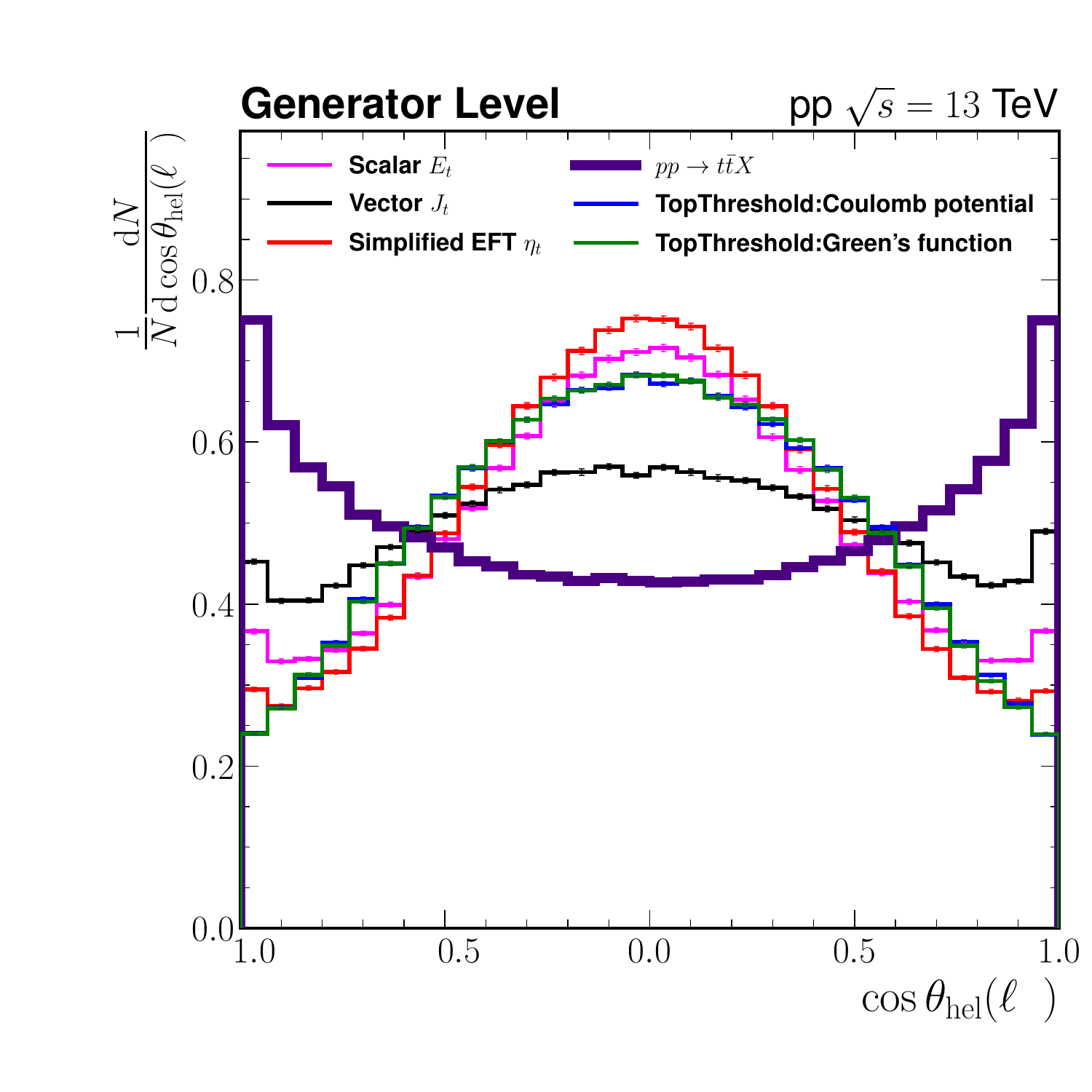}
    \end{subfigure}\hfill
    \begin{subfigure}[t]{0.33\textwidth}
        \includegraphics[width=\linewidth]{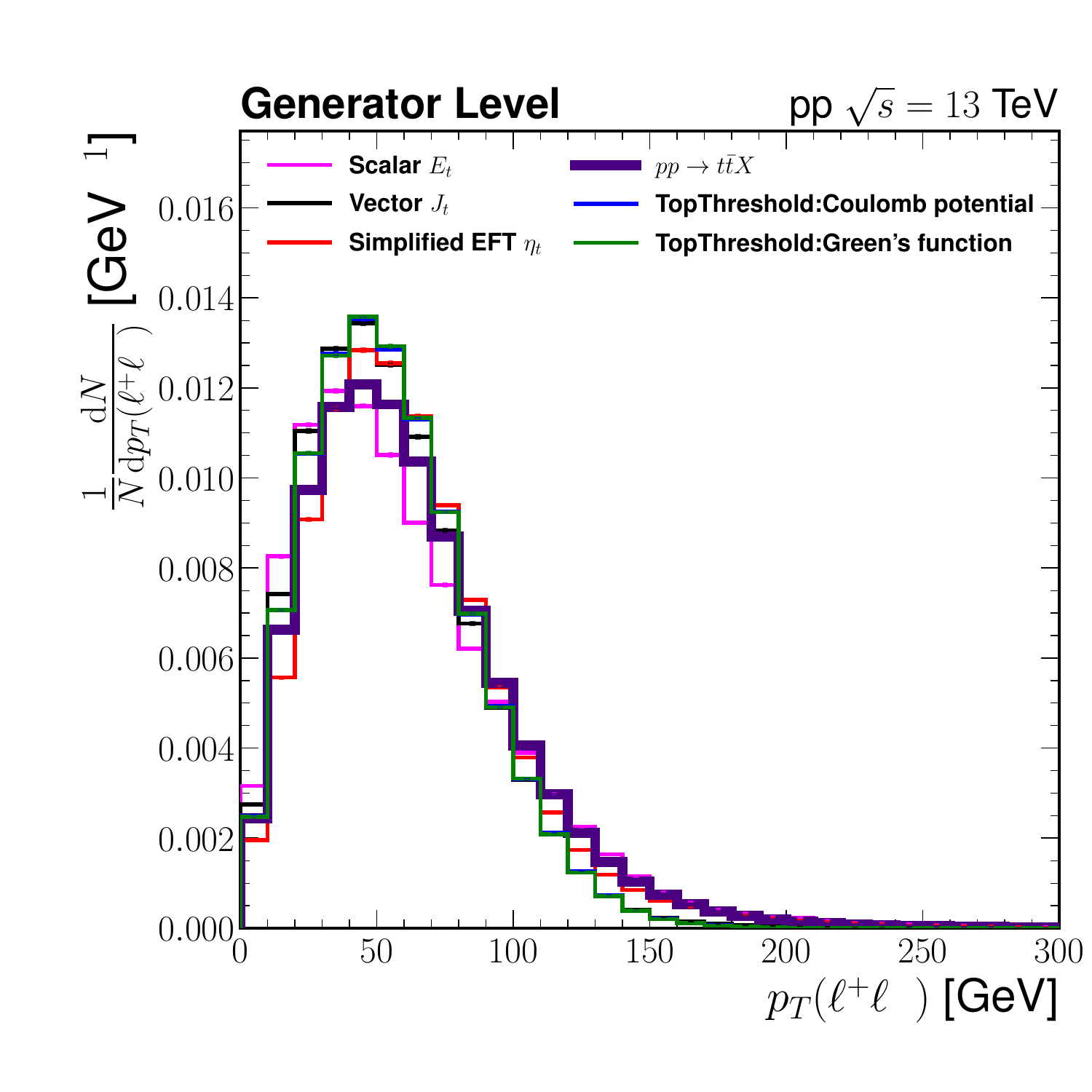}
    \end{subfigure}\hfill
    \begin{subfigure}[t]{0.33\textwidth}
        \includegraphics[width=\linewidth]{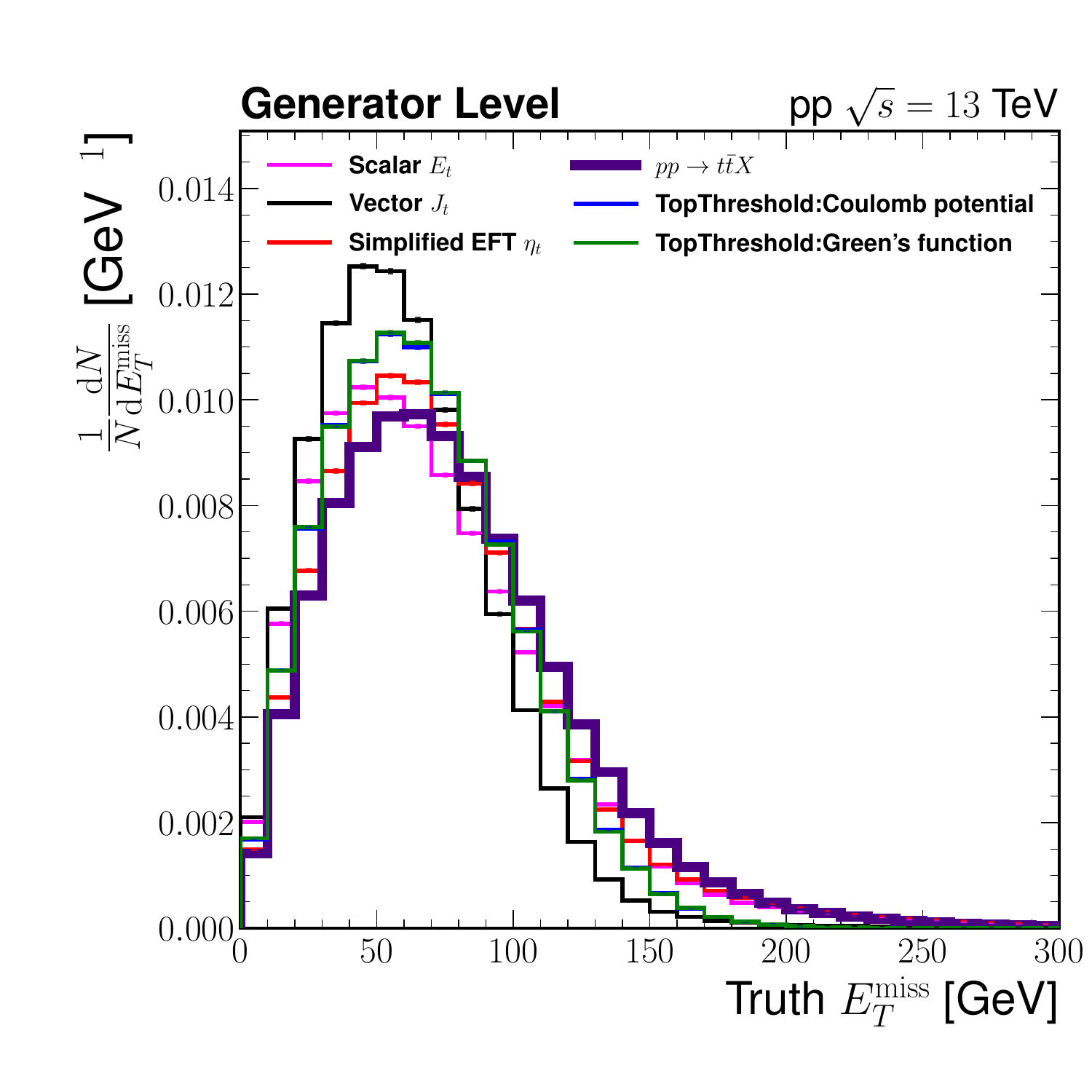}
    \end{subfigure}

    \caption{\justifying Angular and kinematic observables at generator level comparing $t\bar{t}$ background with toponium simplified and EFT model prospects. Variables use events with common final state dilepton topology as EFT model does not produce $t\bar{t}$ decays.}

    \label{fig:truth_lepton}
\end{figure*}

\noindent On the other hand at the right in same Figure, the invariant mass from simplified EFT $\eta_t$ along with scalar $\eta_t$ and vector $J_t$ compound states productions are shown. Scalar and vector toponium models locate the resonance at 341 GeV as it is not restricted by the $t\bar{t}$ limit. It is mostly dominated by $\eta_t$ $\rightarrow$ ZH and $J_t$ $\rightarrow$ ZW$^+$W$^-$ decays.\\ 
\begin{figure*}[h!]
    \centering

    \includegraphics[width=0.49\textwidth]
    {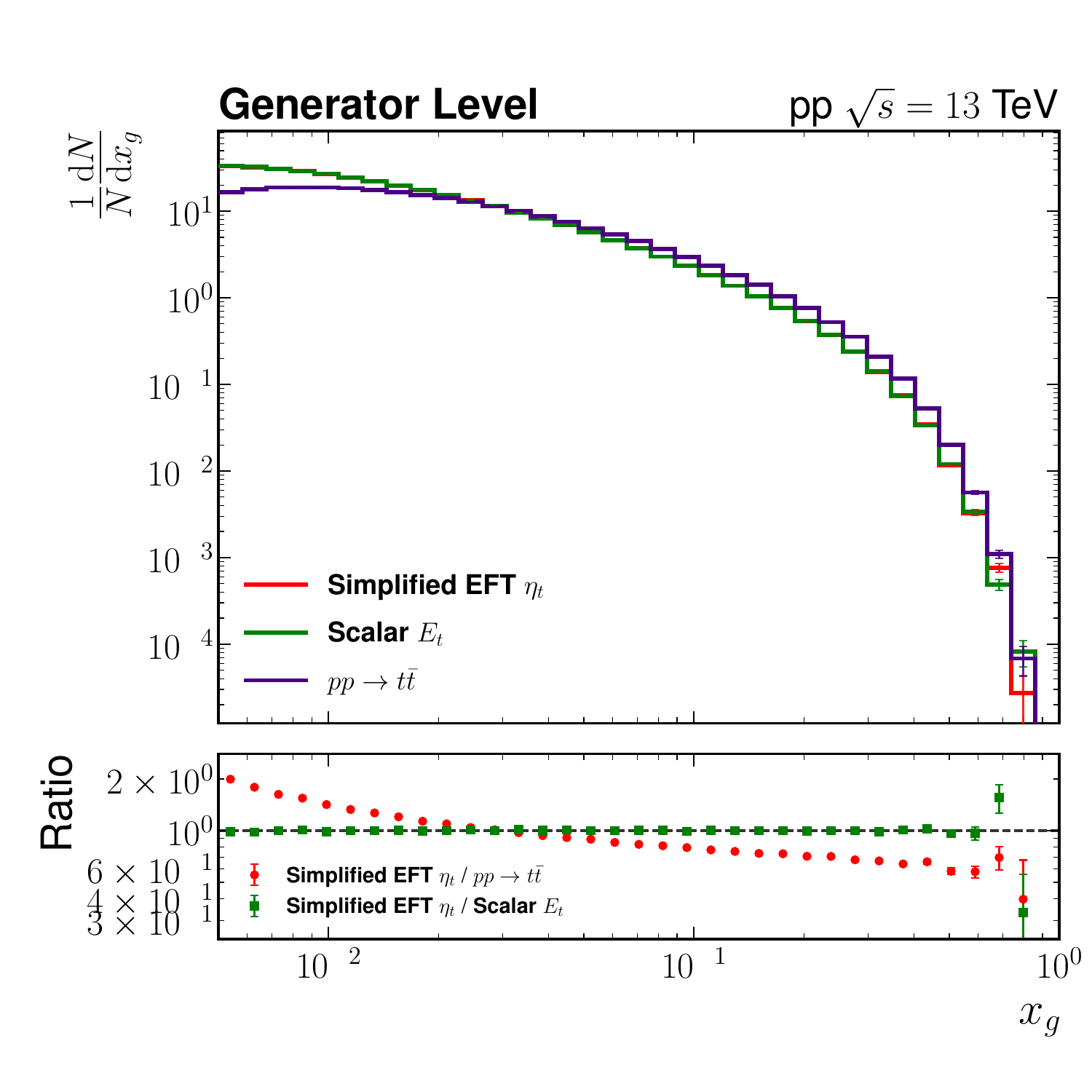}
    \hfill
    \includegraphics[width=0.49\textwidth]
    {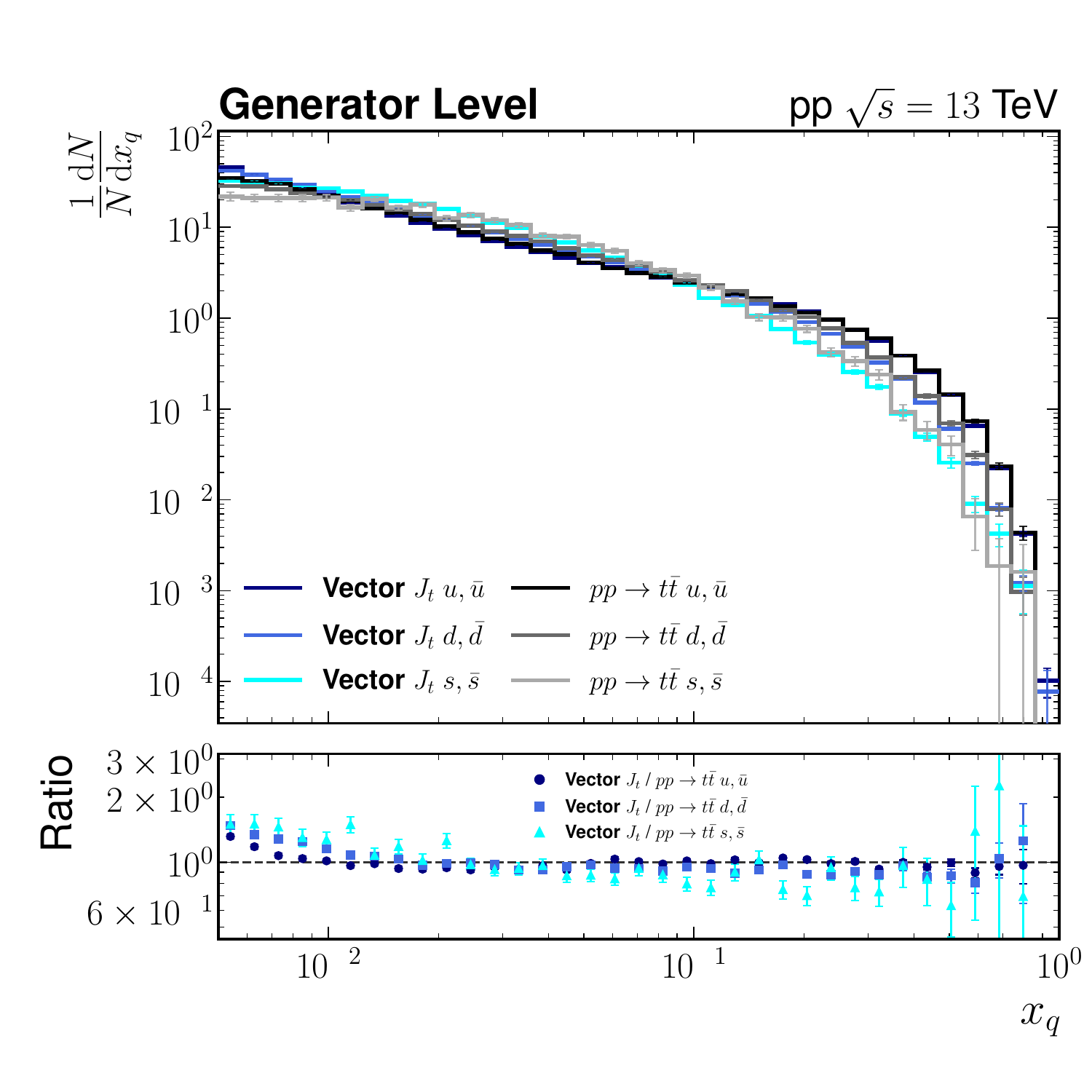}

    \vspace{0.5cm}

    \includegraphics[width=0.49\textwidth]
    {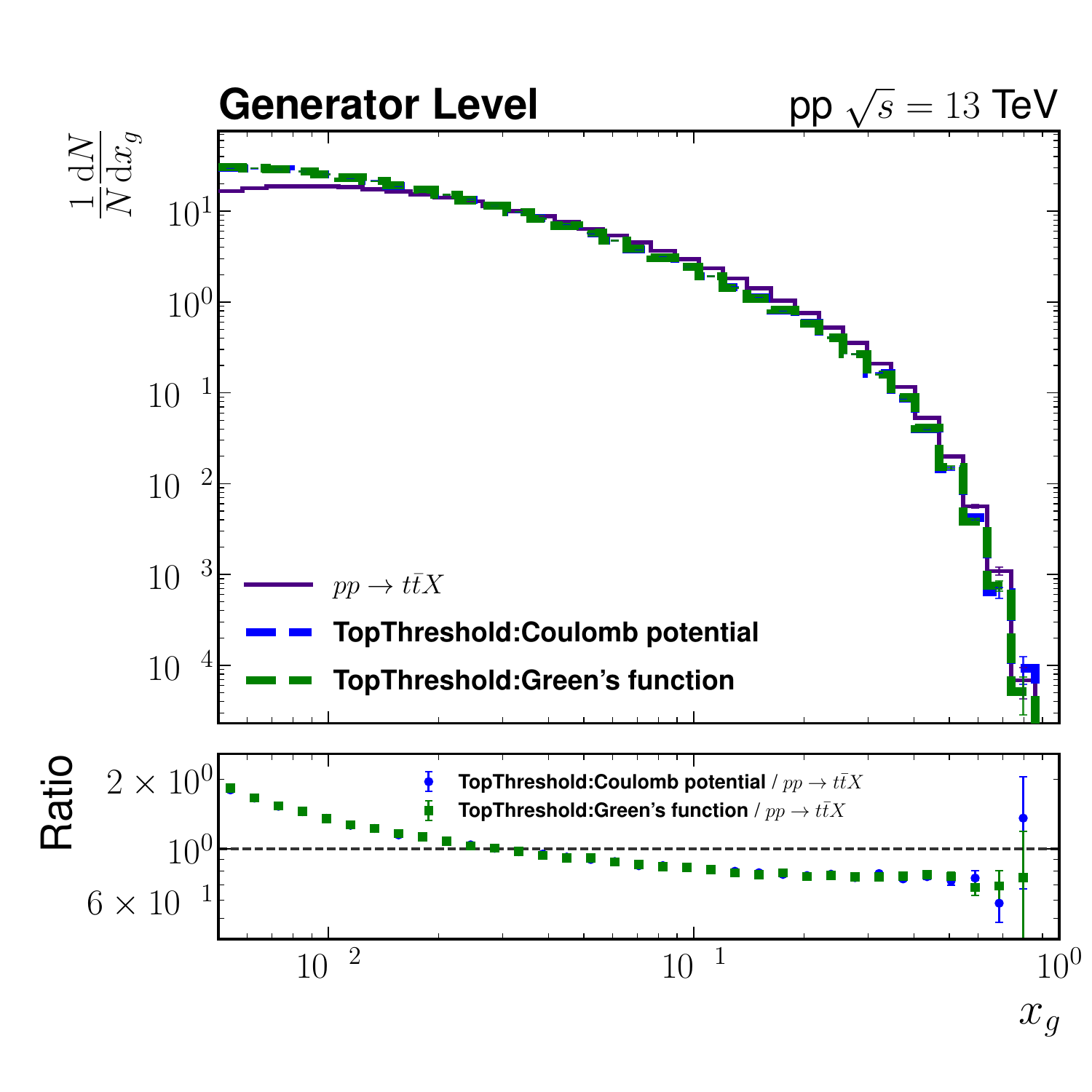}
    \hfill
    \includegraphics[width=0.49\textwidth]
    {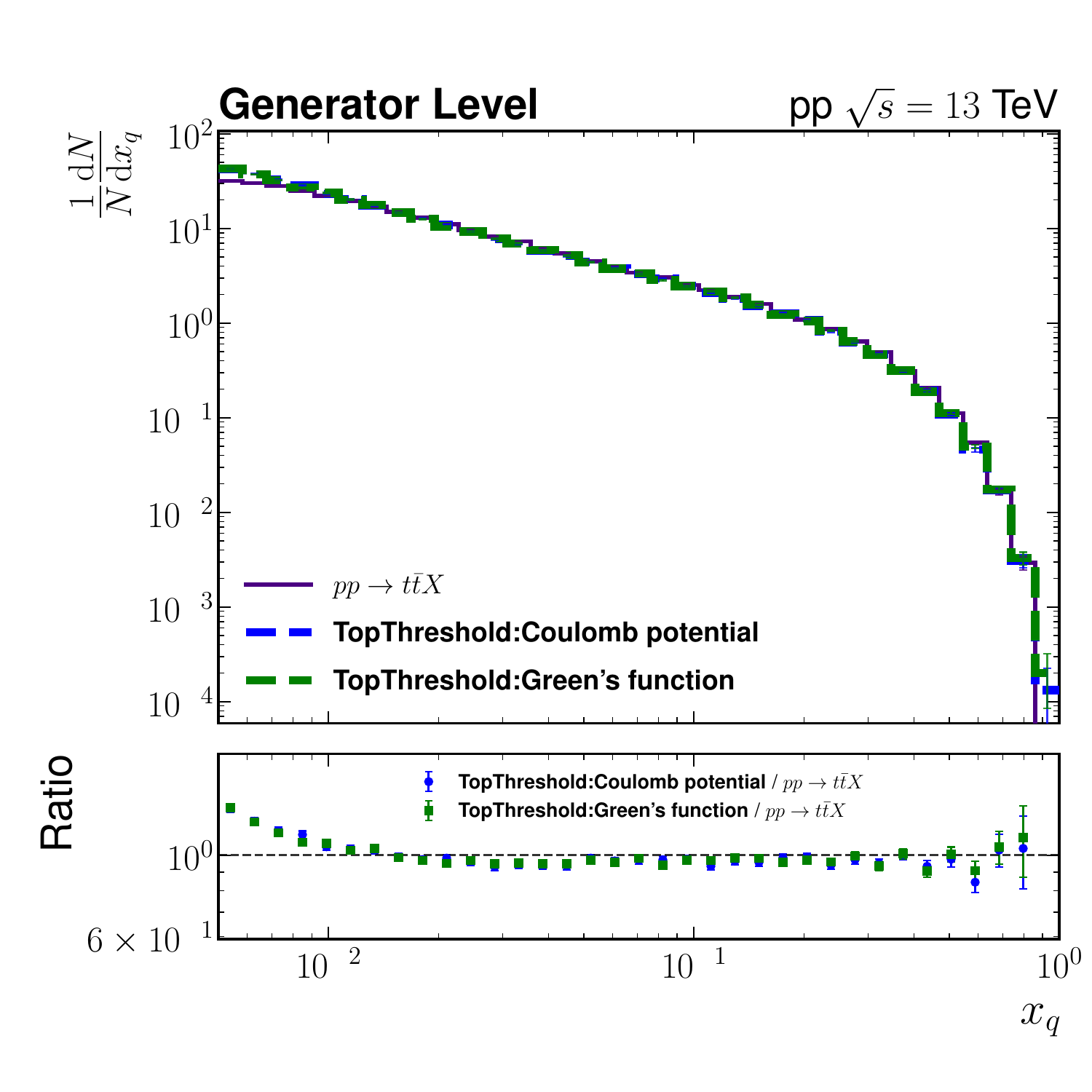}

    \caption{\justifying
    Generator-level gluon initiated production (left) and light-quark at initial state (right) energy-fraction distributions for the  EFT $\eta_t$, Scalar $E_t$, vector $J_t$ and background $pp\to t\bar{t}X$, and NR-QCD effects from Green and Coulomb functions.}
    \label{fig:bjorken_comparison}
\end{figure*}

\noindent Since toponium EFT $\eta_t$ model delivers a $t\bar{t}$ production final state, it makes sense to focus comparisons against standard $t\bar{t}$ production. Figure~\ref{fig:truth_top_variables_ttbar_topponium} shows a set of angular and kinematic variables for signal $\eta_t$ in red curves and $t\bar{t}$ background process in black color at generator level. The azimuthal angular distance between top anti-top pairs $\Delta\phi(t,\bar{t})$ at the top left, is visibly different with $\eta_t$ decays. Such distance is not consistent with back-to-back topology for the signal and averages around $\sim$ 0, while the $t\bar{t}$ background has contributions at $|\Delta\phi|$ $\sim$ $\pi$, consistent with back-to-back production regularly seen in top quark pair production at the LHC. In terms of pseudorapidity angular distance $\Delta\eta(t,\bar{t})$ (top middle), $t\bar{t}$ pairs from $\eta_t$ decays show less $\sigma$ with respect to zero by a factor of about 2.5 with respect background that shows larger angular separation between top quarks.   This effect propagates to the full angular distance between the pair $\Delta$R $=$ $\sqrt{\Delta\phi^2 + \Delta\eta^2}$ at the right with maximum value at $\sim$ $\pi$ for $t\bar{t}$ background and at zero for $\eta_t$ signal and standard deviation of $\sim$1. The spin-sensitive angular variables, built from the lepton directions in the parent-top rest frames, are treated in Sec.~\ref{sec:spin_corr}. In terms of full $t\bar{t}$ transverse momentum (bottom middle) $\eta_t$ decays show a larger component consistent with angularly closer $t\bar{t}$ products, while for total energy (bottom right), signal processes show up with smaller values with mean $\sim$348 GeV and $\sigma\sim$ 228 smaller than background by around 100 GeV.\\

\noindent Figure~\ref{fig:truth_lepton} shows additional kinematic observables at truth level but rather than computing out of $t\bar{t}$ final state variables from final state leptons are produced so NR-QCD (Green function and Coulomb) model prospects (green and blue) can be compared along simplified EFT model (red), Scalar E$_t$ (pink colour), Vector J$_t$ (black color) and background (violet). In terms of angular $\Delta R = \sqrt{\Delta\phi^2 + \Delta\eta^2}$ and azimuthal $\Delta\phi$ distances (top row) EFT $\eta_t$ has more consistency with scalar $E_t$ component, while vector $J_t$ component gets closer to $t\bar{t}$ background. $t\bar{t}$ events with NR-QCD effects from Green and Coulomb functions follow also the background trend. A clear larger reach for background $t\bar{t}$ and Vector $J_t$ is seen in the dilepton system invariant mass. cos$\theta$($l^-$) variable shows a lepton back-to-back topology with peaks at $\pm$1 for $t\bar{t}$ background followed by vector $J_t$ and scalar $E_t$ components. EFT $\eta_t$ and NR-QCD corrections keep larger contributions around 0. In terms of dilepton $l^+l^-$ system transverse momentum p$_\mathrm{T}$ and missing energy E$_\mathrm{T}^{miss}$ background $t\bar{t}$ shows the largest values with Vector $J_t$ and NR-QCD contributions showing the smallest.\\

\noindent To finalize the discussion at generator level, Figure \ref{fig:bjorken_comparison} shows the different energy fractions $x$, with respect to the corresponding colliding proton energy of initial state partons that originated either the $\eta_t$ signal from simplified model or the $t\bar{t}$ background processes. For initial state gluons at the left of the Figure, the energy fractions are consistent between scalar $E_t$ and EFT $\eta_t$ while background $t\bar{t}$ delivers larger $x$ with respect both. NR-QCD from green and coulomb functions show also smaller $x$ with respect background (bottom left). Averages are $<x>$ 0.19 for $t\bar{t}$ background  against 0.18 for signal $\eta_t$. On the other hand at low and high $x$ value the toponium signal shows up larger (up to $\sim$ 1.7 times) and smaller contributions (up to $\sim$ 7 times) respectively as the ratio trend below shows in black color. For light quarks (u, v) in the initial state at the right of the Figure, the average fractions are larger in general with values 0.23 and 0.25 for $t\bar{t}$ and $\eta_t$ signal respectively. In this case the contributions for low $x$ values are smaller for the signal by up to $\sim$7 times and no significant difference at high $x$. One interesting aspect of this comparison is that at low $x$ s-quark contributions are larger for the signal $\eta_t$ but on the other hand the u and d contributions are lower than the background by up to $\sim$7 times. \\

\section{Spin correlation prospects and comparison to $t\bar{t}$ background}
\label{sec:spin_corr}
\noindent Spin correlation measurements are a promising strategy to characterise and separate signal from background $t\bar{t}$ events, provided the toponium predictions are compared with those along the NR-QCD contributions at near the $m_{\bar{t}t}$ threshold limits predictions. For $t\bar{t}$ pairs decaying to a $\ell^+\ell^-$ system, in signal or background, the lepton directions give access to the spin state of the pair. The ATLAS collaboration has used such measurements to observe quantum entanglement between the top and antitop quarks~\cite{ATLAS:2024Entanglement}, a signature of the production mechanism and of the extreme properties of the top quark: a mass of 172~GeV and a width of about 1.4~GeV, which give a lifetime shorter than the hadronisation time scale.\\

\noindent The spin state of a $t\bar{t}$ pair is described by 15 real parameters: the polarisation vectors $\vec{B}_1$ and $\vec{B}_2$ of the top and antitop quark (six numbers) and a $3\times3$ correlation matrix $C_{ij}$ (nine numbers) that quantifies the spin entanglement along pairs of axes. They are defined in the helicity basis $\{\hat{k},\hat{r},\hat{n}\}$ of Ref.~\cite{CMS:2019nrx}: $\hat{k}$ is the top-quark direction in the $t\bar{t}$ rest frame, $\hat{r}$ is in the production plane and $\hat{n}$ is orthogonal to the production
plane. All 15 coefficients follow from the angles $\theta^{i}_1$ and $\theta^{i}_2$ between the axis $i$ and the direction of the $\ell^+$ and of the $\ell^-$, each measured in the rest frame of its parent top or antitop quark:
\begin{equation}
B^{i}_{1,2}=3\langle\cos\theta^{i}_{1,2}\rangle,\qquad
C_{ij}=-9\langle\cos\theta^{i}_{1}\cos\theta^{j}_{2}\rangle,\qquad i,j\in\{k,r,n\}.
\label{eq:BC_definition}
\end{equation}
The opening angle between the two leptons in the same frames gives the coefficient $D$ directly,
\begin{equation}
\frac{1}{\sigma}\frac{d\sigma}{d\cos\phi}=\frac{1}{2}\left(1-D\cos\phi\right),\qquad
D=-\frac{C_{kk}+C_{rr}+C_{nn}}{3}.
\label{eq:cosphi_dist}
\end{equation}
Spin correlations have been measured by CMS~\cite{CMS:2019nrx} and ATLAS~\cite{ATLAS:2020aln}, and the SM predictions agree with them within uncertainties that range from a few percent to $\sim$10\%, depending on the observable.\\

\noindent Two laboratory-frame observables are also measured: $\cos\phi_{\rm lab}$, the same dot product built from the lepton directions in the laboratory frame, and $|\Delta\phi_{\ell\ell}|$, the absolute azimuthal difference between the leptons. They have no direct correspondence with the correlation matrix and are summarised in Table~\ref{tab:observables_coefficients} lists all observables.\\
 
\begin{table}[h!]
\centering
\begin{tabular}{lll}
\toprule
Observable & Measured coefficient & Symmetries \\
\midrule
$\cos\theta^{k}_{1}$, $\cos\theta^{k}_{2}$ & $B^{k}_{1}$, $B^{k}_{2}$ & P-odd, CP-even \\
$\cos\theta^{r}_{1}$, $\cos\theta^{r}_{2}$ & $B^{r}_{1}$, $B^{r}_{2}$ & P-odd, CP-even \\
$\cos\theta^{n}_{1}$, $\cos\theta^{n}_{2}$ & $B^{n}_{1}$, $B^{n}_{2}$ & P-even, CP-even \\
$\cos\theta^{k}_{1}\cos\theta^{k}_{2}$ & $C_{kk}$ & P-even, CP-even \\
$\cos\theta^{r}_{1}\cos\theta^{r}_{2}$ & $C_{rr}$ & P-even, CP-even \\
$\cos\theta^{n}_{1}\cos\theta^{n}_{2}$ & $C_{nn}$ & P-even, CP-even \\
$\cos\phi$ & $D=-(C_{kk}+C_{rr}+C_{nn})/3$ & P-even, CP-even \\
$\cos\phi_{\mathrm{lab}}$, $|\Delta\phi_{\ell\ell}|$ & -- & -- \\
\bottomrule
\end{tabular}
\caption{Observables, measured coefficients and their P and CP properties~\cite{CMS:2019nrx}.}
\label{tab:observables_coefficients}
\end{table}

\noindent The coefficients are extracted at parton level from the top quarks before their decay and from the charged leptons of the dileptonic decays, with no acceptance requirements. Table~\ref{tab:spin_coeff_combined} compares the extracted coefficients with
the CMS measurement~\cite{CMS:2019nrx}, the uncertainties of the samples are statistical only, those of the measurement are total.\\
 
\begin{figure}[!htp]
    \centering
    \includegraphics[width=0.45\linewidth]{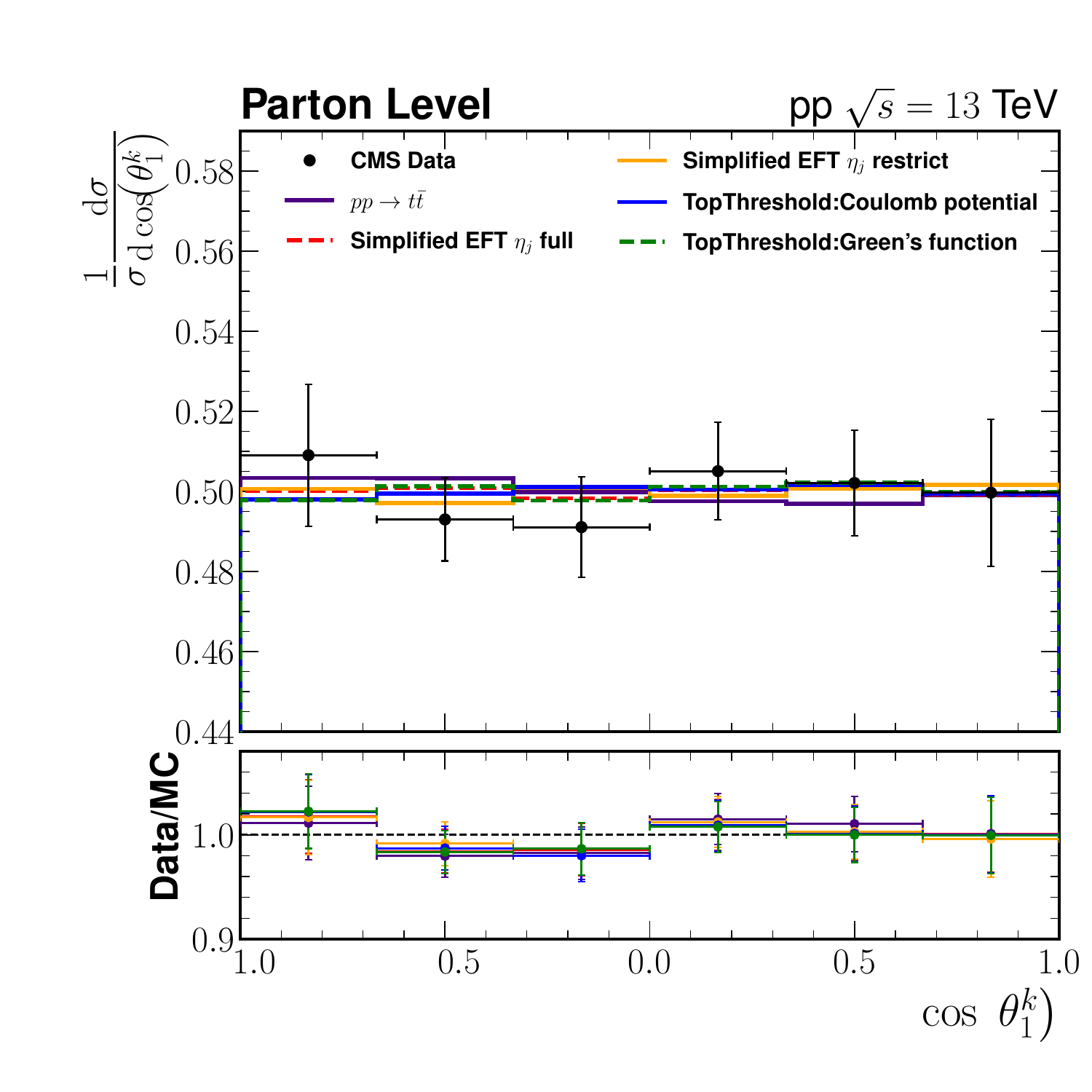}
    \includegraphics[width=0.45\linewidth]{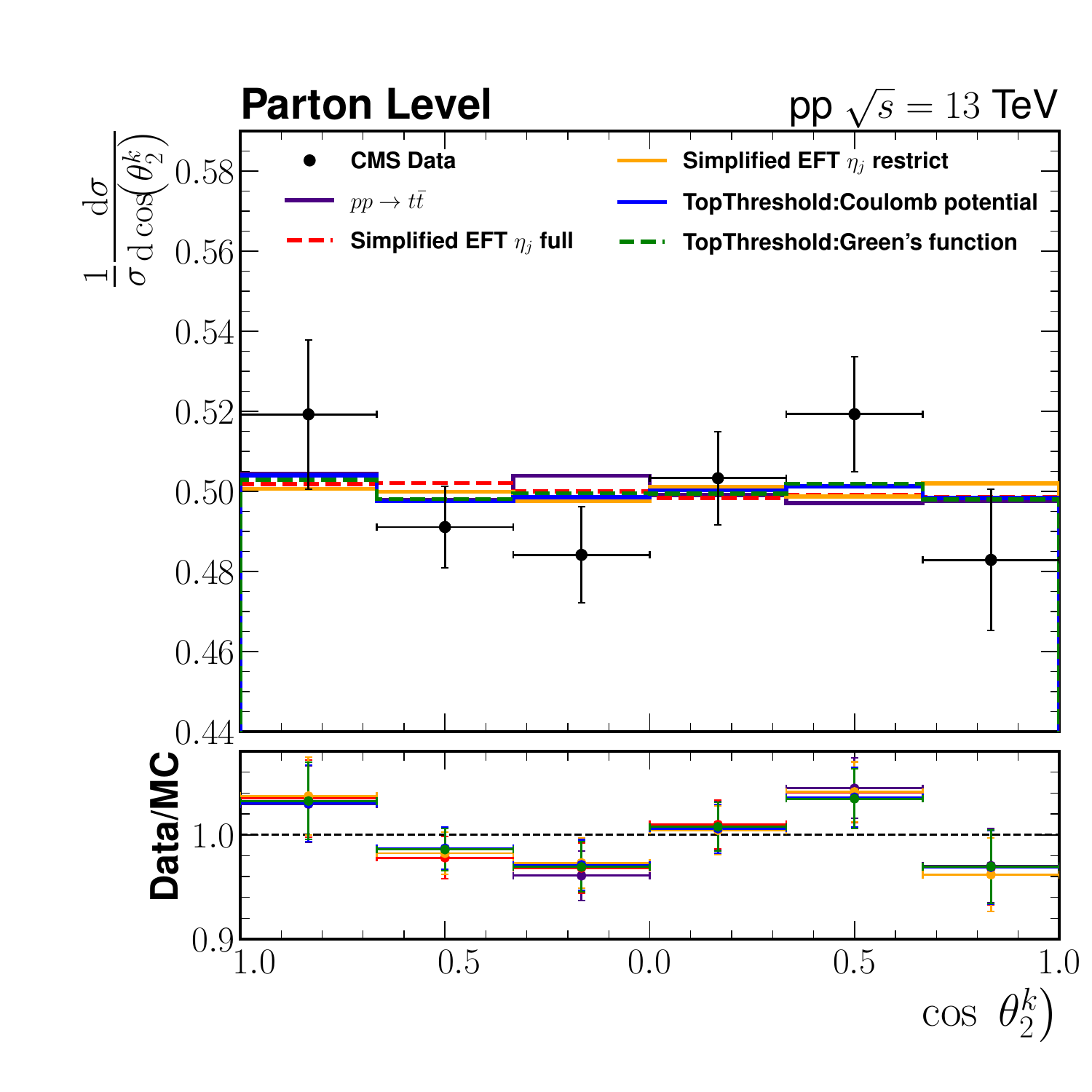}\\
    \includegraphics[width=0.45\linewidth]{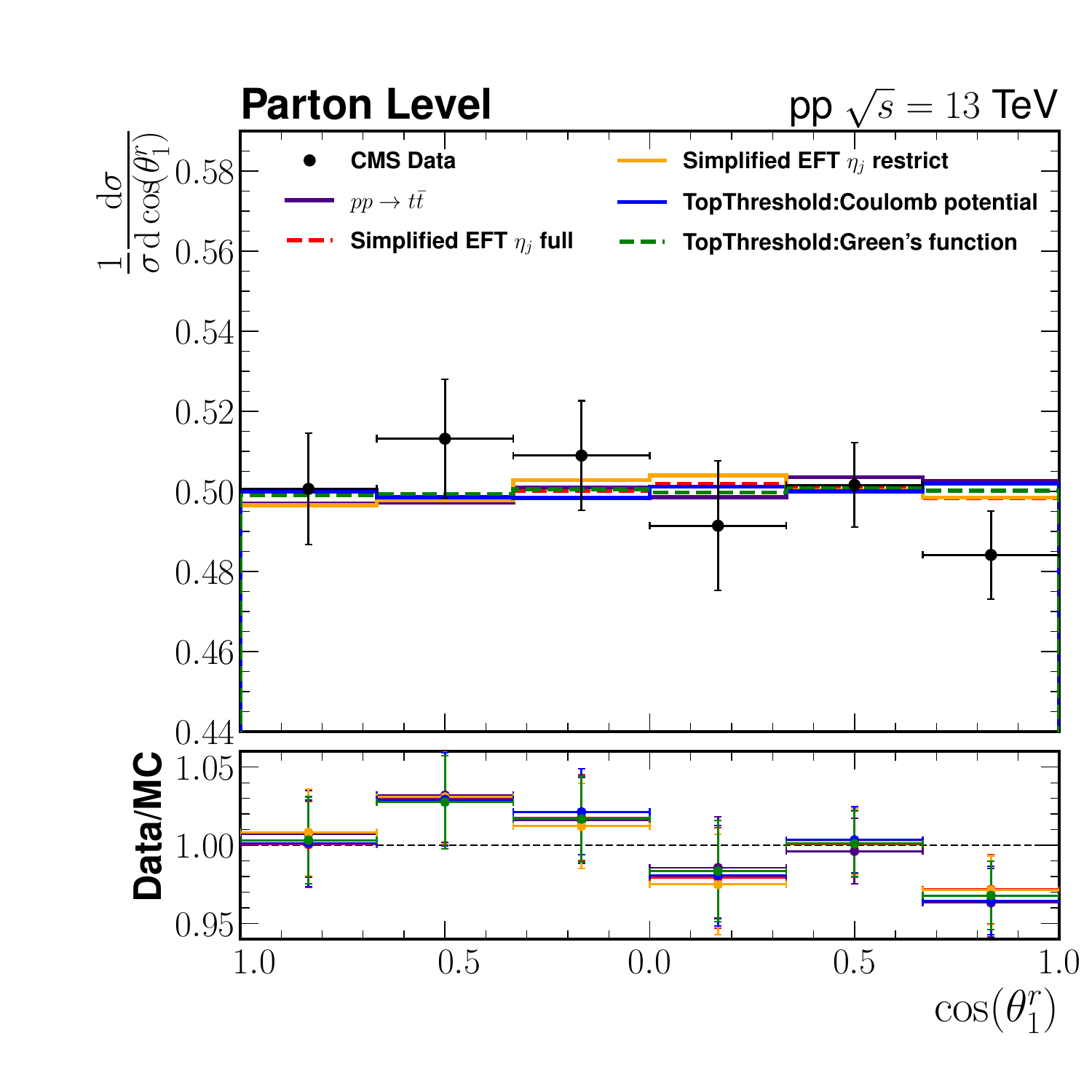}
    \includegraphics[width=0.45\linewidth]{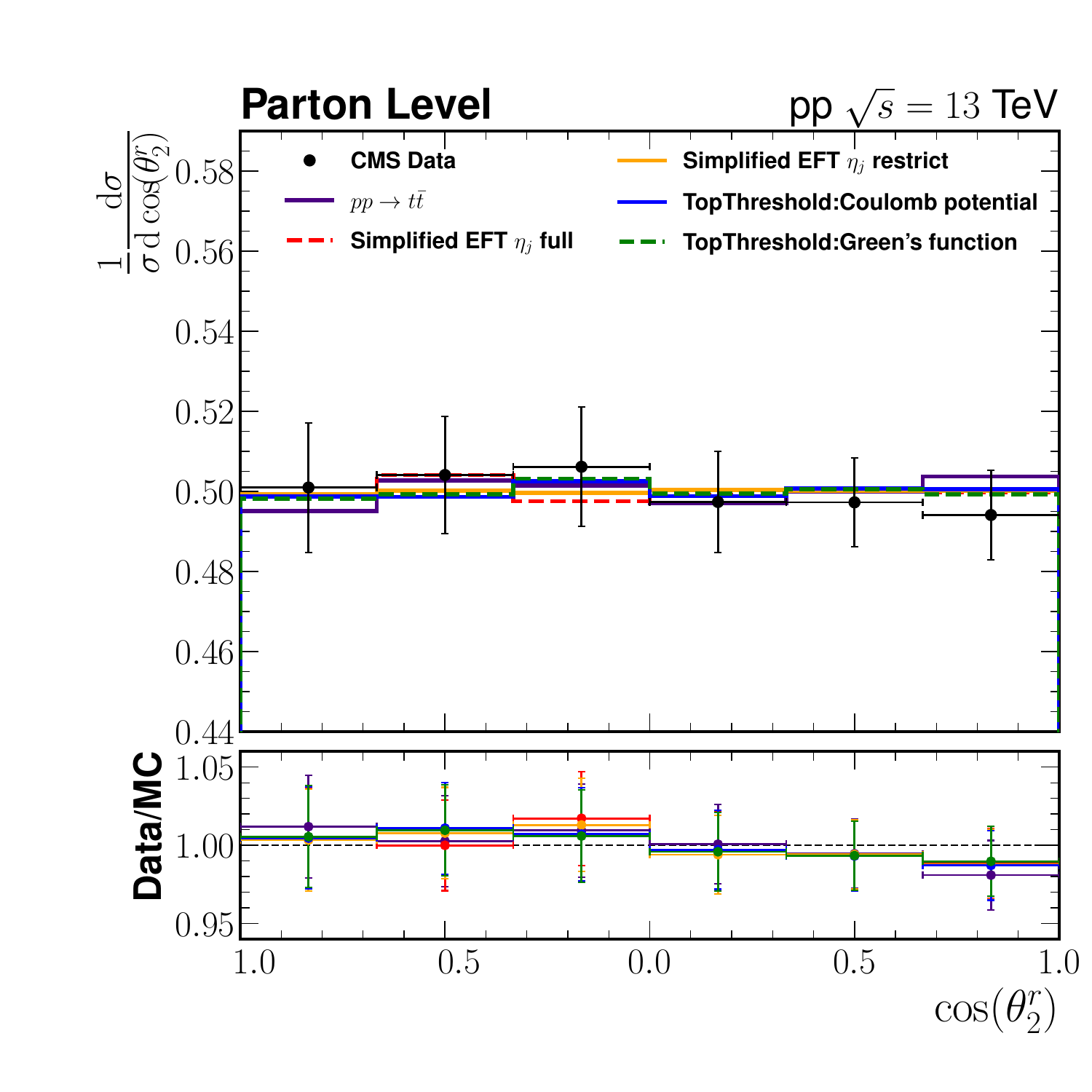}\\
    \includegraphics[width=0.45\linewidth]{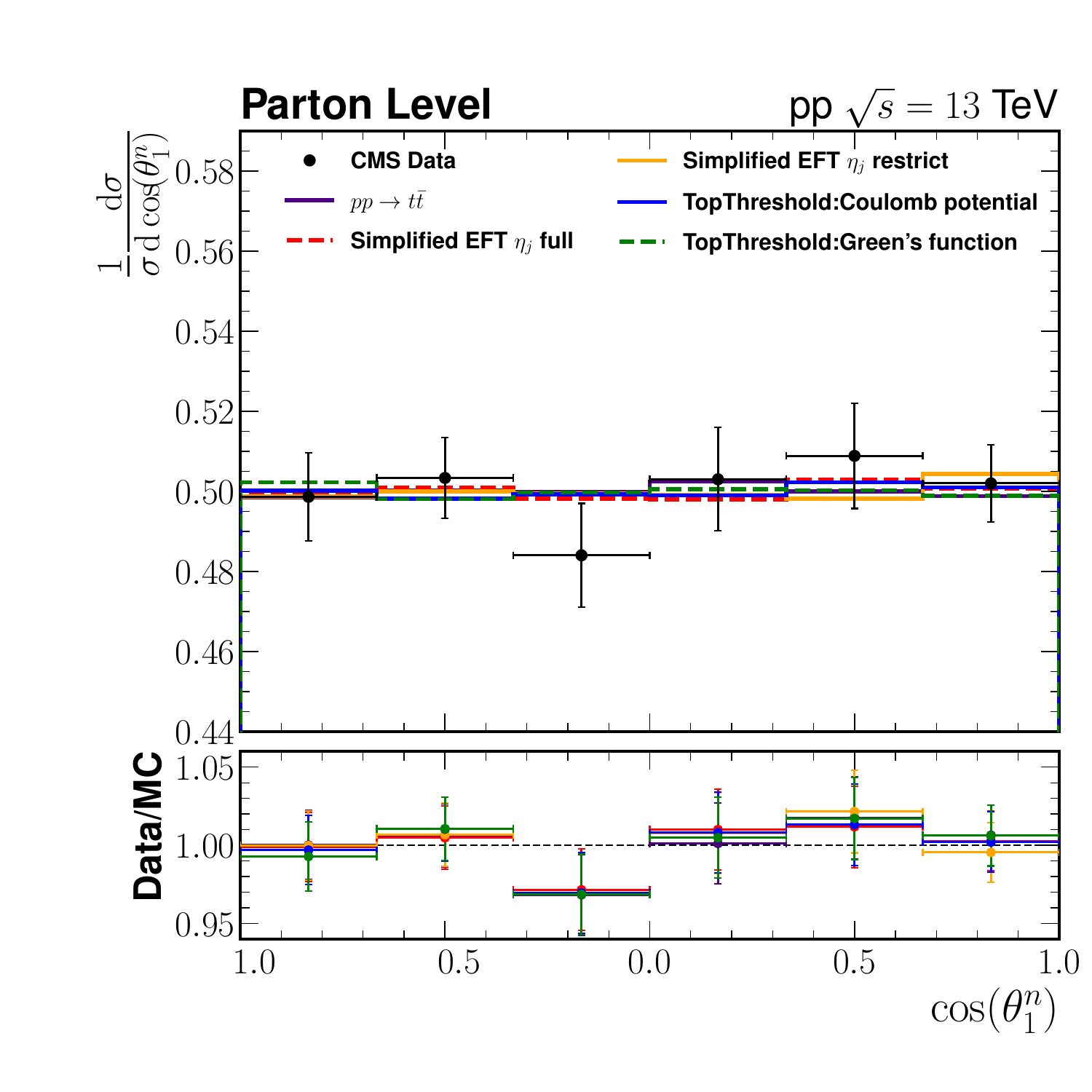}
    \includegraphics[width=0.45\linewidth]{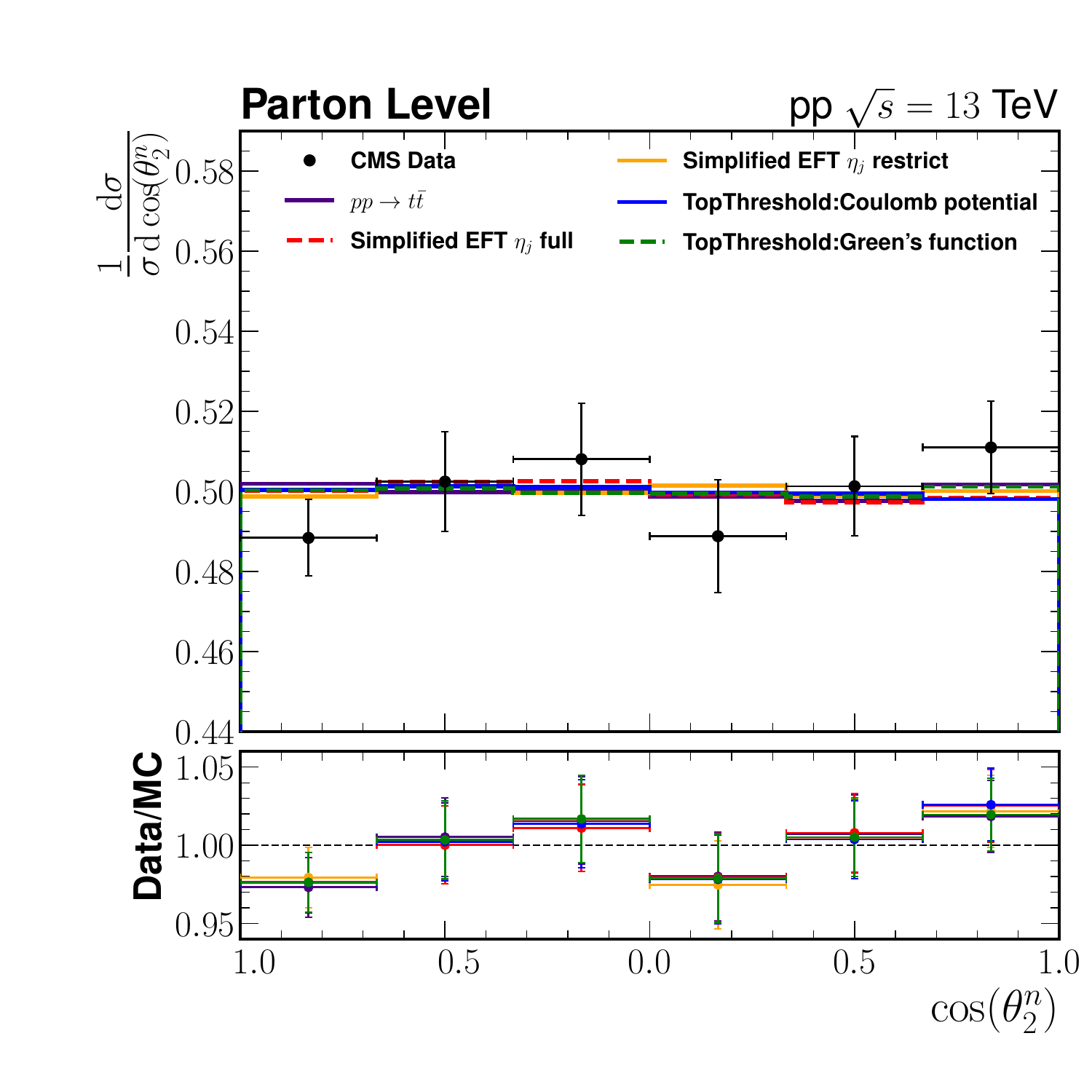}
    \caption{Unfolded CMS data and predictions for $\cos\theta^{i}_{1,2}$ ($i=k,r,n$), normalised to unit area, for the top quark (left column) and the antitop quark (right column). The lower panels show data/prediction.}
    \label{fig:lep_costheta_all}
\end{figure}

\noindent Figure~\ref{fig:lep_costheta_all} shows the six $\cos\theta^{i}_{1,2}$ distributions. They are flat for the CMS data, the $t\bar{t}$ baseline and every signal hypothesis, and the polarisation coefficients in Table~\ref{tab:spin_coeff_combined} are compatible with zero: $|B|<0.005$ in all simulated samples (for the Restricted and Full toponium samples) and $|B|\le0.023$ in data, with uncertainties of 0.013--0.023. This is the expected outcome, since a $J=0$ state cannot be polarised and QCD conserves parity in the continuum. The polarisation observables therefore have no discriminating power and are not used further.

\newpage

\begin{figure}[!ht]
    \centering
    \includegraphics[width=0.45\linewidth]{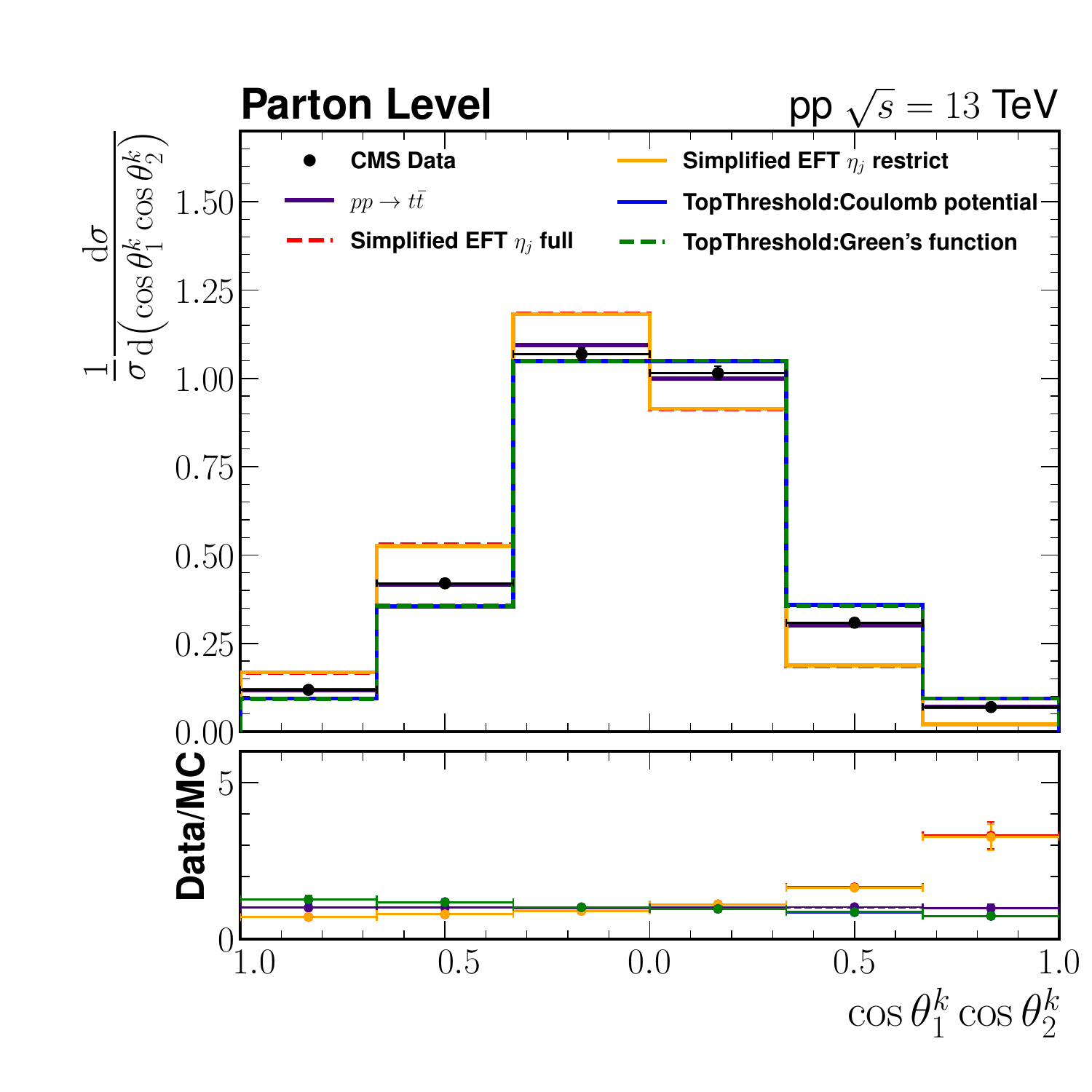}
    \includegraphics[width=0.45\linewidth]{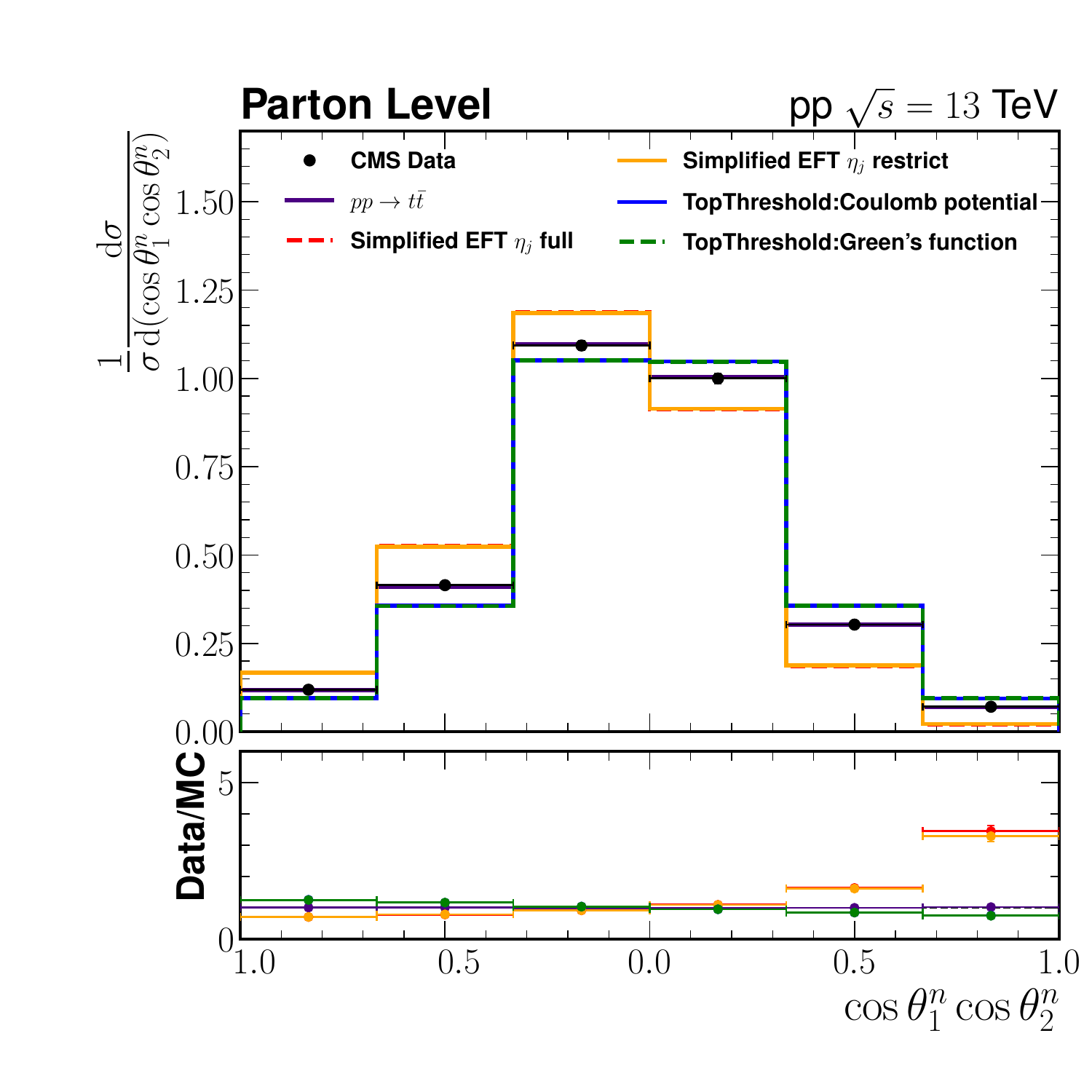}\\
    \includegraphics[width=0.45\linewidth]{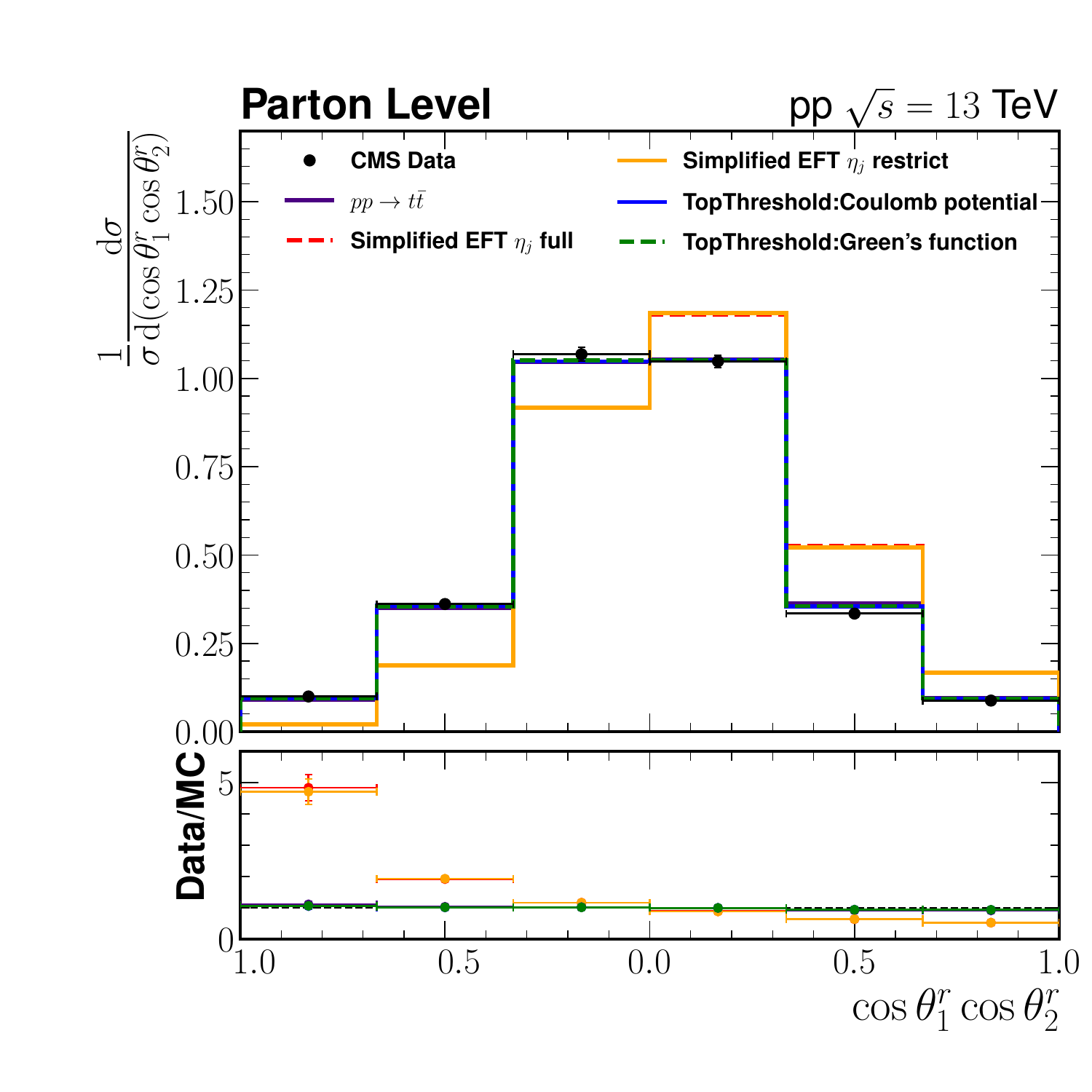}
    \includegraphics[width=0.45\linewidth]{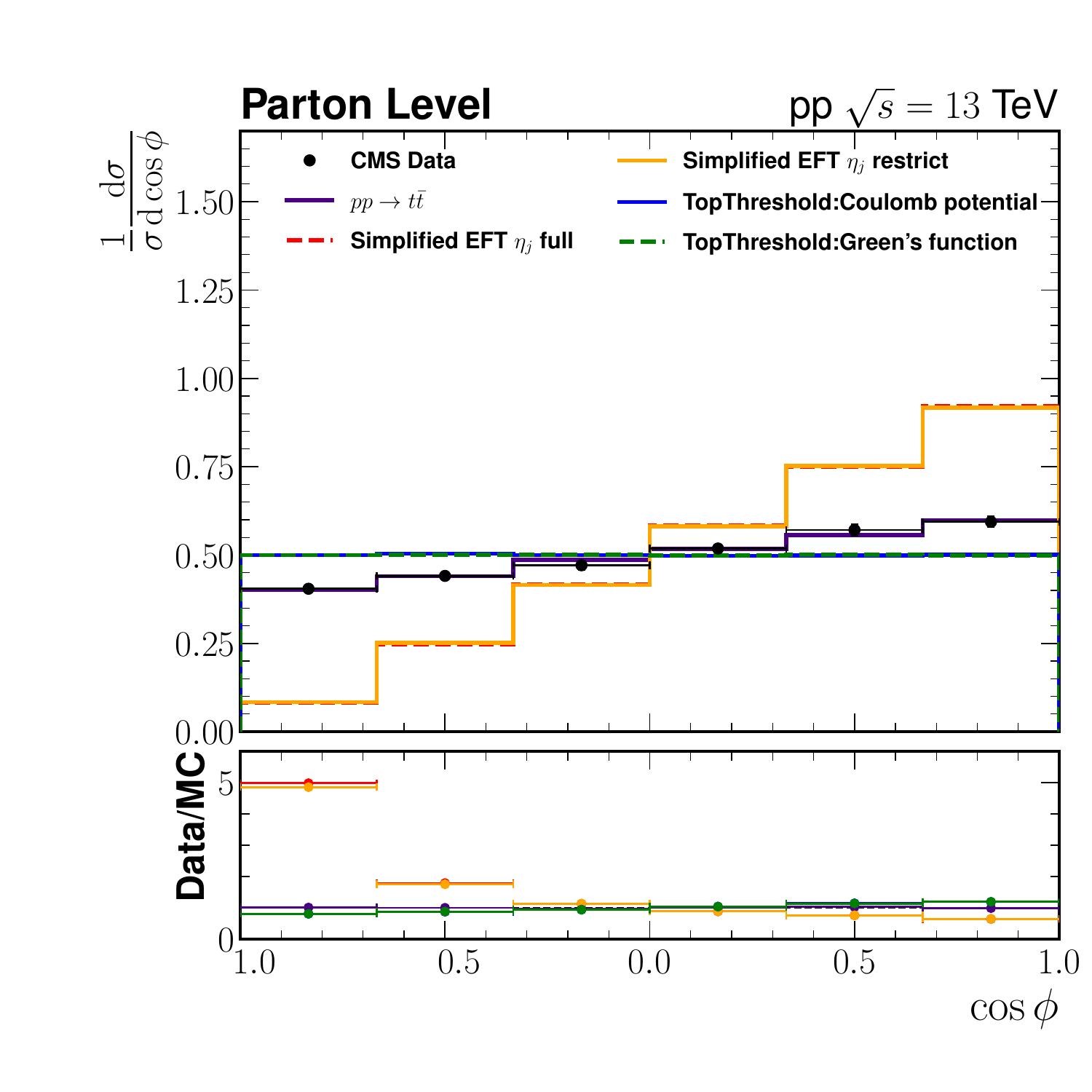}
    \caption{Unfolded CMS data and predictions for $\cos\theta^{i}_{1}\cos\theta^{i}_{2}$ ($i=k,n,r$) and for the opening angle $\cos\phi$ of the leptons in their parent rest frames, normalised to unit area. The lower panels show data/prediction.}
    \label{fig:c_coeff_all}
\end{figure}

\noindent Figure~\ref{fig:c_coeff_all} shows $\cos\theta^{i}_1\cos\theta^{i}_2$ and $\cos\phi$. With respect to the CMS data, the $\eta_t$ prediction for $\hat{k}$ is enhanced at negative products, by factors of 1.4, 1.3 and 1.1 in the three negative bins, and depleted at large positive products, where data/prediction is $\simeq1.6$ and $\simeq3$ in the last two bins. The $\hat{r}$ and $\hat{n}$ distributions are quite similar to the $\hat{k}$ one. In the opening angle the prediction rises linearly from $\simeq0.09$ to $\simeq0.92$ across $-1<\cos\phi<1$, which is $D=-1$, while the data give $D=-0.237\pm0.011$.

\begin{figure}[!ht]
    \centering
    \includegraphics[width=0.49\linewidth]{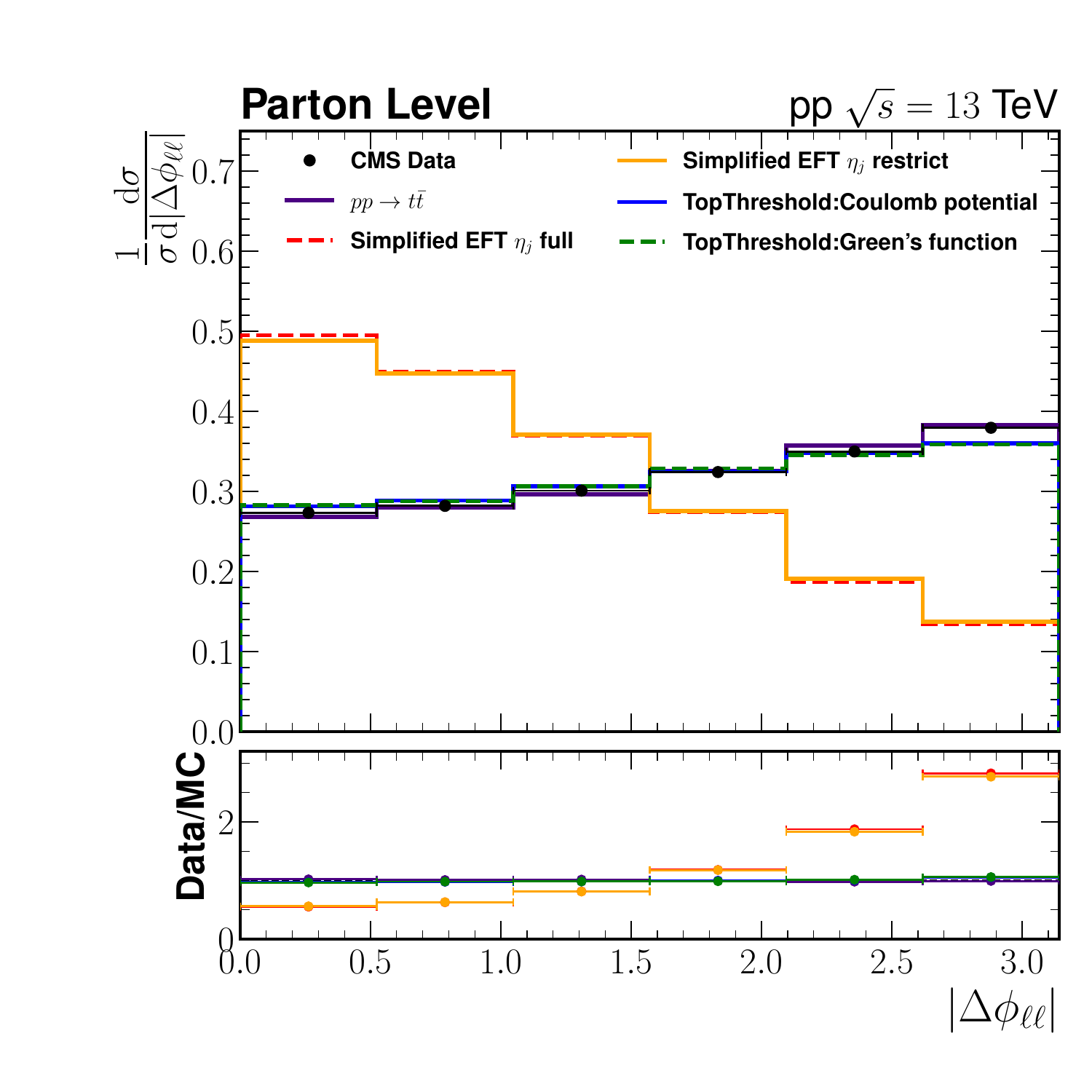}
    \includegraphics[width=0.49\linewidth]{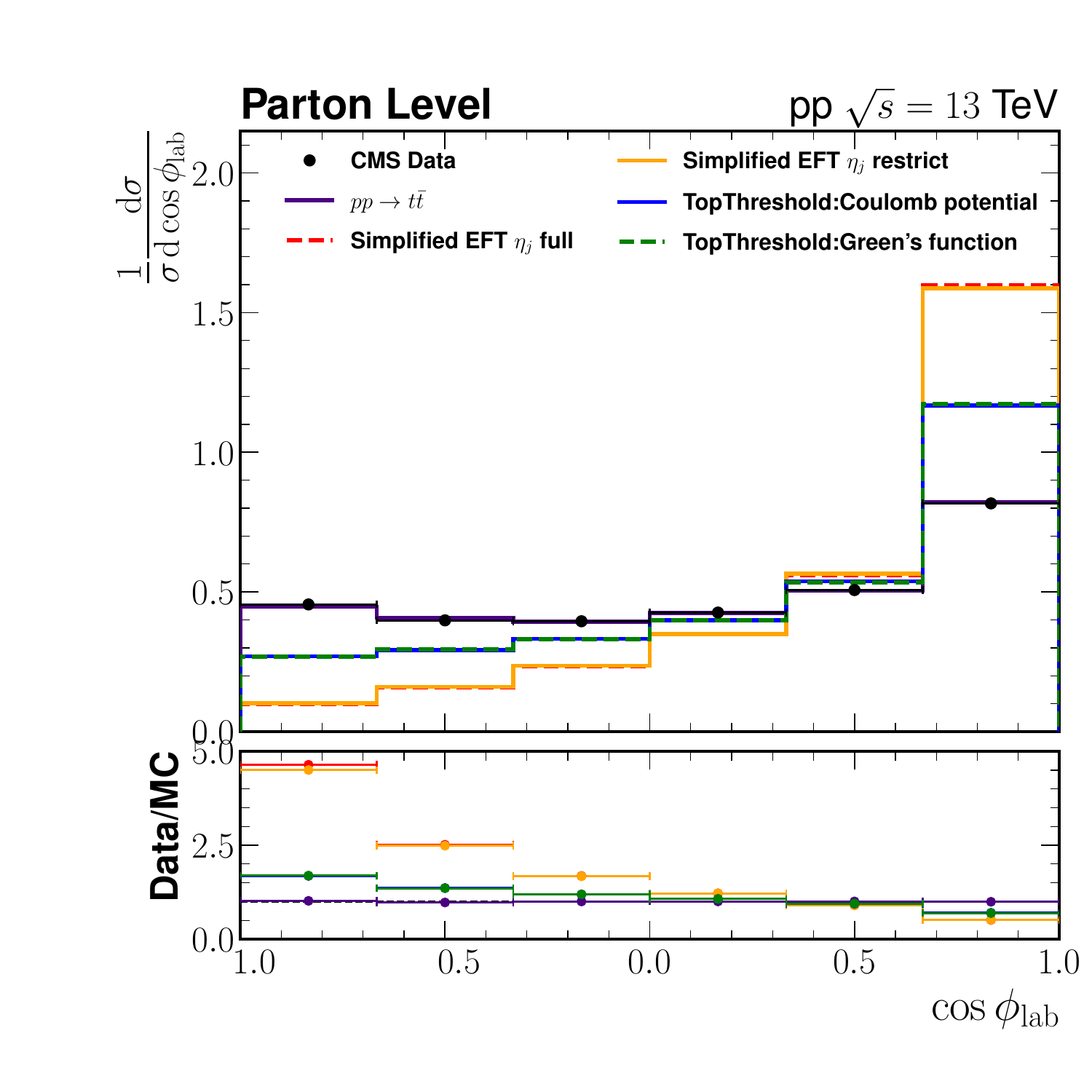}
    
    \caption{Normalised parton-level distributions of $|\Delta\phi_{\ell\ell}|$ (left) and $\cos\phi_{\rm lab}$ (right) for CMS data~\cite{CMS:2019nrx}, the $t\bar{t}$ baseline, the Restricted and Full $\eta_t$ samples and the TopThreshold samples. The lower panels show data/prediction.}
    \label{ttbar_deltaphi}
\end{figure}

\noindent Figure~\ref{ttbar_deltaphi} shows $|\Delta\phi_{\ell\ell}|$ and $\cos\phi_{\rm lab}$. The perturbative $t\bar{t}$ prediction describes the measured $|\Delta\phi_{\ell\ell}|$ distribution, while the standalone NR-QCD predictions exhibit flatter distributions (0.28 to 0.36), within $\sim10\%$ of the data in every bin. The $\eta_t$ prediction has the opposite trend: 0.49 in the first bin, a factor 1.8 above the data, falling to 0.135 in the last, where data/prediction is 2.8. In $\cos\phi_{\rm lab}$ the $\eta_t$ prediction exceeds the data by a factor 1.95 in the last bin (1.60 against 0.82) and falls below them by a factor $\simeq4.5$ in the first (0.10 against 0.455). The NR-QCD predictions reach 1.17 in the last bin, 1.4 times the data, and 0.27 in the first.

\noindent Table~\ref{tab:spin_coeff_combined} reports the fitted coefficients for the toponium EFT model which includes $t\bar{t}$ decays and NLO perturbative baseline background that agrees with the Data reference within $1.6\sigma$ for the three diagonal coefficients: $C_{kk},C_{rr},C_{nn}\to+1$. For the case of the toponium models the diagonal fitted coefficients show a perceptual increase in value ranging from 200\%-2330\% for toponium $t\bar{t}$ decays with respect the perturbative NLO baseline. The increase for the three values has a positive sign. \(D\) coefficient also increases by 329.2\%. The numerical values indicates that characterization of observed excess events via spin correlations would clearly indicate a substantial increase in value if toponium scenario is favoured.\\

\noindent The fitted coefficients results for the case of non-relativistic resummation effect (NR-QCD) are also included in Table~\ref{tab:spin_coeff_combined} so they can also be compared with Data, perturbative $t\bar{t}$ baseline reference and toponium models prospects. Resummation NR-QCD relative magnitudes on matrix diagonal coefficient fits range from 0.6\% to 10.4\%. Indicating a quite reduced amount of correlation is attributed to a NR resummation effect. It is important to highlight that the correlation results with respect perturbative NLO is different for toponium case and NR resummation effect, the first one increases the correlation level of the top quarks, most possible cause the dynamics associated with the bound state and non-preferred back-to-back emission but showing a flat $t\bar{t}$ $\Delta$R distribution with slight preference to smaller values and then aligned top quarks. For the case of NR resummation effect in Table~\ref{tab:spin_coeff_combined}, the correction if any will push towards uncorrelated  $t\bar{t}$ system. Their $\cos\phi$ distribution is flat, and the products $\cos\theta^{i}_1\cos\theta^{i}_2$ are symmetric about zero, i.e.\ $D\approx0$ and no spin correlation is present.\\

\begin{table}[!ht]
\centering

\begingroup
\scriptsize

\setlength{\tabcolsep}{2pt}
\renewcommand{\arraystretch}{1.7}

\resizebox{\linewidth}{!}{%

\begin{tabular}{
@{}l@{\hspace{7pt}}
*{5}{
S[table-format=-1.3]
@{}l
@{\hspace{7pt}}
}
S[table-format=-1.3]
@{}l@{}
}

\toprule

\multicolumn{5}{c}{}
& \multicolumn{4}{c}{Toponium}
& \multicolumn{4}{c}{NR-QCD}
\\

\cmidrule(lr){6-9}
\cmidrule(l){10-13}

Coefficient
& \multicolumn{2}{c}{CMS data}
& \multicolumn{2}{c}{$t\bar{t}$ NLO}
& \multicolumn{2}{c}{Restricted LO}
& \multicolumn{2}{c}{Full LO}
& \multicolumn{2}{c}{Coulomb}
& \multicolumn{2}{c}{Green}
\\

\midrule


$B_1^k$
& 0.005 & \spinunc{0.023}
& -0.008 & \spinunc{0.004}
& 0.002 & \spinunc{0.003}
& 0.000 & \spinunc{0.003}
& -0.005 & \spinunc{0.002}
& 0.001 & \spinunc{0.002}
\\

$B_2^k$
& 0.007 & \spinunc{0.023}
& -0.008 & \spinunc{0.004}
& 0.001 & \spinunc{0.003}
& -0.005 & \spinunc{0.003}
& -0.001 & \spinunc{0.002}
& 0.000 & \spinunc{0.002}
\\

$B_1^r$
& -0.023 & \spinunc{0.017}
& -0.005 & \spinunc{0.004}
& 0.003 & \spinunc{0.003}
& 0.001 & \spinunc{0.003}
& -0.003 & \spinunc{0.002}
& -0.002 & \spinunc{0.002}
\\

$B_2^r$
& -0.010 & \spinunc{0.020}
& 0.006 & \spinunc{0.004}
& 0.001 & \spinunc{0.003}
& -0.001 & \spinunc{0.003}
& -0.002 & \spinunc{0.002}
& 0.002 & \spinunc{0.002}
\\

$B_1^n$
& 0.006 & \spinunc{0.013}
& 0.003 & \spinunc{0.004}
& 0.004 & \spinunc{0.003}
& 0.002 & \spinunc{0.003}
& -0.001 & \spinunc{0.002}
& -0.005 & \spinunc{0.002}
\\

$B_2^n$
& 0.017 & \spinunc{0.013}
& -0.004 & \spinunc{0.004}
& 0.000 & \spinunc{0.003}
& -0.005 & \spinunc{0.003}
& -0.002 & \spinunc{0.002}
& 0.004 & \spinunc{0.002}
\\

\midrule


$C_{kk}$
& 0.300 & \spinunc{0.038}
& 0.338 & \spinunc{0.007}
& 1.003 & \spinunc{0.004}
& 1.012 & \spinunc{0.004}
& -0.001 & \spinunc{0.004}
& -0.007 & \spinunc{0.004}
\\

$C_{rr}$
& 0.081 & \spinunc{0.032}
& 0.041 & \spinunc{0.007}
& 0.998 & \spinunc{0.004}
& 0.999 & \spinunc{0.004}
& -0.003 & \spinunc{0.004}
& -0.004 & \spinunc{0.004}
\\

$C_{nn}$
& 0.329 & \spinunc{0.020}
& 0.325 & \spinunc{0.007}
& 1.001 & \spinunc{0.004}
& 1.012 & \spinunc{0.004}
& 0.001 & \spinunc{0.004}
& -0.001 & \spinunc{0.004}
\\

\midrule


$D$
& -0.237 & \spinunc{0.011}
& -0.235 & \spinunc{0.004}
& -1.001 & \spinunc{0.002}
& -1.008 & \spinunc{0.002}
& 0.001 & \spinunc{0.002}
& 0.004 & \spinunc{0.002}
\\

\bottomrule

\end{tabular}%

}

\endgroup

\caption{
Comparison of top-quark polarization and spin-correlation
coefficients measured by CMS~\cite{CMS:2019nrx}
with the perturbative $t\bar{t}$ NLO baseline,
restricted and full LO toponium samples,
and threshold-only NR-QCD predictions obtained using
the Coulomb and Green-function models.
The quoted MC uncertainties are statistical only.
}

\label{tab:spin_coeff_combined}

\end{table}


\section{Non-relativistic resummation fits to spin and angular distributions}
\label{sec:top_threshold_template_fits}

\noindent Distribution at Figure~\ref{ttbar_deltaphi} can be used to identify at which degree the non-relativistic effect shapes for Green function and Coulomb potentials resummations can improve the modelling agreement with the Data distributions when combined to the $t\bar{t}$ NLO perturbative baseline, here we first describe the combination procedure in order to extract the best fit result for fractional contribution for the additional resummation effect. The contribution from top-threshold production to the
dilepton angular distribution was investigated through a comparison
with the normalized CMS measurement of
\(|\Delta\phi_{\ell\ell}|\) at parton level and
\(\sqrt{s}=13~\mathrm{TeV}\). The inclusive \(pp\to t\bar t\)
prediction was combined separately with the Green-function and
Coulomb threshold descriptions. The threshold model was denoted by
\(X\in\{G,C\}\), where \(G\) and \(C\) represented the Green-function
and Coulomb descriptions, respectively. For each bin \(i\), the
combined template and its finite-simulation covariance were
constructed as
\begin{equation}
\begin{aligned}
T_{X,i}(f_X)
&=(1-f_X)T_{t\bar t,i}
  +f_XT_{\mathrm{thr},X,i},\\
V_{\mathrm{MC},X}(f_X)
&=(1-f_X)^2V_{\mathrm{MC},t\bar t}
  +f_X^2V_{\mathrm{MC},\mathrm{thr},X}.
\end{aligned}
\label{eq:threshold_template_mixture}
\end{equation}
\noindent Here, \(T_{X,i}(f_X)\) denoted the predicted normalized
content of bin \(i\) for threshold model \(X\).
The quantities \(T_{t\bar t,i}\) and
\(T_{\mathrm{thr},X,i}\) represented the normalized bin contents of
the inclusive \(t\bar t\) and threshold templates, respectively.
The parameter \(f_X\), constrained to \(0\leq f_X\leq1\), represented
the threshold fraction, while \(1-f_X\) corresponded to the inclusive
\(t\bar t\) fraction. The matrices \(V_{\mathrm{MC},t\bar t}\) and
\(V_{\mathrm{MC},\mathrm{thr},X}\) described the finite-simulation
statistical covariances of the two normalized templates, and
\(V_{\mathrm{MC},X}(f_X)\) represented their combined covariance at
a given value of \(f_X\). The squared coefficients in the second line
followed from the linear propagation of independent template
uncertainties. Statistical uncertainties from the independent simulation
samples were propagated through the normalization Jacobian, so that
the correlations induced among the normalized bins were retained.
The statistical and systematic uncertainties of the CMS measurement
were included through their full covariance matrices.\\

\noindent To account for the unit-area constraint, the comparison was
performed in five independent shape directions. An orthonormal
projection matrix \(Q\) was constructed with
\(\boldsymbol w^{T}Q=0\), where the bin widths were denoted by
\(w_i=\Delta x_i\). The fitted contribution was obtained by minimizing
\begin{equation}
\chi_X^2(f_X)=
\boldsymbol r_X^{T}
\left[
Q^{T}
\left(
V_{\mathrm{stat}}+V_{\mathrm{syst}}
+V_{\mathrm{MC},X}(f_X)
\right)
Q
\right]^{-1}
\boldsymbol r_X,
\label{eq:threshold_shape_chi2}
\end{equation}
\noindent where the projected data--prediction residual was defined
as \(\boldsymbol r_X=Q^{T}(\boldsymbol d-\boldsymbol T_X)\).
Finite-simulation errors were treated as Gaussian auxiliary template
nuisances and were profiled without a covariance-determinant term.
One parameter was fitted in each case, and four nominal degrees of
freedom were assigned. Confidence intervals were extracted from
\(\Delta\chi_X^2=\chi_X^2-\chi_{X,\min}^2\), for which the
one-parameter asymptotic thresholds \(1\) and \(3.841\) were adopted
at 68.27\% and 95\% confidence, respectively.\\

\begin{figure}[!ht]
    \centering
    \begin{minipage}[t]{0.49\linewidth}
        \vspace{0pt}
        \includegraphics[width=\linewidth]{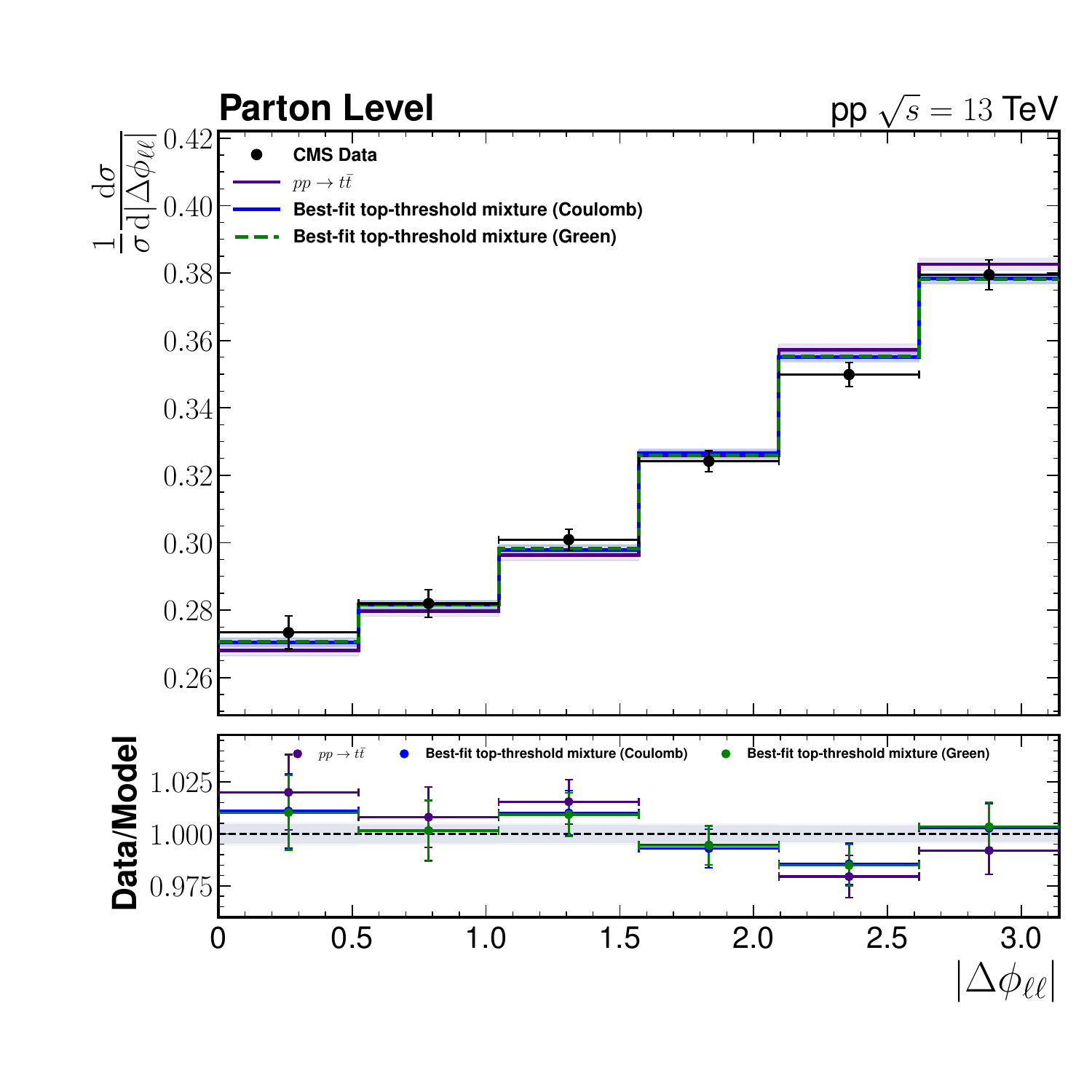}
    \end{minipage}\hfill
    \begin{minipage}[t]{0.49\linewidth}
        \vspace{0pt}
        \includegraphics[width=\linewidth]{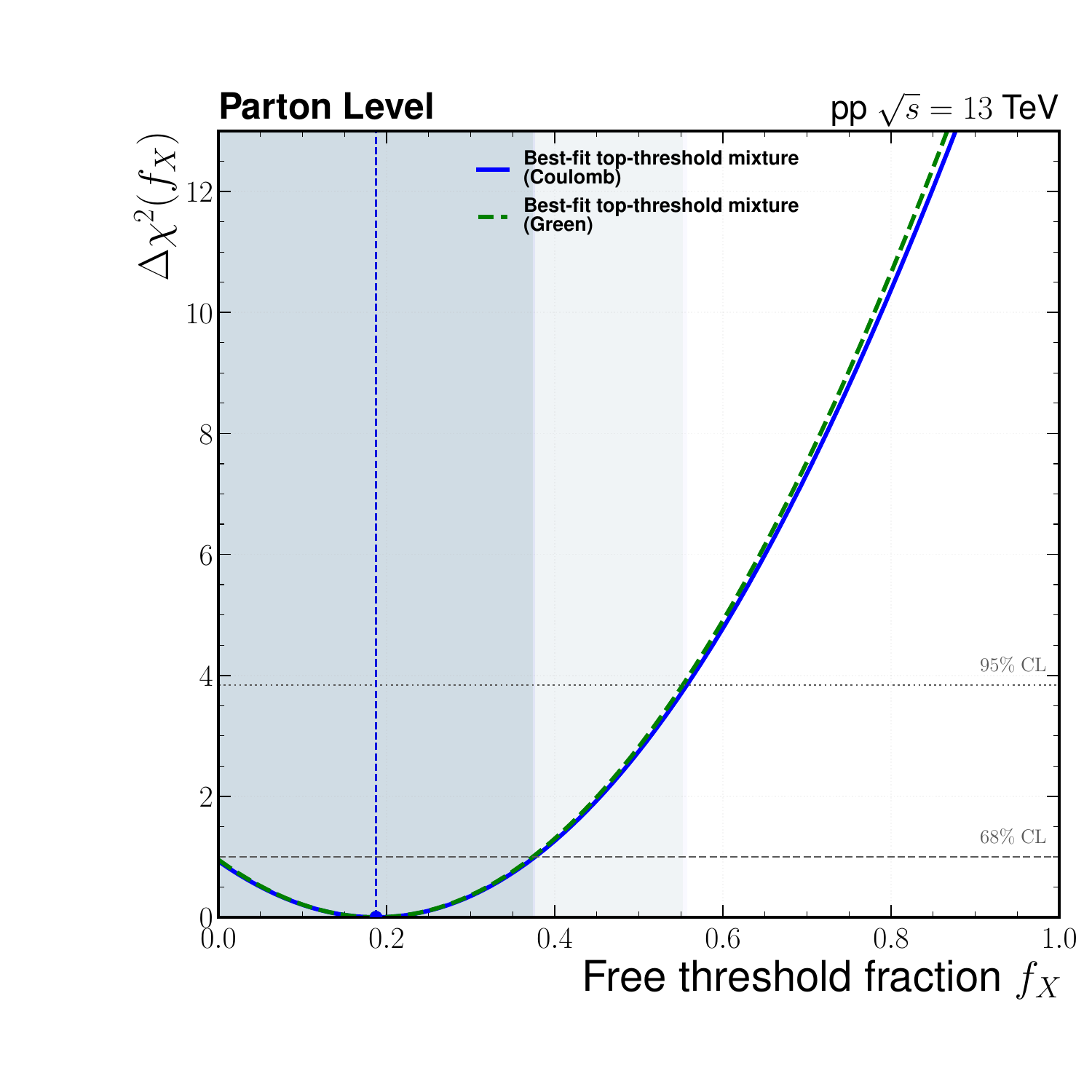}
    \end{minipage}

    \caption{The normalized \(\cos\phi_{\mathrm{lab}}\) distributions
    obtained with free Green and Coulomb threshold fractions were
    compared with CMS data on the left, with Data/Model ratios
    provided below.
    The corresponding \(\Delta\chi^2(f_X)\) profiles were displayed
    on the right, where the 68\% and 95\% confidence thresholds were
    indicated by horizontal lines.}
    \label{fig:threshold_fraction_fit}
\end{figure}

\noindent In the first case, the threshold fraction \(f_X\) was
allowed to vary between zero and one, while the inclusive fraction
was constrained to \(f_{t\bar t}=1-f_X\). Equivalent component
cross sections were then assigned using the fixed total dileptonic
reference \(\sigma_{\mathrm{ref}}=39.340~\mathrm{pb}\),
\begin{equation}
\sigma_X^{\mathrm{eq}}
=f_X\sigma_{\mathrm{ref}},
\qquad
\sigma_{t\bar t}^{\mathrm{eq}}
=(1-f_X)\sigma_{\mathrm{ref}}.
\label{eq:threshold_fixed_total_mapping}
\end{equation}
\noindent Under this prescription, the assigned inclusive contribution
was reduced as the threshold fraction was increased, and their sum
was kept fixed. The fitted distributions were compared with the CMS
measurement in Figure~\ref{fig:threshold_fraction_fit} on the left,
and the fraction dependence of \(\Delta\chi^2\) was displayed on the
right.\\

\begin{figure}[!ht]
    \centering
    \begin{minipage}[t]{0.49\linewidth}
        \vspace{0pt}
        \includegraphics[width=\linewidth]{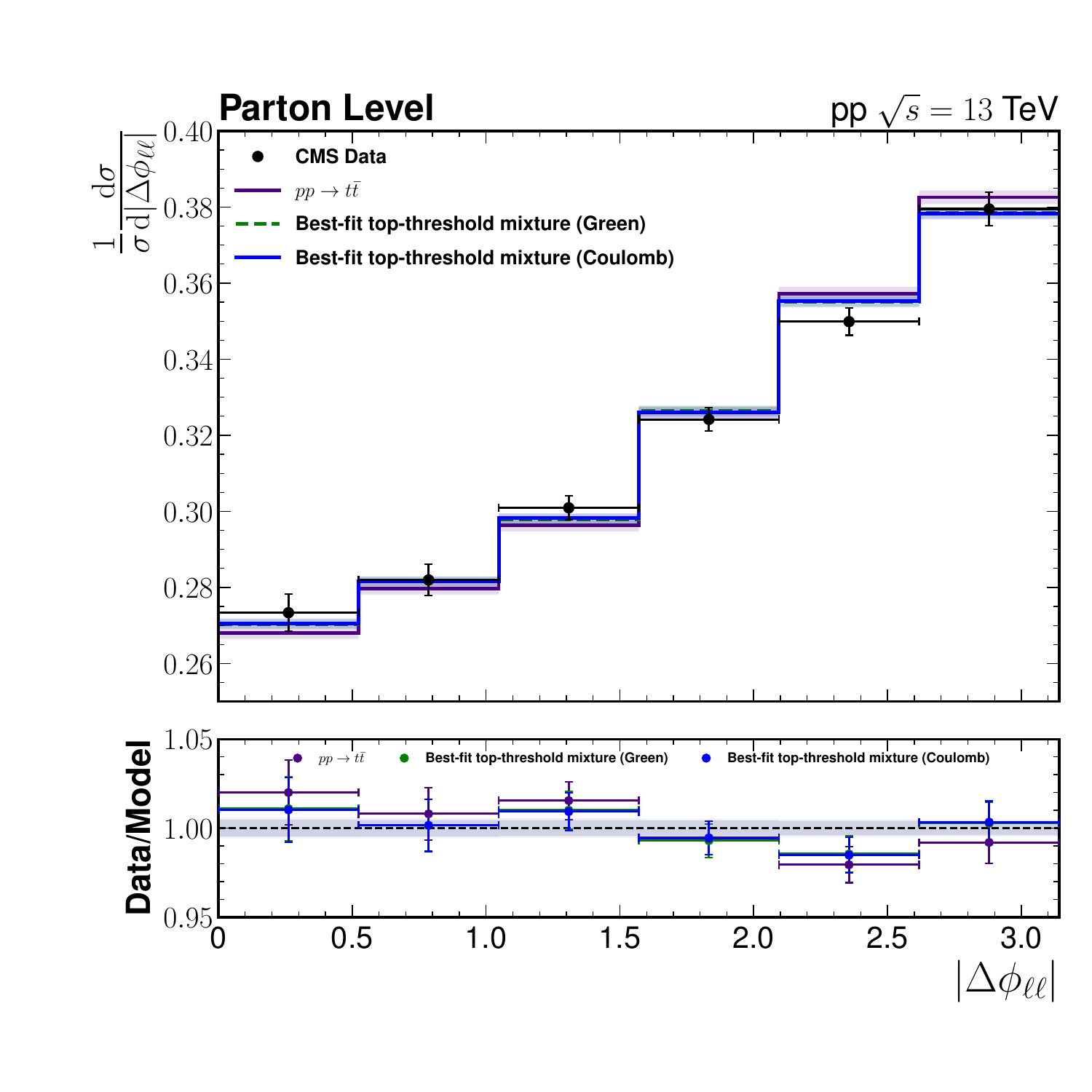}
    \end{minipage}\hfill
    \begin{minipage}[t]{0.49\linewidth}
        \vspace{0pt}
        \includegraphics[width=\linewidth]{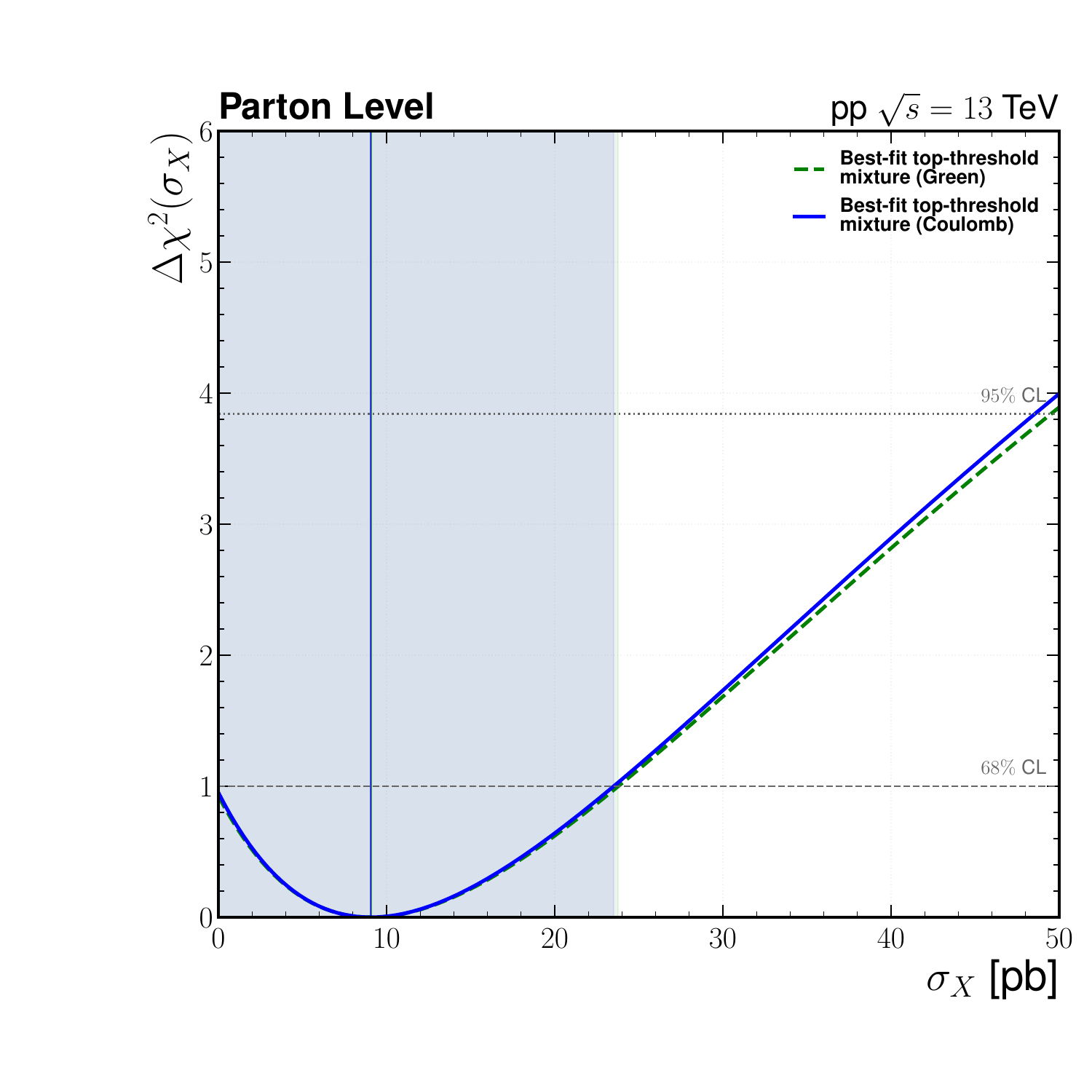}
    \end{minipage}

    \caption{The fitted distributions and Data/Model ratios were
    compared on the left after the inclusive contribution was fixed
    to \(\sigma_{t\bar t}=39.340~\mathrm{pb}\).
    The Green and Coulomb \(\Delta\chi^2(\sigma_X)\) profiles were
    displayed on the right, together with the 68\% and 95\%
    confidence thresholds.}
    \label{fig:threshold_cross_section_fit}
\end{figure}

\noindent In the second case, the inclusive dileptonic $t\bar{t}$ component was
fixed to \(\sigma_{t\bar t}=39.340~\mathrm{pb}\), and the threshold
cross section was allowed to vary with \(\sigma_X\geq0\).
The same normalized mixture was expressed through
\begin{equation}
f_X=
\frac{\sigma_X}{\sigma_{t\bar t}+\sigma_X},
\qquad
\sigma_X=
\sigma_{t\bar t}\frac{f_X}{1-f_X}.
\label{eq:threshold_fixed_ttbar_mapping}
\end{equation}
\noindent The total rate was therefore assigned as
\(\sigma_{t\bar t}+\sigma_X\), and the inclusive contribution was
kept unchanged. With identical templates and covariances, the
fraction fit was reparameterized through this relation, and the same
fitted shape and minimum \(\chi^2\) were retained. The fraction and
cross-section interval endpoints were related by the same transformation.
The resulting distributions and cross-section profiles were compared
in Figure~\ref{fig:threshold_cross_section_fit}.\\

\noindent The fitted threshold contributions and goodness-of-fit
results were summarized in Table~\ref{tab:threshold_fit_summary_dphi}.
The uncertainties obtained from the profile likelihood included the
CMS statistical and systematic covariance matrices together with the
finite-simulation statistics. For the fixed-\(t\bar t\)
parameterization, the profile and normalization uncertainties were
combined in quadrature. The latter included variations from the
renormalization and factorization scales, PDF plus \(\alpha_s\), and
the top-quark mass.\\

\begin{table}[!ht]
\centering
\footnotesize
\renewcommand{\arraystretch}{1.10}

\begin{tabular*}{\linewidth}{@{\extracolsep{\fill}}lcc@{}}
\toprule

\multicolumn{3}{l}{\textbf{Baseline prediction}} \\
\midrule

& $\chi^2/N_{\mathrm{dof}}$
& CMS reference \\

Pure $t\bar t$ dileptonic
& $4.429/5$
& $4.0/5$ \\

\addlinespace[0.4em]
\midrule

\multicolumn{3}{l}{\textbf{Threshold-template fits}} \\
\midrule

Fitted quantity
& Green
& Coulomb \\

\midrule

$\hat f_X$ [\%]
& $18.719^{+18.686}_{-18.719}$
& $18.756^{+18.921}_{-18.756}$ \\

$\sigma_X^{\mathrm{eq}}$ [pb], fixed total
& $\mathbf{7.364}^{+7.351}_{-7.364}$
& $\mathbf{7.379}^{+7.443}_{-7.379}$ \\

$\sigma_X$ [pb], fixed $\sigma_{t\bar t}$
& $\mathbf{9.060}^{+14.448}_{-9.060}$
& $\mathbf{9.082}^{+14.700}_{-9.082}$ \\

$\chi^2/N_{\mathrm{dof}}$
& $3.478/4$
& $3.502/4$ \\

\bottomrule

\end{tabular*}

\caption{Baseline dileptonic $t\bar t$ prediction and threshold-template
fits to the normalized $|\Delta\phi_{\ell\ell}|$ distribution.
The corresponding CMS value of $\chi^2/N_{\mathrm{dof}}=4.0/5$
for the baseline prediction is shown for comparison.
The Green- and Coulomb-template uncertainties correspond to the
68\% profile intervals. The equivalent cross sections in the fixed-total
prescription are obtained using a reference dilepton cross section of
$39.340~\mathrm{pb}$.}

\label{tab:threshold_fit_summary_dphi}
\end{table}

\noindent From Figure~\ref{fig:threshold_cross_section_fit} and Table~\ref{tab:threshold_fit_summary_dphi} it can then be concluded that non-relativistic resummation effect when combined with perturbative baseline improves the agreement with the $t\bar{t}$ spin correlation Data distributions by reducing its $\chi^2/N_\mathrm{dof}$ value by a factor of $\sim$21.5\%, delivering a fitted cross-section for the additional resummation effect within a range from 7.3 to 9.1 pb in consistency with Data excess beyond the perturbative baseline recently reported by the LHC experiments, excluding a contribution $>$ 50 pb. Though a Data measurement restricted to low $m_{t\bar{t}}$ values $<$ 400 GeV is required, as the combination procedure here is performed over the whole mass spectrum while it introduces an extra effect that affects only the low mass area near to the threshold. The reported exclusion limits for the resummation effect can then be recalculated from same procedure but restricting angular measurement to low mass $t\bar{t}$ events or categorizing by  $m_{t\bar{t}}$ classes. Comparable fractions, cross sections and goodness-of-fit values were obtained with the Green and Coulomb templates, and no statistically significant preference between them was established. The difference between the cross sections obtained under the two normalization prescriptions resulted from whether the total rate or only the inclusive \(t\bar t\) contribution was kept fixed. Since a normalized distribution primarily constrained the relative threshold fraction, the absolute cross section remained dependent on the adopted normalization prescription and its phase-space interpretation.\\

\begin{table}[htbp]
\centering

\begingroup
\footnotesize
\setlength{\tabcolsep}{3pt}
\renewcommand{\arraystretch}{1.15}

\resizebox{\textwidth}{!}{%
\begin{tabular}{@{}lcccc@{}}

\toprule

Observable
& CMS MG5\_aMC@NLO
& $t\bar{t}$ NLO
& Coulomb fit
& Green fit \\

\midrule


$\cos\theta_1^k$
& 0.20 & 0.31 & 0.25 & 0.24 \\

$\cos\theta_2^k$
& 1.00 & 1.05 & 1.25 & 1.17 \\

$\cos\theta_1^r$
& 0.88 & 0.51 & 0.64 & 0.64 \\

$\cos\theta_2^r$
& 0.10 & 0.25 & 0.10 & 0.16 \\

$\cos\theta_1^n$
& 0.36 & 0.34 & 0.43 & 0.43 \\

$\cos\theta_2^n$
& 0.62 & 0.87 & 0.92 & 0.53 \\

$\cos\theta_1^{k^*}$
& 0.26 & 0.35 & 0.40 & 0.37 \\

$\cos\theta_2^{k^*}$
& 0.32 & 0.40 & 0.44 & 0.39 \\

$\cos\theta_1^{r^*}$
& 0.30 & 0.26 & 0.32 & 0.32 \\

$\cos\theta_2^{r^*}$
& 0.12 & 0.09 & 0.10 & 0.11 \\

\midrule


$\cos\theta_1^k\cos\theta_2^k$
& 0.64 & 0.66 & 0.64 & 0.63 \\

$\cos\theta_1^r\cos\theta_2^r$
& 0.34 & 0.56 & 0.70 & 0.70 \\

$\cos\theta_1^n\cos\theta_2^n$
& 0.06 & 0.10 & 0.12 & 0.12 \\

\midrule


$\cos\theta_1^r\cos\theta_2^k
+\cos\theta_1^k\cos\theta_2^r$
& 0.32 & 0.40 & 0.45 & 0.48 \\

$\cos\theta_1^r\cos\theta_2^k
-\cos\theta_1^k\cos\theta_2^r$
& 0.62 & 1.00 & 1.24 & 0.94 \\

$\cos\theta_1^n\cos\theta_2^r
+\cos\theta_1^r\cos\theta_2^n$
& 0.34 & 0.32 & 0.40 & 0.40 \\

$\cos\theta_1^n\cos\theta_2^r
-\cos\theta_1^r\cos\theta_2^n$
& 0.38 & 0.32 & 0.39 & 0.39 \\

$\cos\theta_1^n\cos\theta_2^k
+\cos\theta_1^k\cos\theta_2^n$
& 0.80 & 0.73 & 0.91 & 0.81 \\

$\cos\theta_1^n\cos\theta_2^k
-\cos\theta_1^k\cos\theta_2^n$
& 0.48 & 0.36 & 0.45 & 0.45 \\

\midrule


$\cos\phi$
& 0.14 & 0.29 & 0.37 & 0.37 \\

$\cos\phi_{\mathrm{lab}}$
& 1.52 & 1.65 & 2.07 & 2.07 \\

$|\Delta\phi_{\ell\ell}|$
& 0.80 & 0.89 & 0.88 & 0.87 \\

\midrule


\textbf{All observables}
& \textbf{0.82}
& \textbf{0.74}
& \textbf{0.75}
& \textbf{0.75} \\

\bottomrule

\end{tabular}%
}

\endgroup

\caption{
Comparison of $\chi^2/\mathrm{ndof}$ values for the
22 normalized spin-correlation and angular observables.
The CMS MG5\_aMC@NLO results are compared with
our perturbative $t\bar{t}$ NLO prediction and
the fitted Coulomb and Green threshold mixtures.
The individual fixed predictions use five degrees
of freedom, while the individual mixture fits use
four nominal degrees of freedom.
The global fixed predictions use 110 degrees
of freedom and the global mixture fits use
109 nominal degrees of freedom.
}

\label{tab:spin_chi2_comparison}

\end{table}

\noindent Table~\ref{tab:spin_chi2_comparison} includes a shape agreement comparison for all angular distributions on the other hand after executing same combination procedure between NLO perturbative $t\bar{t}$ and NR resummation alternatives with Green function and Coulomb potential. $\chi^2$/N$_{dof}$ indicator shows a reduction in its value in selected cases such as $\cos\theta_i^{r, n, k}$ cases and $\cos\theta_1^k\cos\theta_2^k$ with improvements ranging from 4.5\% to 60\%. In average the general improvement considering all variables simultaneously is $\sim$1.35\%.\\


\section{Dileptonic top-pair reconstruction and detector-level observables}
\label{sec:detector_reconstruction_observables}

\noindent An approximate reconstruction motivated by the
dileptonic \(t\bar t\) reconstruction procedures used by
CMS~\cite{CMS:2019nrx,CMS:2025kzt} was applied to
\textsc{Delphes}~3.5.0~\cite{delphes} samples simulated
with the CMS detector card. Events were selected using
reconstructed objects according to
Table~\ref{tab:det_selection}.\\

\begin{table}[!ht]
\centering
\footnotesize
\renewcommand{\arraystretch}{1.10}
\setlength{\tabcolsep}{4pt}
\def\softline{%
    \arrayrulecolor{black!20}%
    \specialrule{0.25pt}{0.4pt}{0.4pt}%
    \arrayrulecolor{black}%
}
\begin{tabular*}{\linewidth}{
    @{}
    >{\raggedright\arraybackslash}p{\linewidth}
    @{}
}
\toprule
Event and object selection \\
\midrule
Exactly one missing-momentum entry with finite magnitude
and azimuth, and \(p_T^{\mathrm{miss}}\geq0\). \\
\softline
At least two electron or muon candidates with finite
kinematics, charge \(\pm1\), \(p_T>20~\mathrm{GeV}\)
and \(|\eta|<2.4\). \\
\softline
At least two candidates after rejecting electrons
with \(1.4442<|\eta|<1.5660\). \\
\softline
At least two candidates with finite
\(0\leq I_{\mathrm{rel}}\leq0.15\), where
\(I_{\mathrm{rel}}\) is the surrounding scalar transverse-momentum
sum, excluding the lepton, divided by its transverse momentum. \\
\softline
Exactly two leptons satisfying the preceding requirements. \\
\softline
Opposite lepton charges. \\
\softline
Leading-lepton \(p_T>25~\mathrm{GeV}\). \\
\softline
\(m_{\ell\ell}>20~\mathrm{GeV}\). \\
\softline
For same-flavour pairs,
\(|m_{\ell\ell}-m_Z|>15~\mathrm{GeV}\),
with \(m_Z=91.1876~\mathrm{GeV}\). \\
\softline
For same-flavour pairs,
\(p_T^{\mathrm{miss}}>40~\mathrm{GeV}\). \\
\softline
At least two jets with finite kinematics, nonnegative masses,
\(p_T>30~\mathrm{GeV}\) and \(|\eta|<2.4\). \\
\softline
At least two jets after requiring
\(\Delta R(j,\ell^\pm)\geq0.4\) for both leptons. \\
\softline
At least one \(b\)-tagged jet among the remaining jets. \\
\softline
A finite positive reconstruction score, with real neutrino
momenta, positive energies and satisfied kinematic constraints. \\
\bottomrule
\end{tabular*}
\caption{Sequential event selection. Same-flavour requirements
apply only to \(e^+e^-\) and \(\mu^+\mu^-\).
The reconstruction score is independent of the generator weight.}
\label{tab:det_selection}
\end{table}

\noindent Both lepton--jet associations were tested for each
jet pair with the largest available number of \(b\) tags.
The labels \(b\) and \(\bar b\) denote the jets assigned to
\(\ell^+\) and \(\ell^-\), respectively, without jet-charge
identification. Massless neutrino momenta were reconstructed
by imposing
\begin{equation}
\begin{aligned}
\vec p_{T,\nu}+\vec p_{T,\bar\nu}
&=\vec p_T^{\,\mathrm{miss}},\\
(p_{\ell^+}+p_\nu)^2&=m_{W^+,h}^2,
& (p_{\ell^-}+p_{\bar\nu})^2&=m_{W^-,h}^2,\\
(p_b+p_{\ell^+}+p_\nu)^2&=m_t^2,
& (p_{\bar b}+p_{\ell^-}+p_{\bar\nu})^2&=m_t^2.
\end{aligned}
\label{eq:det_constraints}
\end{equation}
Here, \(m_t=172.5~\mathrm{GeV}\), and \(h=1,\ldots,100\)
labels independent pairs of trial \(W\) masses drawn from
a relativistic Breit--Wigner distribution in squared mass,
with central mass \(80.4~\mathrm{GeV}\), width
\(2.1~\mathrm{GeV}\) and \(0<m_W<m_t\).
The same trials were used for all associations, without
additional smearing of reconstructed objects.\\

\noindent For each association and trial, the real,
positive-energy solution satisfying the constraints
with the smallest pair mass was retained. Solutions were
weighted by the product of approximate leading-order
lepton--jet mass densities for the two decay branches,
and the association with the largest summed weight was
selected. Its representative solution was the accepted
trial closest to the weighted mean of the six laboratory
top and antitop momentum components, using their squared
Euclidean distance. Top and neutrino four-momenta were
taken together from this trial.
Equal distances to the weighted mean were resolved
by choosing the trial with the smaller pair mass.
Equal association scores were resolved using the
smaller representative pair mass.
Any remaining ties followed the existing iteration order. Events without a valid
positive-score solution were rejected. The imposed top
masses restrict the reconstructed spectrum to
\(m_{t\bar t}\geq345~\mathrm{GeV}\).\\

\noindent Table~\ref{tab:det_observables24} lists the
24 classification inputs, retaining the angular conventions
introduced above. The set \(\mathcal J\) contains the selected
jets, with \(j_1\) and \(j_2\) ordered by decreasing \(p_T\).
The superscript \(\mathrm{av}\) denotes four-momenta averaged
over accepted trials of the selected association using the
reconstruction weights. For \(m_{T2}^{\ell\ell}\),
\(\ell_1=\ell^+\), \(\ell_2=\ell^-\), and
\(\vec q_{T,i}\) are trial invisible transverse momenta;
\(m_T\) is the lepton--invisible transverse mass evaluated
with zero invisible mass. These trial momenta are independent
of the fitted neutrinos, labelled \(\nu_1=\nu\) and
\(\nu_2=\bar\nu\).\\

\begin{table*}[!ht]
\centering
\footnotesize
\renewcommand{\arraystretch}{1.04}
\setlength{\tabcolsep}{3pt}
\def\softline{%
    \arrayrulecolor{black!20}%
    \specialrule{0.25pt}{0.4pt}{0.4pt}%
    \arrayrulecolor{black}%
}
\begin{tabular*}{\textwidth}{
    @{\extracolsep{\fill}}
    >{\raggedright\arraybackslash}p{0.27\textwidth}
    >{\raggedright\arraybackslash}
    p{\dimexpr0.73\textwidth-2\tabcolsep\relax}
    @{}
}
\toprule
Observable & Physical interpretation \\
\midrule

\(H_T^{\mathrm{vis}}\)
& Scalar transverse-momentum sum of the two selected
leptons and all selected jets. \\
\softline

\(H_T^\ell\)
& Scalar transverse-momentum sum of the two selected leptons. \\
\softline

\(\cos\!\left(\phi_{\ell\ell}^{t\bar t\,\mathrm{RF}}\right)\)
& Opening-angle cosine between the two lepton momenta,
both evaluated in the reconstructed \(t\bar t\) rest frame. \\
\softline

\(\Delta R_{b\bar b}\)
& Separation of the two assigned jets in laboratory
pseudorapidity and azimuth. \\
\softline

\(\Delta R_{\ell\ell}\)
& Separation of the two selected leptons in laboratory
pseudorapidity and azimuth. \\
\softline

\(p_T(j_1)\)
& Transverse momentum of the leading selected jet. \\
\softline

\(p_T(j_2)\)
& Transverse momentum of the subleading selected jet. \\
\softline

\(M_{\bar t}^{\mathrm{av}}\)
& Invariant mass of the reconstruction-weighted mean
antitop four-vector over accepted trials. \\
\softline

\(M_t^{\mathrm{av}}\)
& Invariant mass of the reconstruction-weighted mean
top four-vector over accepted trials. \\
\softline

\(m_{t\bar t}\)
& Invariant mass of the reconstructed top--antitop pair. \\
\softline

\(m_{\ell\ell}\)
& Invariant mass of the two selected leptons. \\
\softline

\(m_{T2}^{\ell\ell}\)
& Dilepton stransverse mass computed from the measured
leptons and missing transverse momentum with zero invisible
test mass, independently of the reconstructed neutrinos. \\
\softline

\(p_T(\nu_1)\)
& Reconstructed neutrino transverse momentum
in the positive-lepton branch. \\
\softline

\(p_T(\nu_2)\)
& Reconstructed antineutrino transverse momentum
in the negative-lepton branch. \\
\softline

\(p_T^{\mathrm{recoil}}/H_T^{\mathrm{vis}}\)
& Magnitude of the vector sum of the reconstructed top
and antitop transverse momenta, divided by
\(H_T^{\mathrm{vis}}\). \\
\softline

\(E_t^{\mathrm{rec}}\)
& Reconstructed top-quark energy in the laboratory frame. \\
\softline

\(p_T(t)\)
& Reconstructed top-quark transverse momentum. \\
\softline

\(\Delta R_{t\bar t}\)
& Separation of the reconstructed top and antitop
in laboratory pseudorapidity and azimuth. \\

\bottomrule
\end{tabular*}
\caption{Physical interpretation of 18 classification inputs.
The previously defined \(C_{nn}\), \(C_{rr}\),
\(\cos\theta_1^k\), \(\cos\theta_2^k\), \(\cos\phi\)
and \(\cos\phi_{\mathrm{lab}}\) are also evaluated at detector
level, completing the 24 inputs.}
\label{tab:det_observables24}
\end{table*}

\begin{figure*}[!ht]
    \centering

    \begin{subfigure}[t]{0.32\textwidth}
        \includegraphics[width=\linewidth]{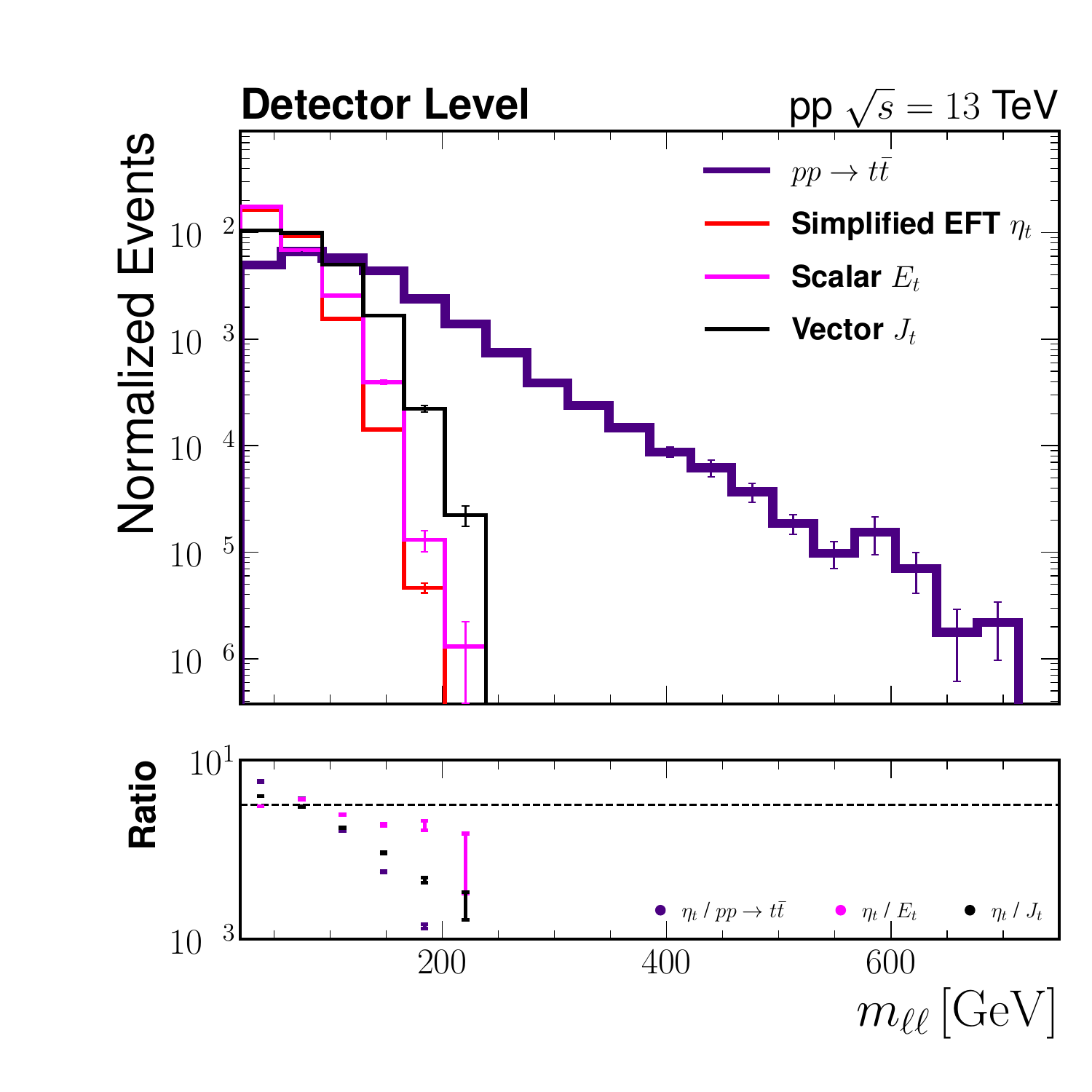}
    \end{subfigure}\hfill
    \begin{subfigure}[t]{0.32\textwidth}
        \includegraphics[width=\linewidth]{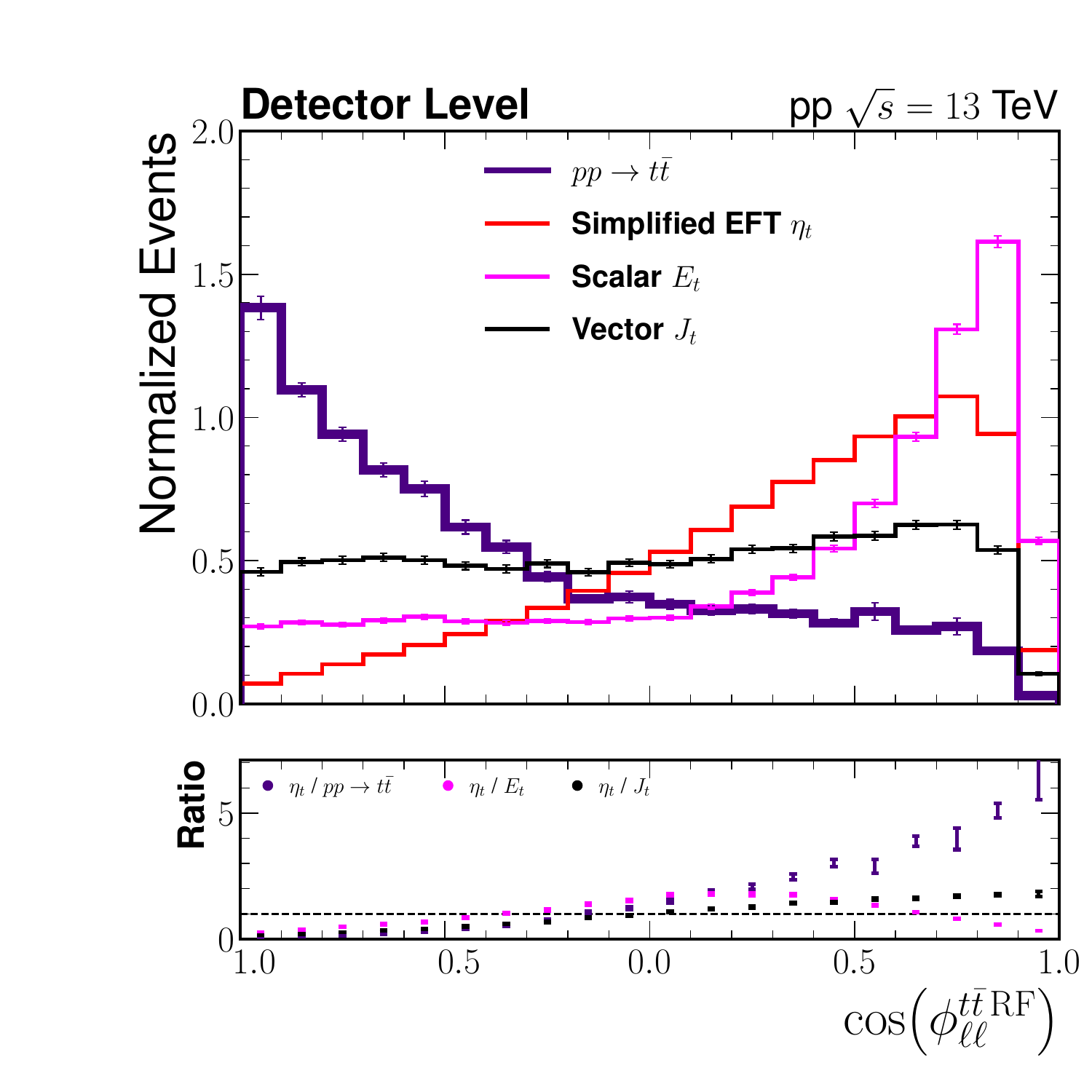}
    \end{subfigure}\hfill
    \begin{subfigure}[t]{0.32\textwidth}
        \includegraphics[width=\linewidth]{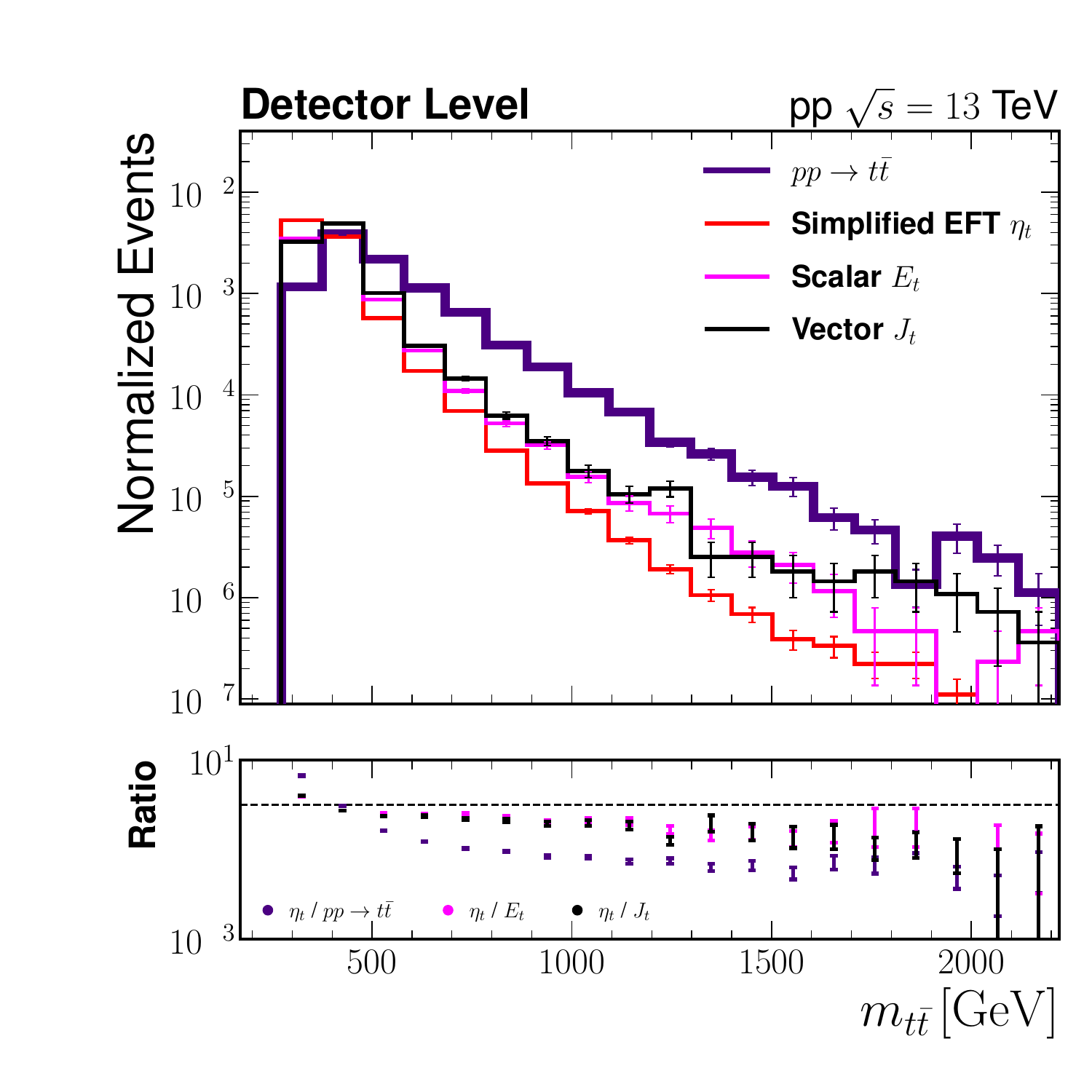}
    \end{subfigure}

    \par\medskip

    \begin{subfigure}[t]{0.32\textwidth}
        \includegraphics[width=\linewidth]{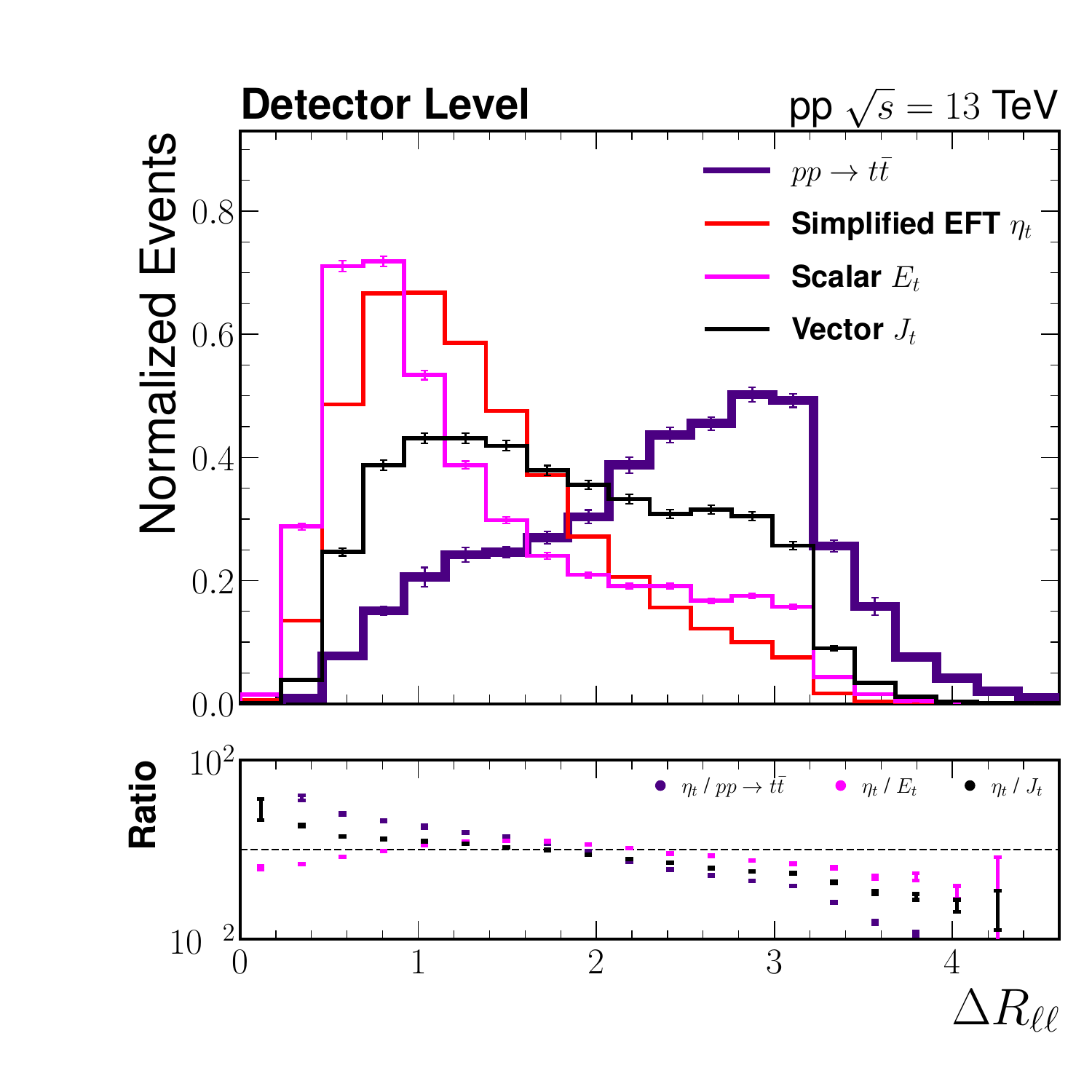}
    \end{subfigure}\hfill
    \begin{subfigure}[t]{0.32\textwidth}
        \includegraphics[width=\linewidth]{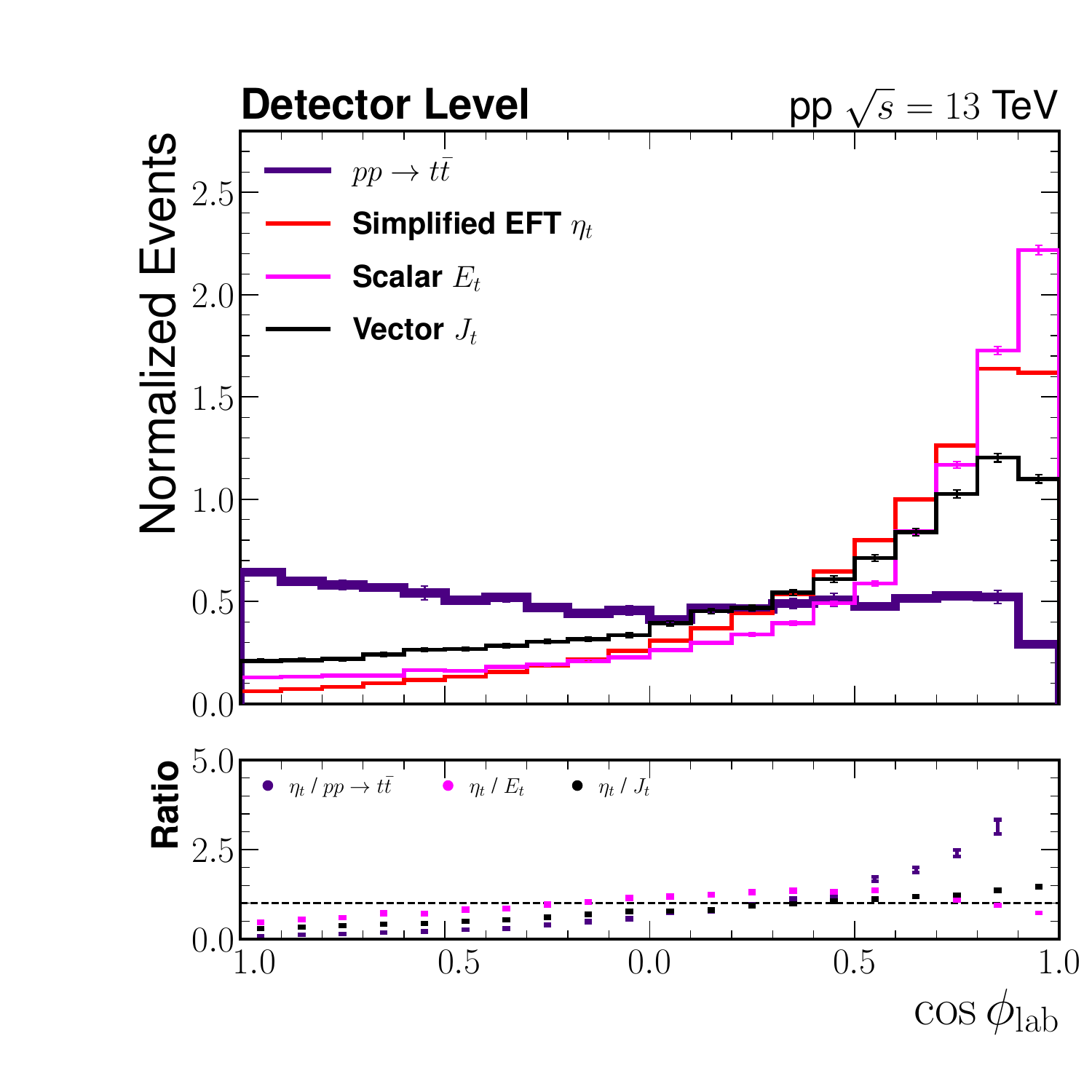}
    \end{subfigure}\hfill
    \begin{subfigure}[t]{0.32\textwidth}
        \includegraphics[width=\linewidth]{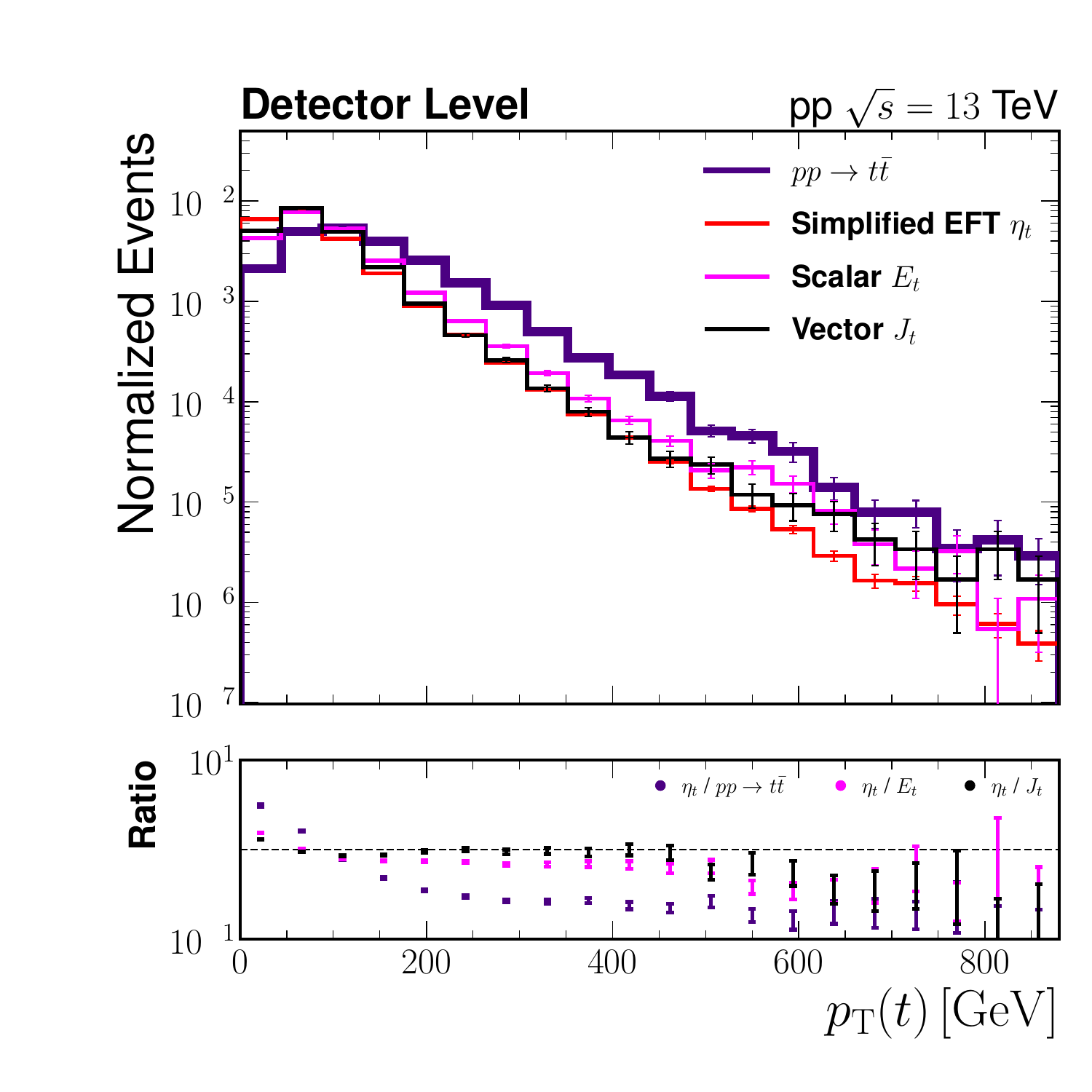}
    \end{subfigure}

    \par\medskip

    \begin{subfigure}[t]{0.32\textwidth}
        \includegraphics[width=\linewidth]{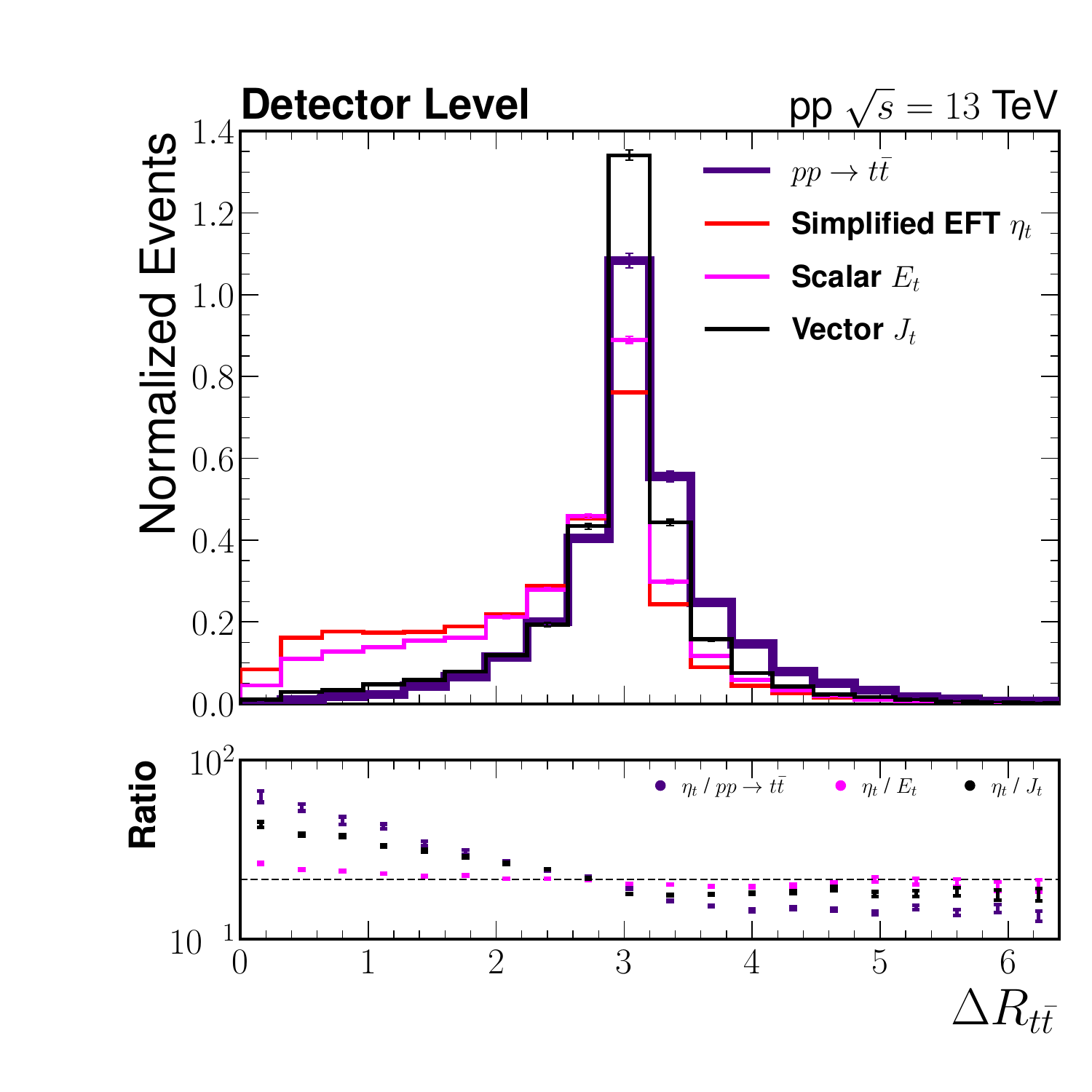}
    \end{subfigure}\hfill
    \begin{subfigure}[t]{0.32\textwidth}
        \includegraphics[width=\linewidth]{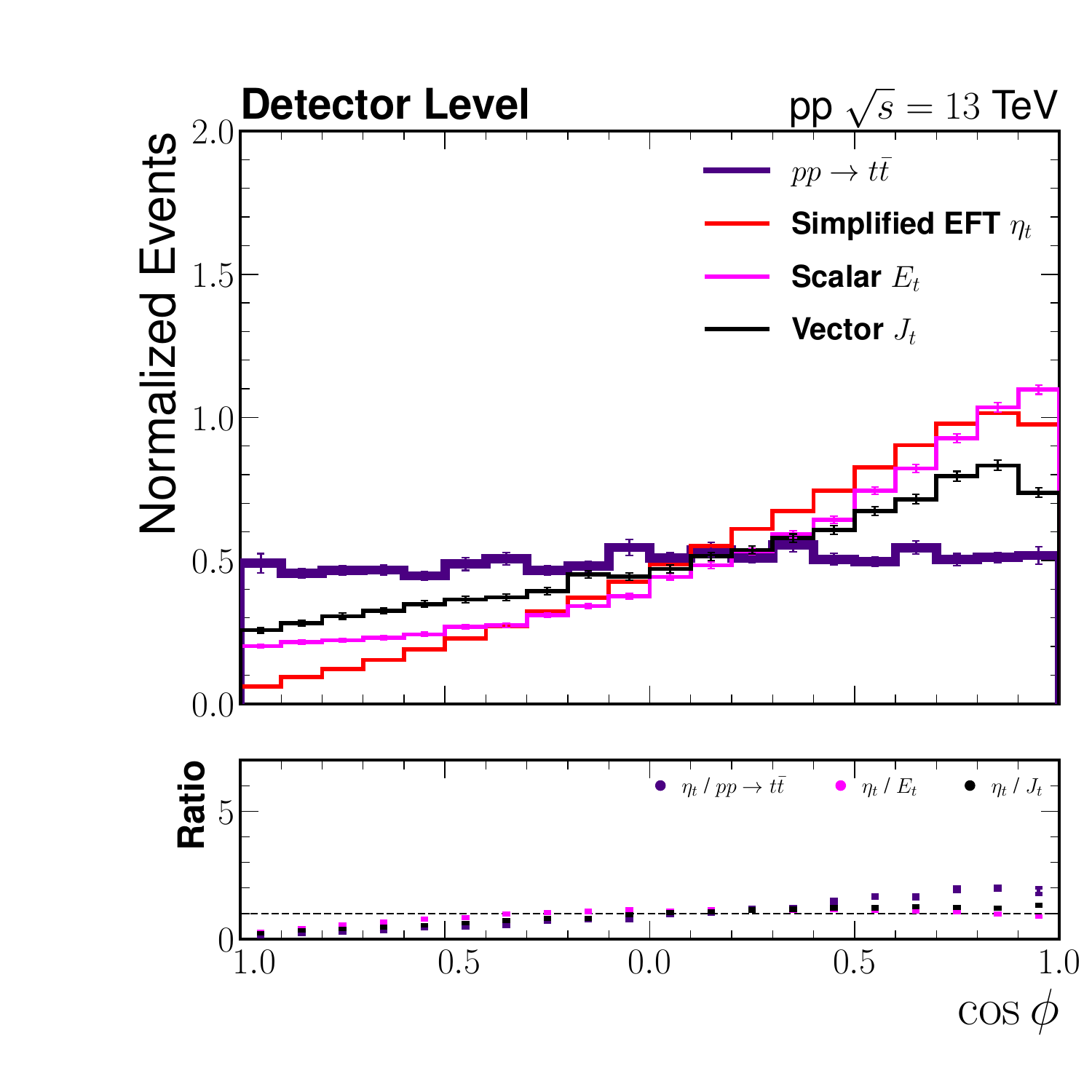}
    \end{subfigure}\hfill
    \begin{subfigure}[t]{0.32\textwidth}
        \includegraphics[width=\linewidth]{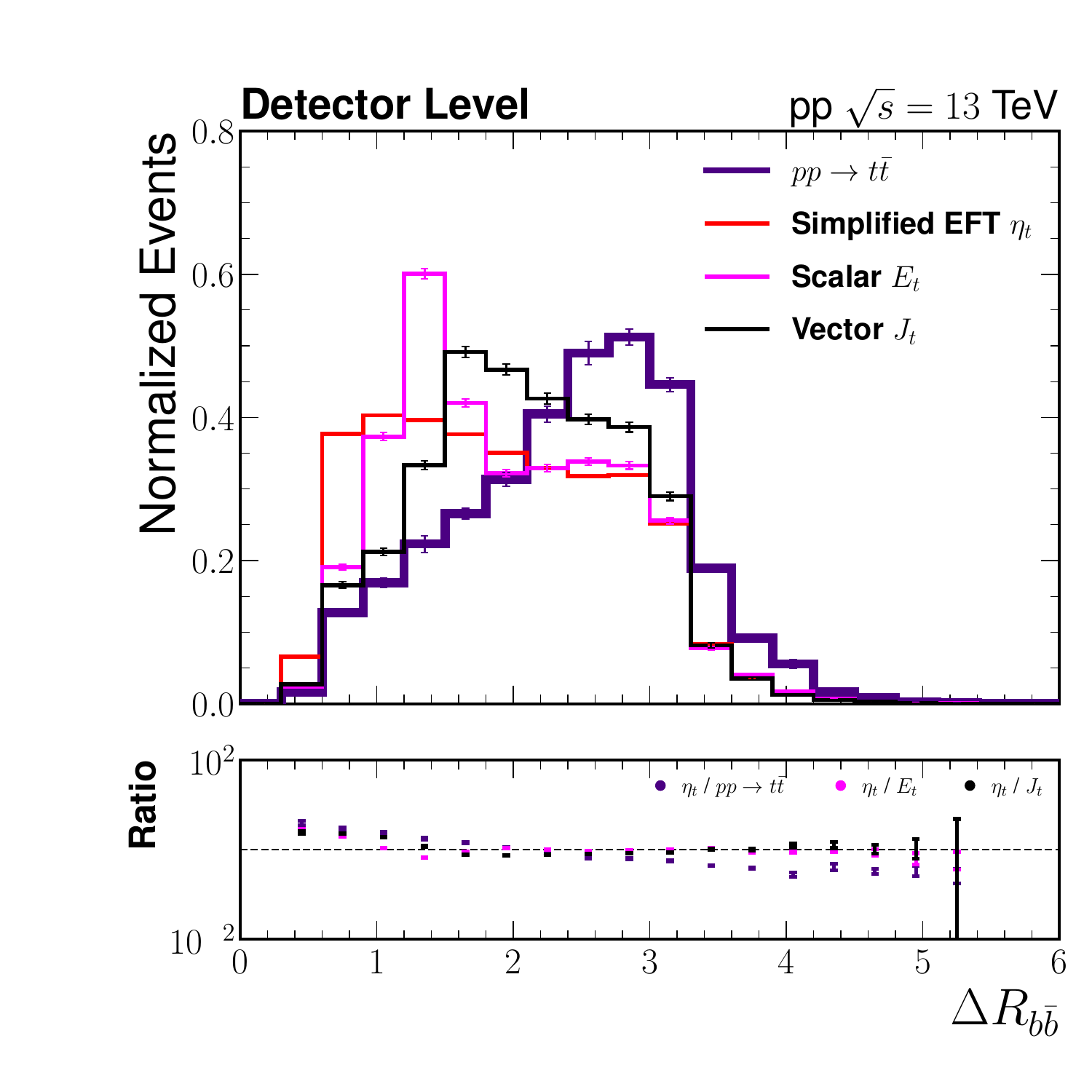}
    \end{subfigure}

    \caption{Detector-level distributions for Simplified EFT
    \(\eta_t\), NRQCD \(E_t\), NRQCD \(J_t\) and
    inclusive \(t\bar t\). From left to right:
    \(m_{\ell\ell}\), \(\cos\!\left(\phi_{\ell\ell}^{t\bar t\,\mathrm{RF}}\right)\),
    \(m_{t\bar t}\) (top);
    \(\Delta R_{\ell\ell}\), \(\cos\phi_{\mathrm{lab}}\),
    \(p_T(t)\) (middle);
    \(\Delta R_{t\bar t}\), \(\cos\phi\),
    \(\Delta R_{b\bar b}\) (bottom).}
    \label{fig:training_top9}
\end{figure*}

\noindent Figure~\ref{fig:training_top9} compares selected
distributions after applying the same selection and
reconstruction to all samples. For channels without
intermediate top quarks, reconstructed top quantities
describe the imposed \(t\bar t\) hypothesis.\\

\begin{figure*}[!ht]
    \centering

    \begin{subfigure}[t]{0.49\textwidth}
        \centering
        \includegraphics[width=\linewidth]{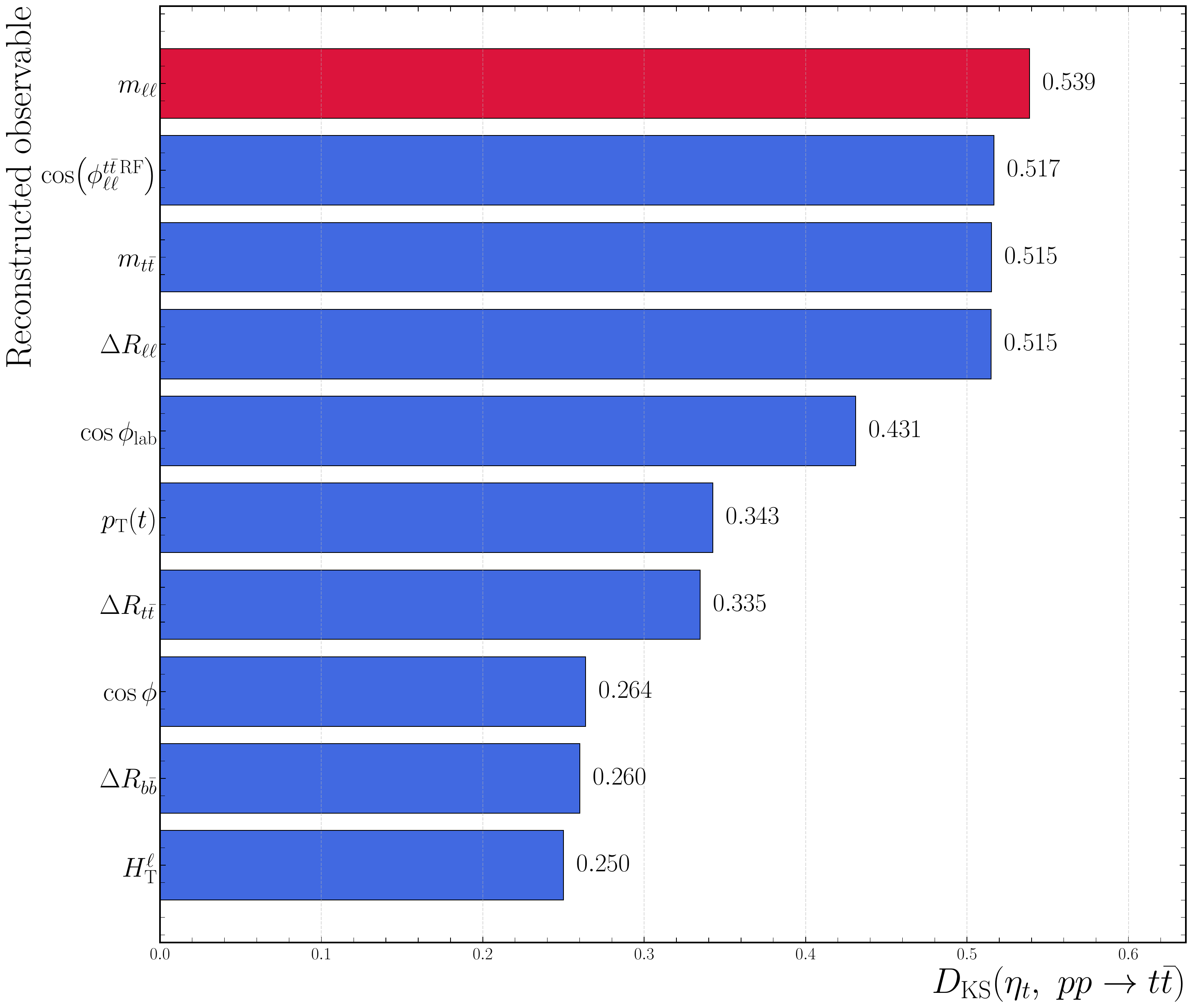}
    \end{subfigure}\hfill
    \begin{subfigure}[t]{0.49\textwidth}
        \centering
        \includegraphics[width=\linewidth]{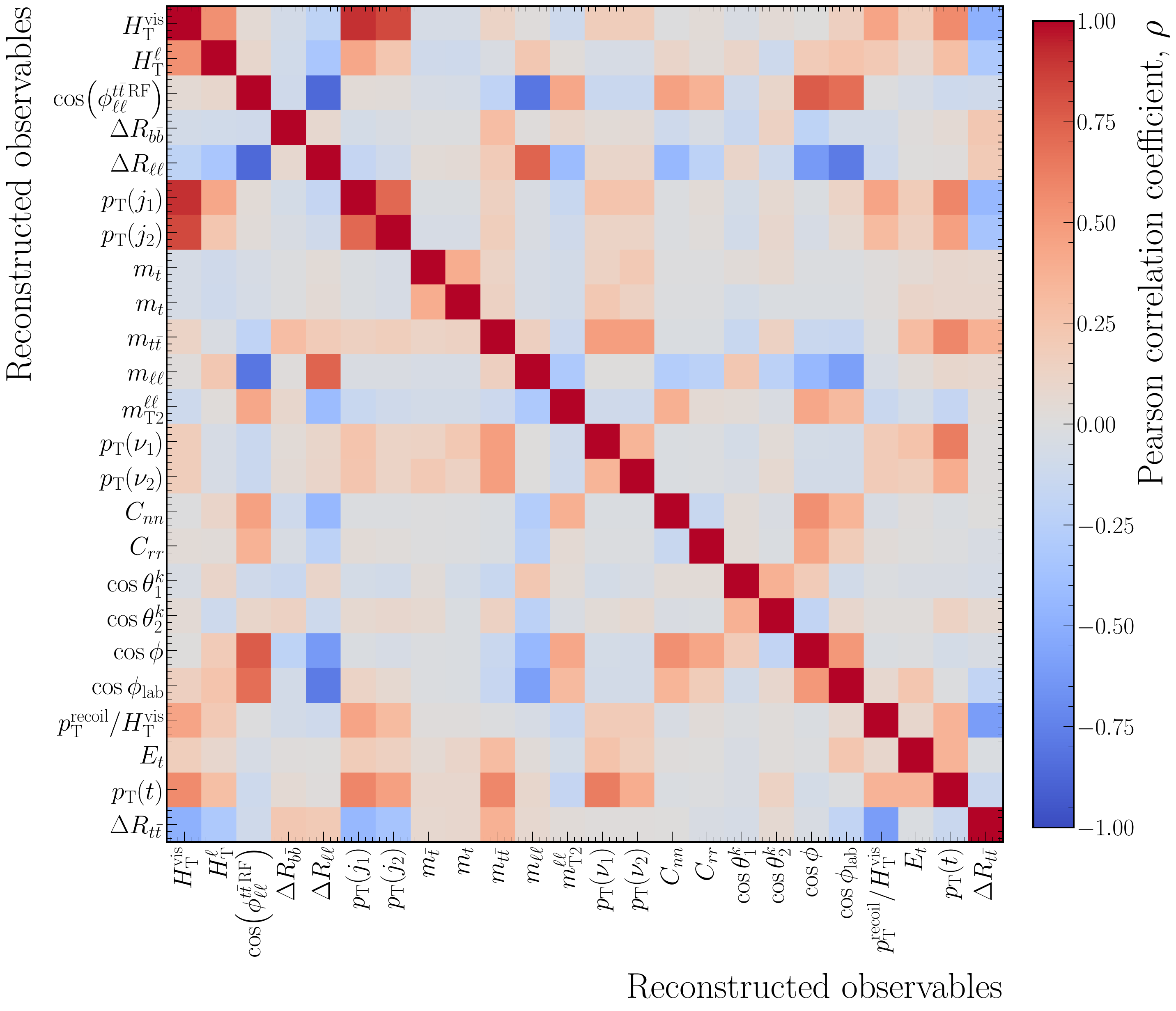}
    \end{subfigure}

    \caption{Ten largest KS statistics among the 24 inputs
    for Simplified EFT \(\eta_t\) versus \(t\bar t\) (left),
    and the Simplified EFT \(\eta_t\) Pearson correlation
    matrix (right).}
    \label{fig:det_ks_corr}
\end{figure*}

\noindent Figure~\ref{fig:det_ks_corr} summarizes
signal--background separation using the
Kolmogorov--Smirnov (KS) statistic and Pearson correlations
among the observables in the Simplified EFT \(\eta_t\) sample.
Both summaries are unweighted, with finite values selected
separately for each observable pair in the correlation matrix.

\section{Neural-network discrimination and expected sensitivity}
\label{sec:nn_classification}

\noindent A neural network (NN) and an XGBoost classifier~\cite{chen2016xgboost}
combined the 24 observables in Table~\ref{tab:det_observables24}
to discriminate Simplified EFT \(\eta_t\) signal from inclusive
\(pp\to t\bar t X\) background. Both assigned scores between
zero and one, with larger values indicating more signal-like
events. Generator weights and reconstruction bookkeeping
quantities were excluded from the inputs.\\

\noindent The NN was implemented in TensorFlow~\cite{tensorflow2015-whitepaper} with
two fully connected hidden layers of 64 and 32 neurons,
LeakyReLU activations, batch normalization, a dropout fraction
of 0.45 and a sigmoid output. Regularization used an \(L_2\)
coefficient of \(3\times10^{-4}\) and Gaussian input noise
with standard deviation 0.02. Training employed AdamW with
an initial learning rate of \(3\times10^{-4}\), binary
cross-entropy with label smoothing of 0.01, and batches
of 1024 events. A maximum of 300 epochs was allowed, with
learning-rate reduction and early stopping based on the
validation AUC. The stopping patience was 15 epochs, and
the parameters giving the highest validation AUC were restored.\\

\noindent XGBoost used a binary logistic objective, a maximum
tree depth of two, a learning rate of 0.03 and up to 1000
boosting rounds. Each tree sampled 80\% of training events
and 90\% of input observables. The \(L_1\) and \(L_2\)
regularization coefficients were 0.10 and 10, respectively.
Early stopping used the validation AUC with a patience
of 60 rounds.\\

\noindent Stratified five-fold cross-validation was performed
on a balanced sample containing 50\% signal and 50\% background.
In each fold, approximately 68\% of the full sample was used
for training, 12\% for validation and 20\% for testing.
Missing inputs were replaced by training-subset medians.
NN inputs were additionally centred and scaled using
training medians and interquartile ranges, whereas XGBoost
inputs were not rescaled. These transformations were applied
unchanged to validation and test subsets. Training was unweighted.\\

\begin{figure*}[!ht]
    \centering

    \begin{subfigure}[t]{0.48\textwidth}
        \centering
        \includegraphics[width=\linewidth]
        {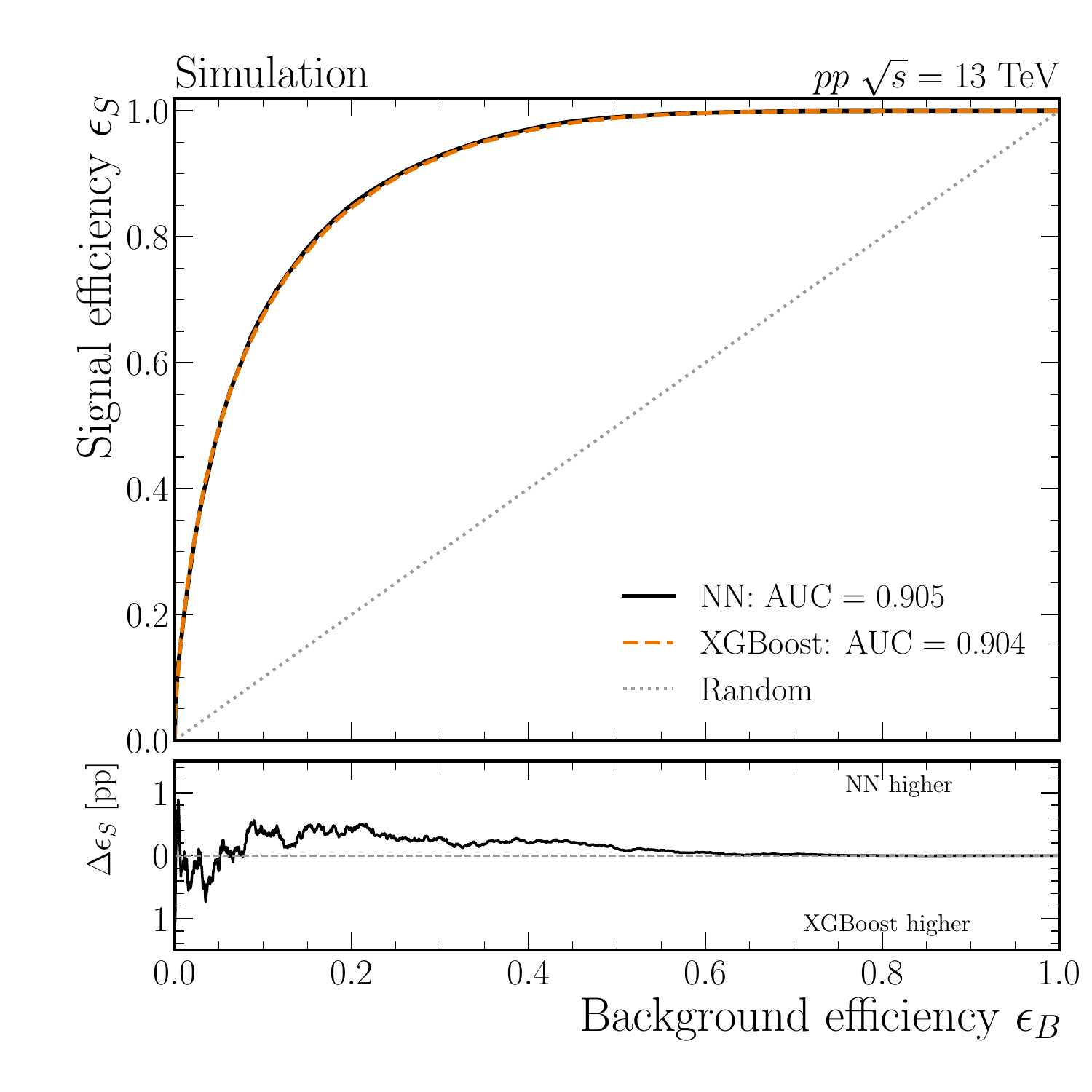}
    \end{subfigure}\hfill
    \begin{subfigure}[t]{0.48\textwidth}
        \centering
        \includegraphics[width=\linewidth]
        {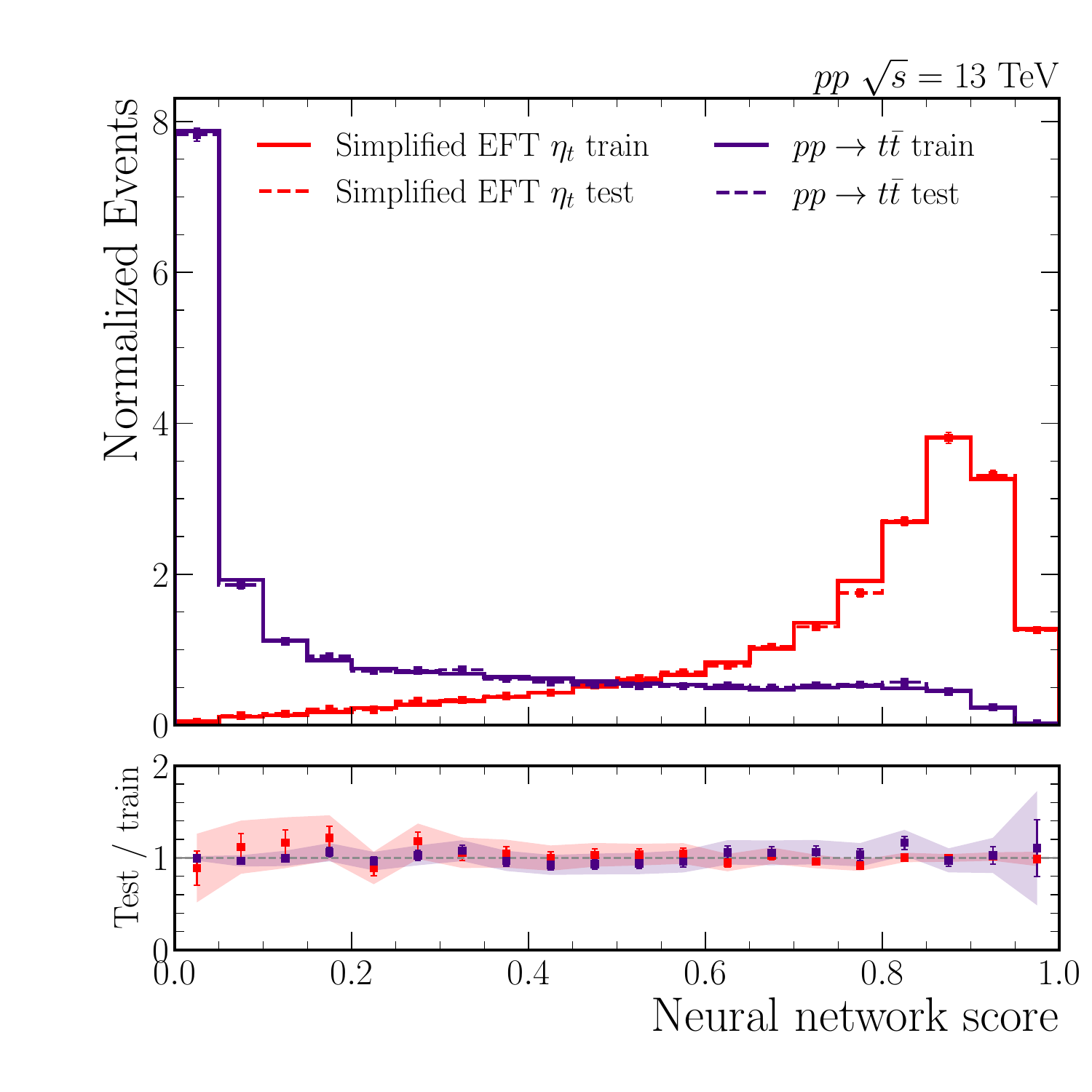}
    \end{subfigure}

    \caption{NN and XGBoost ROC curves, with their
    signal-acceptance difference at common background
    acceptance below (left). NN training and test score
    distributions, normalized separately to unit area,
    with their comparison below (right).
    All distributions are unweighted.}
    \label{fig:nn_training_results}
\end{figure*}

\noindent Figure~\ref{fig:nn_training_results} compares
classification performance and provides a training--test
diagnostic of possible overtraining.
Table~\ref{tab:nn_performance} reports the corresponding
metrics. The NN showed a small AUC advantage over XGBoost,
without an assigned statistical significance for the difference.\\

\begin{table*}[!ht]
    \centering
    \footnotesize
    \renewcommand{\arraystretch}{1.12}
    \setlength{\tabcolsep}{4pt}

    \begin{tabular*}{\textwidth}
        {@{\extracolsep{\fill}}lccc@{}}
        \toprule
        Metric
        & NN fold mean \(\pm\) SD
        & NN OOF
        & XGBoost OOF \\
        \midrule

        AUC
        & \(0.905372\pm0.000678\)
        & 0.905365
        & 0.904185 \\

        Accuracy
        & \(0.825469\pm0.000399\)
        & 0.825469
        & 0.823094 \\

        Precision
        & \(0.799067\pm0.001226\)
        & 0.799064
        & 0.801508 \\

        Signal efficiency (recall)
        & \(0.869614\pm0.001828\)
        & 0.869614
        & 0.858890 \\

        Specificity
        & \(0.781323\pm0.002077\)
        & 0.781323
        & 0.787298 \\

        \(F_1\)
        & \(0.832847\pm0.000408\)
        & 0.832848
        & 0.829208 \\

        \bottomrule
    \end{tabular*}

    \caption{Unweighted classification metrics.
    SD denotes the sample standard deviation across test
    folds, not a confidence interval. Out-of-fold (OOF)
    results pool predictions for events excluded from the
    corresponding model's training and validation subsets.
    Metrics other than AUC use a score threshold of 0.5;
    specificity is the background rejection efficiency.}
    \label{tab:nn_performance}
\end{table*}

\noindent The NN score was used to estimate sensitivity
at \(\sqrt{s}=13~\mathrm{TeV}\) and an integrated luminosity
of \(138~\mathrm{fb}^{-1}\). Event normalization used both
the adopted cross sections and their uncertainties.
The signal benchmark was
\(8.8^{+1.2}_{-1.4}~\mathrm{pb}\), taken from the CMS
simplified-toponium fit~\cite{CMS:2025kzt}.
For the background, the central cross section was
\(833.9~\mathrm{pb}\). Its uncertainties from renormalization
and factorization scales
(\({}^{+20.5}_{-30.0}~\mathrm{pb}\)),
PDFs and \(\alpha_s\) (\(\pm21.0~\mathrm{pb}\)),
and the top-quark mass
(\({}^{+23.2}_{-22.5}~\mathrm{pb}\))
were combined in quadrature separately for the upward
and downward variations, giving
\(833.9^{+37.4}_{-43.0}~\mathrm{pb}\).\\

\noindent Nominal yields were obtained from the luminosity,
central cross sections and adopted efficiencies of
\(0.636\%\) for background and \(5.25\%\) for signal,
giving approximately \(7.32\times10^5\) background
and \(6.38\times10^4\) signal events before any
NN-score requirement. Signed generator weights
were retained when constructing the NN-score templates,
using 10 equal-width bins between zero and one.
The cross-section uncertainties were propagated as
normalization variations of these templates.
The balanced class proportions used for training did not
determine the expected physical signal-to-background ratio.\\

\noindent A binned likelihood was constructed with Combine~\cite{combine},
with expected counts
\(\lambda_i(\mu,\boldsymbol\theta)
=\mu s_i(\boldsymbol\theta)+b_i(\boldsymbol\theta)\).
Here, \(i\) labels NN-score bins, \(s_i\) and \(b_i\) are
the signal and background yields, \(\mu\) is the signal
strength relative to the benchmark normalization, and
\(\boldsymbol\theta\) denotes the nuisance parameters.
The cross-section uncertainties entered the likelihood
as asymmetric lognormal normalization factors.
A 1.6\% luminosity uncertainty was correlated between
processes, and bin-wise simulation statistical uncertainties
were included. Additional detector and modeling shape
uncertainties were not considered.\\

\begin{figure*}[!ht]
    \centering

    \begin{subfigure}[t]{0.48\textwidth}
        \centering
        \includegraphics[width=\linewidth]
        {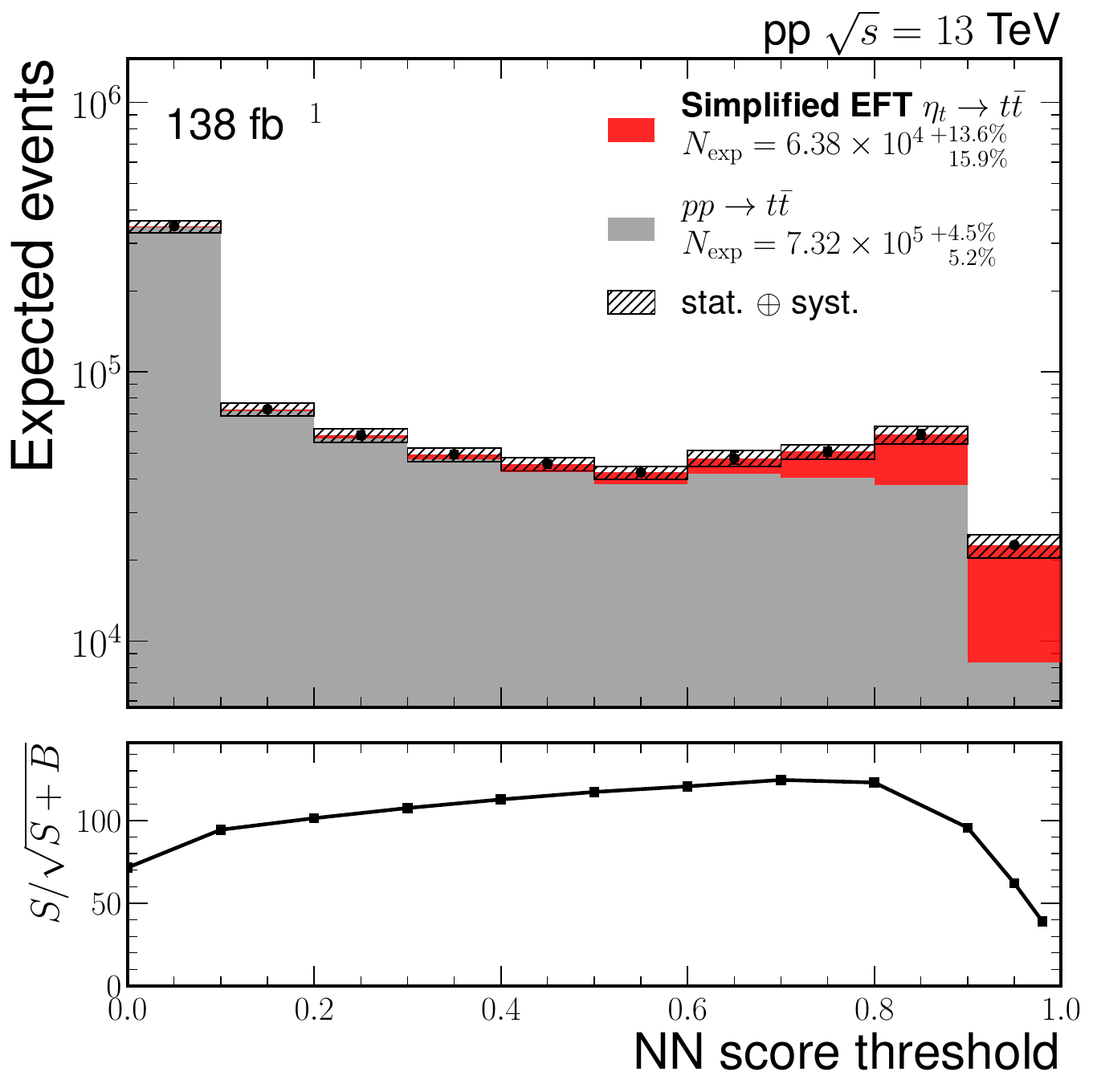}
    \end{subfigure}\hfill
    \begin{subfigure}[t]{0.48\textwidth}
        \centering
        \includegraphics[width=\linewidth]
        {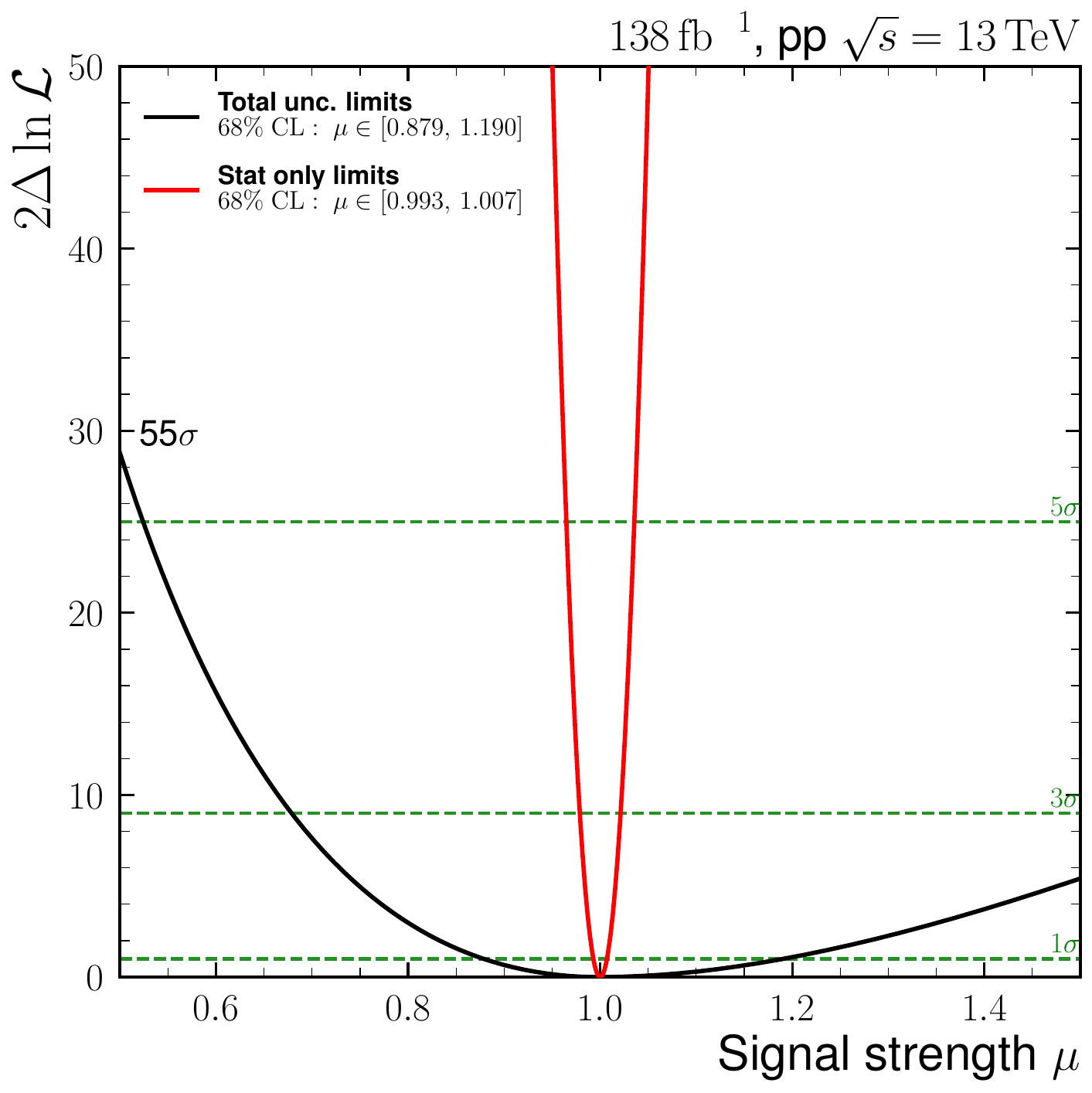}
    \end{subfigure}

    \caption{Expected NN-score yields (left), with the
    counting diagnostic \(S/\sqrt{S+B}\) below, where
    \(S\) and \(B\) denote yields retained above each score
    threshold. Asimov profile likelihood scans (right)
    with configured total uncertainties (black) and
    statistical uncertainties only (red).
    The counting diagnostic does not include nuisance profiling.}
    \label{fig:nn_expected_sensitivity}
\end{figure*}

\noindent An Asimov sample was generated from the nominal
signal-plus-background expectation at \(\mu=1\).
Figure~\ref{fig:nn_expected_sensitivity} compares the
profile likelihood scan with a statistical-only configuration.
The approximate 68\% intervals were \([0.879,\,1.190]\)
and \([0.993,\,1.007]\), respectively, illustrating the
impact of the configured uncertainties on the expected
signal-strength precision. For the nominal Asimov
significance calculation, the background cross-section
uncertainty was symmetrized by averaging its upward and
downward relative variations, giving approximately 4.82\%.

\section{Conclusions and outlook}

\noindent Motivated by recent confirmation of excess events identified at near to the $m_{t\bar{t}}$ threshold region by ATLAS and CMS experiments at the LHC, two independent theoretical models introducing the toponium state production following an Effective Field Theory (EFT) approach with $\eta_t$ production and a spin categorization by scalar $E_t$ and vector $J_t$ toponium states are implemented within Monte Carlo frameworks so their production and phenomenologies can be studied and directly compared against the perturbative $pp\rightarrow t\bar{t}$ baseline and recent Data measurements. Additionally, from its recent availability within the PYTHIA8 generator, a non-relativistic resummation effect (NR-QCD) providing independent Green's function and Coulomb potential corrections that do not deliver a toponium state but could add important corrections to the $t\bar{t}$ system dynamics beyond the perturbative calculation capabilities is also implemented.\\

\noindent The different processes are categorized following $t\bar{t}$ spin state formalism from polarisation vectors and correlation matrix according to standard procedure for spin correlation measurement. cos($\theta_i^{k, r, n}$) individual distributions do not show a special discrepancy between the different processes predictions and Data. Distributions with cos($\theta_1^{k, r, n}$)cos($\theta_2^{k, r, n}$) products show a clear distinct pattern for toponium EFT $\eta_t$ and scalar E$_t$ contributions with respect $pp\rightarrow t\bar{t}$ background and NR-QCD prospects. Differences between these toponium models with respect to Data reach factors of up to $\sim$5 at distributions tail areas. 
Fitted diagonal spin correlation C$_{kk}$, C$_{rr}$, C$_{nn}$ matrix terms show a perceptual increase in value ranging from 200\%-2330\% for toponium $t\bar{t}$ decays with respect the perturbative NLO baseline. Resummation NR-QCD relative magnitudes on matrix diagonal coefficient fits range from 0.6\% to 10.4\% and when combined with perturbative baseline improves the agreement with the $t\bar{t}$ spin correlation Data distributions by reducing its $\chi^2/N_\mathrm{dof}$ value by a factor of $\sim$21.5\%, delivering a fitted cross-section for the additional resummation effect within a range from 7.3 to 9.1 pb in consistency with Data excess beyond the perturbative baseline recently reported by the LHC experiments, excluding a contribution $>$ 50 pb.\\ 

\noindent Since the NR resummation $+$ NLO perturbative combination procedure is taking as reference spin correlation measurements that were $m_{t\bar{t}}$ inclusive, a large uncertainty over the resummation contribution percentage is obtained since it will only affect the near to the threshold $\sim$ 344 GeV mass events. It is then need to have a experimental reference of spin correlation measurements restricted to a sample of events satisfying for example $|m_{t\bar{t}} - 344 GeV|< 10$ so a more precise inspection and exclusion limits on the resummation effect contribution can be established, and confirm whether it can justify the observed excess events by their own.\\ 

\noindent A classifier that makes use of a total of 24 detector level observables, achieves an area under ROC curve (AUC) of approximately $0.905$ leading to signal to background significances around 100. The most discriminating observables between $\eta_t$ and background events are $m_{l^+l^-}$ invariant mass, $\Delta R(l^+l^-)$ angular distance and cos($\theta_{l^+l^-}$) in $t\bar{t}$ frame. The filter can substantially enhance the excess events especially if they favour the toponium scenario by rejecting at least 80\% of the background. Such performance translates into an expected significance over 5$\sigma$ favoring enough sensitivity to either observe or rule out the toponium signal with 13 TeV Run 2 Data.\\

\noindent These results indicate that efficient isolation  of observed $t\bar{t}$ excess events from background is feasible to search for toponium signal, with expected enough sensitivity for potential observation. In case the excess Data favours the toponium models, it will remain with significant yield after execution of neural network filters. Otherwise no discovery consistent with toponium model would be achieved. On the other hand, since a characterisation via spin correlation measurements is shown to give clear signs of discrepancy between toponium $\eta_t$ and scalar E$_t$ productions and $t\bar{t}$ background. If excess events favour the toponium scenario they would not only react positively to the selection filters, they would also show distinct patterns with much higher level of spin correlation fitted coefficients as shown above that would reinforce the bound state nature of the events.\\

\noindent Additionally, since non-relativistic effects from Green's function and Coulomb potential formalisms improve the Data modelling when added to the perturbative baseline, such resummation corrections should be added to the theoretical references in further experimental measurements. Preliminary comparisons shown in this work indicated that it is possible that the observed excess events could be justified by a NR-QCD effect only with no bound state production. In either case further analysis of LHC Run 2 and Run 3 Data would benefit from all the discussed alternatives, so a more precise interpretation of the nature of the recently observed Data excess is reached soon.\\


\section*{Data Availability Statement}
This article has no associated data or the data will not be deposited.

\section*{Code Availability Statement}
This article has no associated code or the code will not be deposited.

\section*{Open Access}
This article is distributed under the terms of the Creative Commons
Attribution License (CC-BY~4.0), which permits any use, distribution
and reproduction in any medium, provided the original author(s)
and source are credited.


\section*{Bibliography}
\bibliographystyle{iopart-num}
\bibliography{refs}

\end{document}